\pdfoutput=1

\documentclass[11pt,twoside,a4paper,cmspaper,final,collab]{cms-tdr}
\def\svnVersion{167929}\def\cmsCernNo{2012-240}\def\cmsCernDate{\today}\def\cmsMessage{Submitted to Physical Review D}

\begin{document}\cmsNoteHeader{SUS-12-003}

\hyphenation{had-ron-i-za-tion}
\hyphenation{cal-or-i-me-ter}
\hyphenation{de-vices}

\RCS$Revision: 148132 $
\RCS$HeadURL: svn+ssh://svn.cern.ch/reps/tdr2/papers/SUS-12-003/trunk/SUS-12-003.tex $
\RCS$Id: SUS-12-003.tex 148132 2012-09-18 11:22:30Z alverson $
\newlength\cmsFigWidth
\newlength\cmsFigWidthTwo
\newlength\cmsFigWidthThree
\newlength\cmsFigWidthFigEight
\newlength\cmsFigWidthFigSix
\ifthenelse{\boolean{cms@external}}{\setlength\cmsFigWidth{0.85\columnwidth}}{\setlength\cmsFigWidth{0.6\textwidth}}
\ifthenelse{\boolean{cms@external}}{\setlength\cmsFigWidthTwo{0.85\columnwidth}}{\setlength\cmsFigWidthTwo{0.45\textwidth}}
\ifthenelse{\boolean{cms@external}}{\setlength\cmsFigWidthThree{0.85\columnwidth}}{\setlength\cmsFigWidthThree{0.32\textwidth}}
\ifthenelse{\boolean{cms@external}}{\setlength\cmsFigWidthFigEight{0.68\columnwidth}}{\setlength\cmsFigWidthFigEight{0.32\textwidth}}
\ifthenelse{\boolean{cms@external}}{\setlength\cmsFigWidthFigSix{0.95\columnwidth}}{\setlength\cmsFigWidthFigSix{0.8\textwidth}}
\ifthenelse{\boolean{cms@external}}{\providecommand{\cmsLeft}{top}}{\providecommand{\cmsLeft}{left}}
\ifthenelse{\boolean{cms@external}}{\providecommand{\cmsRight}{bottom}}{\providecommand{\cmsRight}{right}}
\ifthenelse{\boolean{cms@external}}{%
\newcommand{\scotchrule[1]}{\centering\begin{ruledtabular}\begin{tabular}{#1}}
\newcommand{\donescotchrule}{\end{tabular}\end{ruledtabular}}
}{
\newcommand{\scotchrule[1]}{\centering\begin{tabular}{#1}\hline}
\newcommand{\donescotchrule}{\hline\end{tabular}}
}
\cmsNoteHeader{SUS-12-003} % This is over-written in the CMS environment: useful as preprint no. for export versions

\def\b {\cPqb}
\def\met {\MET}
\def\dphii {\ensuremath{\Delta\phi_i}\xspace}
\def\dphin {\ensuremath{\Delta\hat{\phi}_{\mathrm{min}}}\xspace}
\def\dphinres {\ensuremath{\sigma_{\Delta\phi}}\xspace}
\def\dphinresi {\ensuremath{\sigma_{\Delta\phi,i}}\xspace}
\def\dphimin {\ensuremath{\Delta\phi_{\mathrm{min}}}\xspace}
\def\ptres {\ensuremath{\sigma_{\pt}}\xspace}
\def\metti {\ensuremath{\sigma_{T_i}}\xspace}
\def\tbar {{\cPaqt}\xspace}
\def\nttsig {\ensuremath{N^{t\bar t}_{SIG}}\xspace}
\def\rsl {\ensuremath{R_{SL}}\xspace}
\def\nttsb  {\ensuremath{N^{t\bar t}_{SB}}\xspace}
\def\WJets{\ensuremath{{\PW} \rightarrow \ell \nu}\xspace}
\def\ZJets{\ensuremath{\Z/\gamma^* \rightarrow \ell^{+}\ell^{-}}\xspace}
\newcommand{\MHT}{\ensuremath{H_{\mathrm T}^{\mathrm{miss}}}\xspace}
\def\zmumu {\ensuremath{\Z \rightarrow {\Pgmp\Pgmm}}\xspace}
\def\znunu {\ensuremath{\Z \rightarrow \nu \overline{\nu}}\xspace}
\def\zee {\ensuremath{\mathrm{\Z \rightarrow e^+ e^-}}\xspace}
\def\zll {\ensuremath{\Z \rightarrow \ell^+ \ell^-}\xspace}
\def\minDeltaPhiN {\dphin}
\def\vsmtvs {\rule[-0.2cm]{0cm}{0.50cm}}
\def\smtvs {\rule[-0.3cm]{0cm}{0.75cm}}
\def\mtvs {\rule[-0.5cm]{0cm}{1.3cm}}
\def\dthetat{\ensuremath{\Delta\theta_{\mathrm{T}}}\xspace}
\def\rdphin{\ensuremath{N (\dphin \ge 4.0) / N (\dphin < 4.0)}\xspace}
\def\num-rdphin{\ensuremath{N (\dphin \ge 4.0) /}}
\def\den-rdphin{\ensuremath{N (\dphin < 4.0)}\xspace}
\def\Zinvisible{\znunu}
\def\tonebbbb{T1bbbb\xspace}
\def\ttwobb{T2bb\xspace}
\def\tonetttt{T1tttt\xspace}
\def\ttwott{T2tt\xspace}
\def\cls{CL$_{\rm s}$\xspace}
\def\mt2{M$_{\rm T2}$\xspace}
\def\transmass{\ensuremath{M_{\mathrm T}}\xspace}
\newcommand\WIDE{{Needs wide SB update.}}
\newcommand\LUMI{{Needs lumi update.}}
\def\lsim{\mathrel{\rlap{\lower4pt\hbox{\hskip1pt$\sim$}}
    \raise1pt\hbox{$<$}}}                % less than or approx. symbol
\def\gsim{\mathrel{\rlap{\lower4pt\hbox{\hskip1pt$\sim$}}
    \raise1pt\hbox{$>$}}}                % greater than or approx. symbol

\title{Search for supersymmetry in events with b-quark jets
and missing transverse energy in pp collisions at \texorpdfstring{7\TeV}{7 TeV}
}

\date{\today}

\abstract{
Results are presented from a search for physics beyond the standard model
based on events with large missing transverse energy,
at least three jets,
and at least one, two, or three b-quark jets.
The study is performed using a sample
of proton-proton collision data collected at $\sqrt{s}=7$\TeV
with the CMS detector at the LHC in 2011.
The integrated luminosity of the sample is 4.98\fbinv.
The observed number of events is found to be consistent with
the standard model expectation,
which is evaluated using control samples in the data.
The results are used to constrain cross sections for the
production of supersymmetric particles decaying to
b-quark-enriched final states
in the context of simplified model spectra.
}

\hypersetup{%
pdfauthor={CMS Collaboration},%
pdftitle={Search for supersymmetry in events with b-quark jets
and missing transverse energy in pp collisions at 7 TeV
},%
pdfsubject={CMS},%
pdfkeywords={CMS, physics, SUSY, btagging}}

\maketitle %maketitle comes after all the front information has been supplied

\section{Introduction}

Many extensions of the standard model (SM)
predict that events in high-energy proton-proton collisions
can contain large missing transverse energy (\met)
and multiple, high-transverse momentum (\pt) jets.
For example,
in R-parity-conserving~\cite{bib-rparity}
models of supersymmetry (SUSY)~\cite{bib-susy},
SUSY particles are created in pairs.
Each member of the pair initiates a decay chain
that terminates with the lightest SUSY particle (LSP)
and SM particles.
If the LSP only interacts weakly,
as in the case of a dark-matter candidate,
it escapes detection,
potentially yielding significant~\met.
Furthermore,
in some scenarios~\cite{Martin:1997ns},
the SUSY partners of the bottom and top quarks can
be relatively light,
leading to the enhanced production of events with
bottom-quark jets ({\cPqb}~jets).
Events of this type,
with {\cPqb}~jets and large~\met,
represent a distinctive topological signature
that is the subject of a search described in this paper.

We present a search for new physics (NP) in events with large \MET,
no isolated leptons,
three or more high-\pt jets,
and at least one, two, or three {\cPqb} jets.
The analysis is based on a sample of proton-proton collision data collected
at $\sqrt{s}=7$\TeV with the Compact Muon Solenoid (CMS) detector
at the CERN Large Hadron Collider (LHC) in 2011,
corresponding to an integrated luminosity of 4.98\fbinv.
Recent searches for NP in a similar final state are presented in
Refs.~\cite{bib-cms-ra1b,bib-atlas-susyb,bib-atlas-susyb-2,bib-cms-mt2-2011,bib-atlas-susyb-3}.
Our analysis is characterized by a strong reliance on
techniques that use control samples in data to evaluate the SM background.

The principal sources of the SM background are
events with top quarks,
comprising \ttbar pair and single-top-quark events,
events with a {\PW} or {\Z} boson accompanied by jets,
and non-top multijet events produced purely through
strong-interaction processes.
We hereafter refer to this last class of events
as ``QCD'' background.
Diboson ({$\PW\PW$}, {$\Z\Z$}, or {$\PW\Z$}) events represent
a smaller source of background.
For events with a {\PW} boson or a top quark,
significant \MET can arise if a {\PW} boson decays into a
charged lepton and a neutrino.
The neutrino provides a source of genuine~\MET.
Similarly, significant \MET can arise in events with a {\Z} boson
if the {\Z} boson decays to two neutrinos.
For QCD background events,
significant \MET arises primarily from the mismeasurement of jet~\pt.
A smaller component of the QCD background
arises from events with semileptonic decays of {\cPqb} and c quarks.

\begin{figure}[thb]
\begin{center}
\includegraphics[width=\cmsFigWidthTwo]{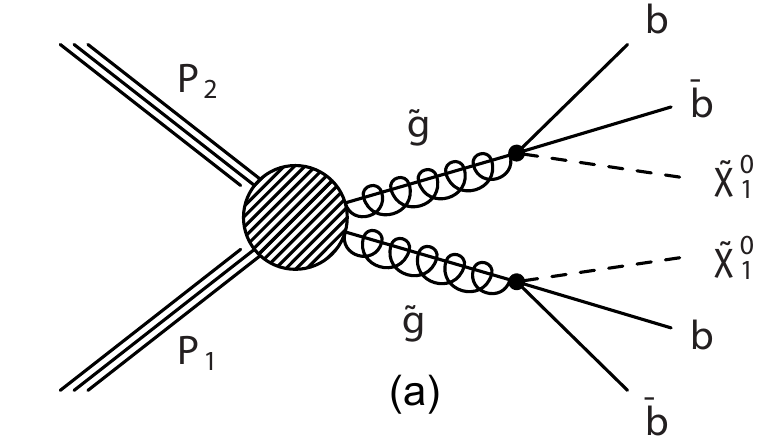}
\includegraphics[width=\cmsFigWidthTwo]{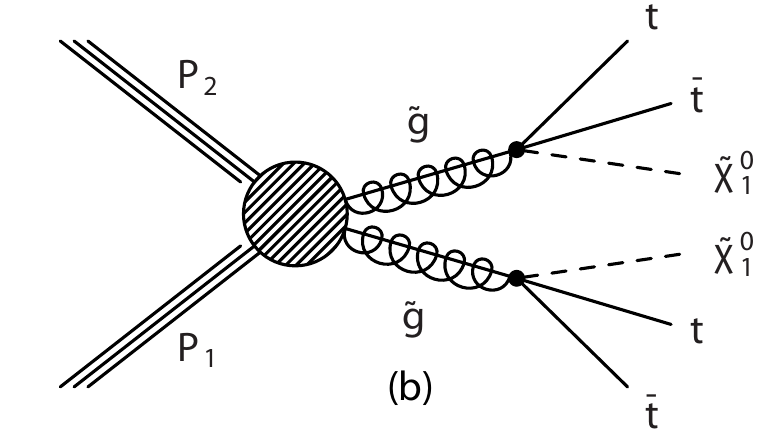}
\caption{
Event diagrams for the
(a)~\tonebbbb and (b)~\tonetttt simplified models.
}
\label{fig-sms}
\end{center}
\end{figure}

We interpret our results in the context of
simplified model spectra (SMS)~\cite{bib-sms-1,bib-sms-2,bib-sms-3,bib-sms-4},
which provide a general framework to characterize NP signatures.
They include only a few NP particles and focus on generic topologies.
We consider the SMS scenarios denoted \tonebbbb and \tonetttt.
Event diagrams are shown in Fig.~\ref{fig-sms}.
These two models are characterized by {\cPqb}-jet-enriched final states,
large jet multiplicities,
and large \met values,
making our analysis sensitive to their production.
For convenience,
we express SMS phenomenology using SUSY nomenclature.
In \tonebbbb (\tonetttt),
pair-produced gluinos~\sGlu each decay into two {\cPqb}-quark jets
({\cPqt}-quark jets) and the LSP,
taken to be the lightest neutralino~$\PSGcz$.
The LSP is assumed to escape detection,
leading to significant~\met.
If the SUSY partner of the bottom quark (top quark) is much lighter
than any other squark,
with the gluino yet lighter,
gluino decays are expected to be dominated by the three-body process
shown in Fig.~~\ref{fig-sms}(a) [Fig.~~\ref{fig-sms}(b)].

As benchmark NP scenarios,
we choose the \tonebbbb and \tonetttt models with
gluino mass $m_{\sGlu} = 925$\gev and LSP mass $m_{\mathrm{LSP}}=100$\gev,
with normalization to the
next-to-leading order (NLO) plus next-to-leading-logarithm (NLL) cross
section~\cite{bib-nlo-nll-01,bib-nlo-nll-02,bib-nlo-nll-03,bib-nlo-nll-04,bib-nlo-nll-05}.
These two benchmark models lie near the
boundary of our expected sensitivity.

In Sections~\ref{sec-detector}-\ref{sec-selection}
we describe the detector and event selection.
Section~\ref{sec-dphin} introduces the \dphin variable,
used in the evaluation of the QCD background.
Our techniques to evaluate the SM background from control samples
in data are presented in Section~\ref{sec-background}.
In Section~\ref{sec-likelihood}
we describe our analysis framework,
based on a likelihood method that
simultaneously determines the SM background and
tests the consistency of NP models with the data,
taking into account possible NP contamination of control
sample regions.
The interpretation of our results
is presented in Section~\ref{sec-simplified}.
A summary of the analysis is given in
Section~\ref{sec-summary}.

\section{Detector and trigger}
\label{sec-detector}

A detailed description of the CMS detector is
given elsewhere~\cite{bib-cms-detector}.
The CMS coordinate system is defined with the origin at the center
of the detector and the $z$ axis along the direction of the
counterclockwise beam.
The transverse plane is perpendicular to the beam axis,
with $\phi$ the azimuthal angle (measured in radians),
$\theta$ the polar angle,
and $\eta=-\ln[\tan(\theta/2)]$ the pseudorapidity.
A superconducting solenoid
provides an axial magnetic field of 3.8~T.
Within the field volume are a silicon pixel and strip tracker,
a crystal electromagnetic calorimeter,
and a brass-scintillator hadron calorimeter.
Muons are detected with gas-ionization chambers embedded in the
steel flux-return yoke outside the solenoid.
The tracker covers the region $|\eta|<2.5$
and the calorimeters $|\eta|<3.0$.
The region $3<|\eta|<5$ is instrumented with a forward calorimeter.
The near-hermeticity of the detector permits accurate
measurements of energy balance in the transverse plane.

The principal trigger used for the analysis selects events based on
the quantities \HT and \MHT, where \HT is the scalar sum of the
transverse energy of jets and \MHT the modulus of the corresponding
vector sum.
Due to increasing beam collision rates,
trigger conditions varied over the period of data collection.
The most stringent trigger requirements were $\HT>350$\gev and $\MHT>110$\gev.
The efficiency of the \HT component
for the final event selection
is measured from data to be 86\% (99\%)
for \HT values of 400\gev (500\gev).
The efficiency of the \MHT component is 98\% for $\MET>250$\gev.
Appropriate corrections are applied to account for trigger inefficiencies and
uncertainties in the various control and search regions of the analysis.

\section{Event selection}
\label{sec-selection}

Physics objects are defined using the particle flow
(PF) method~\cite{bib-cms-pf},
which is used to reconstruct and identify charged and neutral hadrons,
electrons (with associated bremsstrahlung photons), muons, tau leptons,
and photons,
using an optimized combination of information from CMS subdetectors.
The PF objects serve as input for jet reconstruction,
based on the anti-\kt algorithm~\cite{bib-antikt}
with distance parameter~0.5.
Jet corrections~\cite{bib-cms-ptres} are applied to account for
residual effects of non-uniform
detector response in both \pt and~$\eta$.
The missing transverse energy \MET is defined as the modulus
of the vector sum of the transverse momenta of all PF objects.
The \MET vector is the negative of the same vector sum.

The basic event selection criteria are as follows:
\begin{itemize}
\item at least one well-defined primary event vertex~\cite{CMS-PAS-TRK-10-005};
\item at least three jets with $\pt>50$\gev and $|\eta|<2.4$;
\item a lepton veto defined by requiring that there be
   no identified, isolated electron or muon candidate~\cite{CMS-PAS-EGM-10-004,bib-cms-muon}
   with $\pt>10$\gev;
   electron candidates are restricted to $|\eta|<2.5$
   and muon candidates to $|\eta|<2.4$;
\item $\dphin>4.0$, where the \dphin variable is described
   in Section~\ref{sec-dphin}.
\end{itemize}
Electrons and muons are considered isolated if the scalar sum of the
transverse momenta of
charged hadrons, photons, and neutral hadrons surrounding the lepton
within a cone of radius $\sqrt{(\Delta\eta)^2+(\Delta\phi)^2}=0.3$,
divided by the lepton \pt value itself,
is less than 0.20 for electrons and 0.15 for muons.

To identify {\cPqb} jets,
we use the combined-secondary-vertex algorithm
at the medium working point~\cite{bib-cms-btagging}.
This algorithm combines information about secondary vertices,
track impact parameters, and jet kinematics,
to separate {\cPqb} jets from
light-flavored-quark, charm-quark, and gluon jets.
To increase sensitivity to NP scenarios,
which often predict soft {\cPqb} jets,
we use all tagged {\cPqb} jets with $\pt>30\gev$.
The nominal {\cPqb}-jet-tagging efficiency is about 75\%
for jets with a \pt value of 100\gev,
as determined from a sample of {\cPqb}-jet-enriched
dijet events~\cite{bib-cms-btagging}
(for {\cPqb} jets with $\pt\approx 30$\gev,
this efficiency is about 60\%).
The corresponding misidentification rate is about 1.0\%.
We correct the simulated efficiencies for {\cPqb}-jet tagging and
misidentification to match the
efficiencies measured with control samples in the data.
The {\cPqb}-tagging correction factor depends slightly on the jet \pt
and has a typical value of 0.95.
The uncertainty on this correction factor varies from 0.03 to 0.07 for
{\cPqb} jets with \pt from 30 to 670\gev,
and is taken to be 0.13 for {\cPqb} jets with $\pt>670$\gev.

We define five signal regions,
which partially overlap,
to enhance sensitivity in different kinematic regimes.
The five regions
correspond to different minimum requirements on \HT, \MET,
and the number of {\cPqb} jets.
\HT is calculated using jets with $\pt>50$\gev and $|\eta|<2.4$.
The five regions,
denoted 1BL, 1BT, 2BL, 2BT, and 3B, are specified in Table~\ref{tab-sig-regions}
and were chosen without considering the
data to avoid possible bias.
The regions are selected based on expected signal and background
event yields in simulation,
to provide maximal sensitivity for discovery
of the NP scenarios considered in this paper or,
in the case of non-discovery,
to best set limits on their parameters.
Throughout this paper,
we use the generic designation ``SIG'' to refer to
any or all of these five signal regions.

\begin{table}[tb]
\topcaption{
The definition of the signal (SIG) regions.
The minimum requirements on \HT, \met,
and the number of tagged {\cPqb} jets $N_{\cPqb jets}$ are given.
The designations {1\b}, {2\b}, and {3\b} refer to the
minimum $N_{\cPqb jets}$ value,
while ``loose'' and ``tight'' refer to less restrictive and
more restrictive selection requirements, respectively,
for \HT and \met.
}
\scotchrule[lc|ccc]
\multicolumn{2}{c|}{Signal region}  & \HT [\GeVns{}] & \met [\GeVns{}]  & $N_\text{\cPqb jets}$ \\
\hline
1\b-loose  & 1BL &  $>400$      &  $>250$        & $\geq1$ \\
1\b-tight  & 1BT &  $>500$      &  $>500$        & $\geq1$ \\
2\b-loose  & 2BL &  $>400$      &  $>250$        & $\geq2$ \\
2\b-tight  & 2BT &  $>600$      &  $>300$        & $\geq2$ \\
3\b        & 3B  &  $>400$      &  $>250$        & $\geq3$ \\
\donescotchrule
\label{tab-sig-regions}
\end{table}

\begin{figure}[thbp]
\begin{center}
\includegraphics[width=\cmsFigWidthThree]{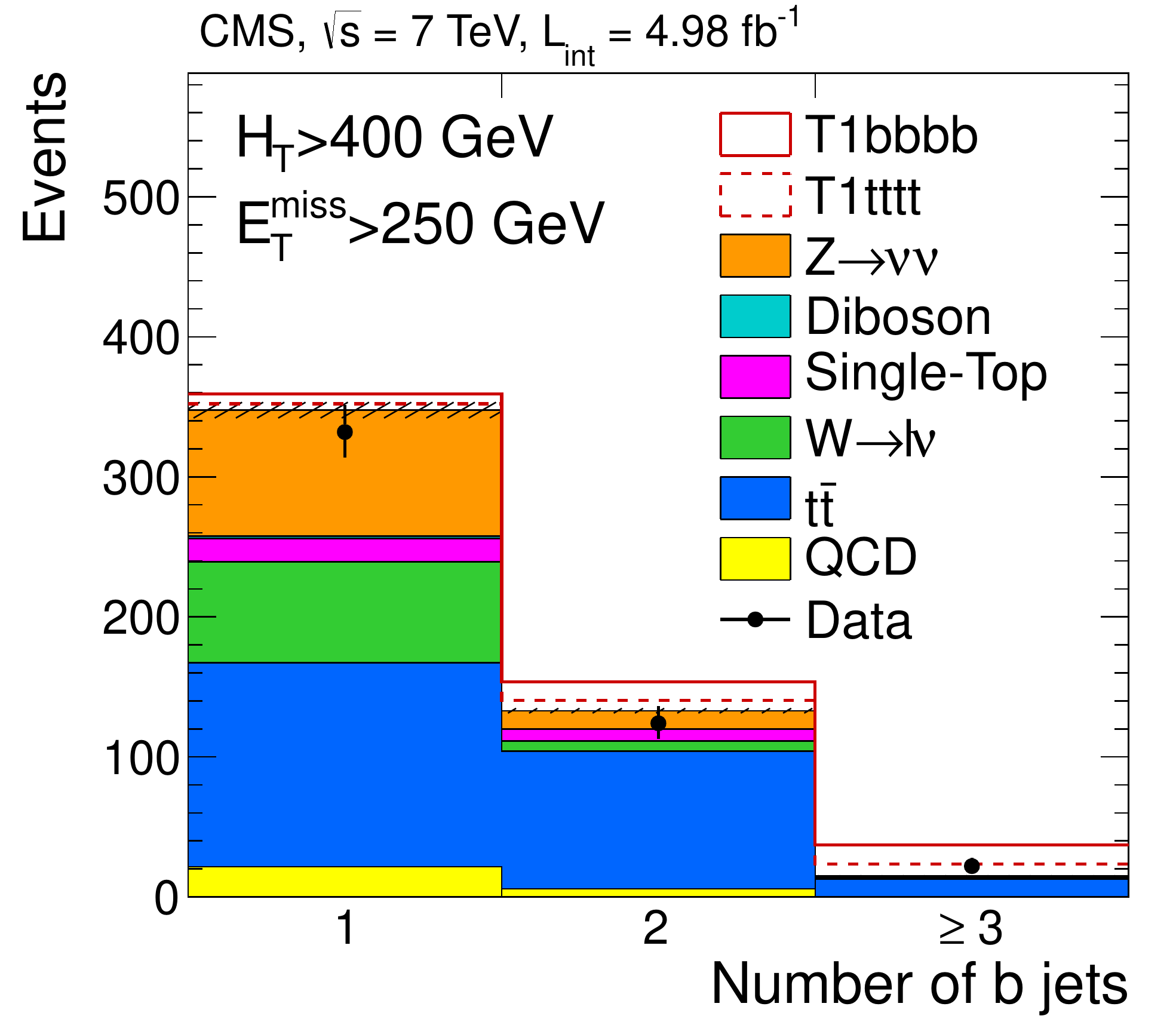}
\includegraphics[width=\cmsFigWidthThree]{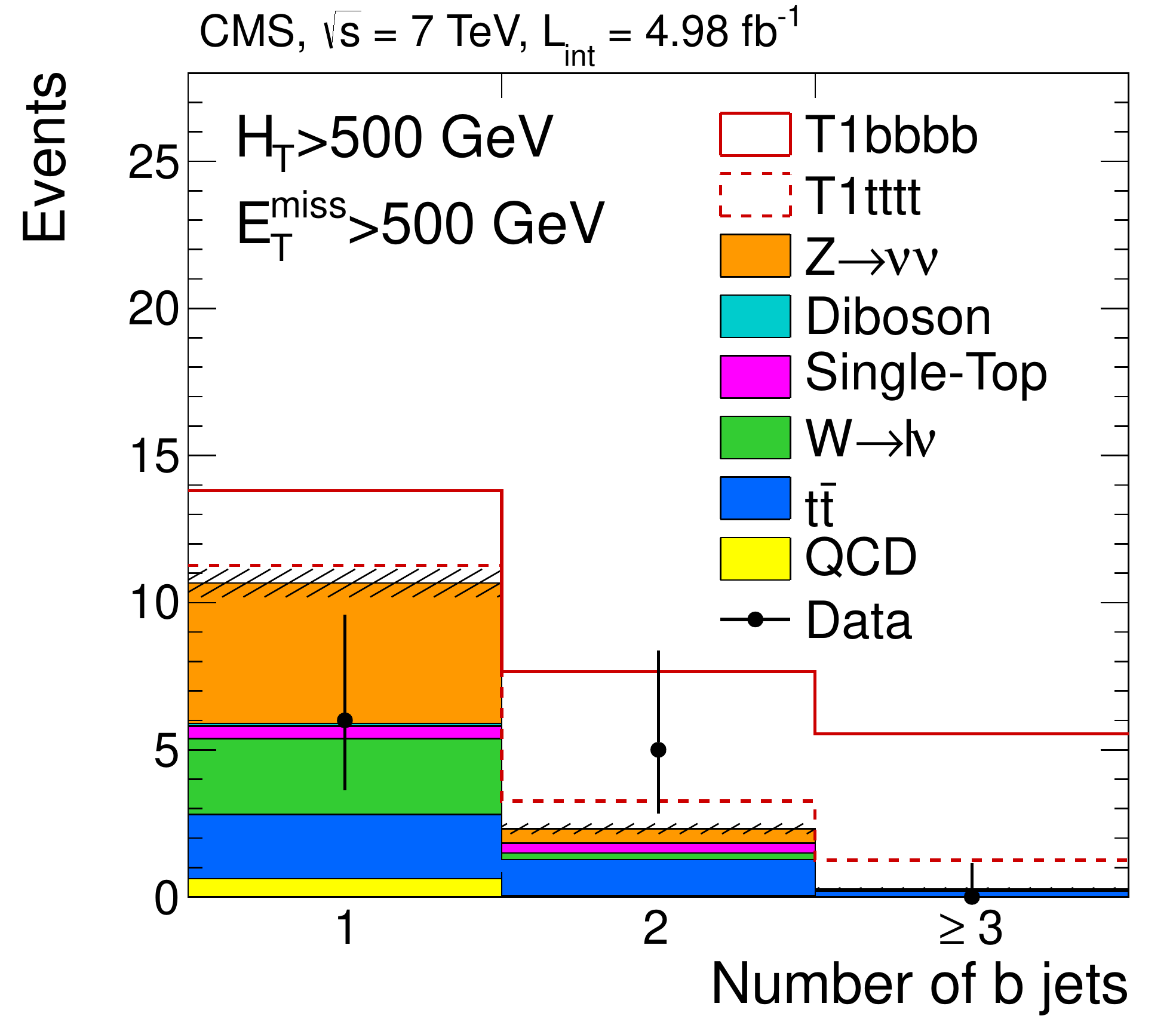}
\includegraphics[width=\cmsFigWidthThree]{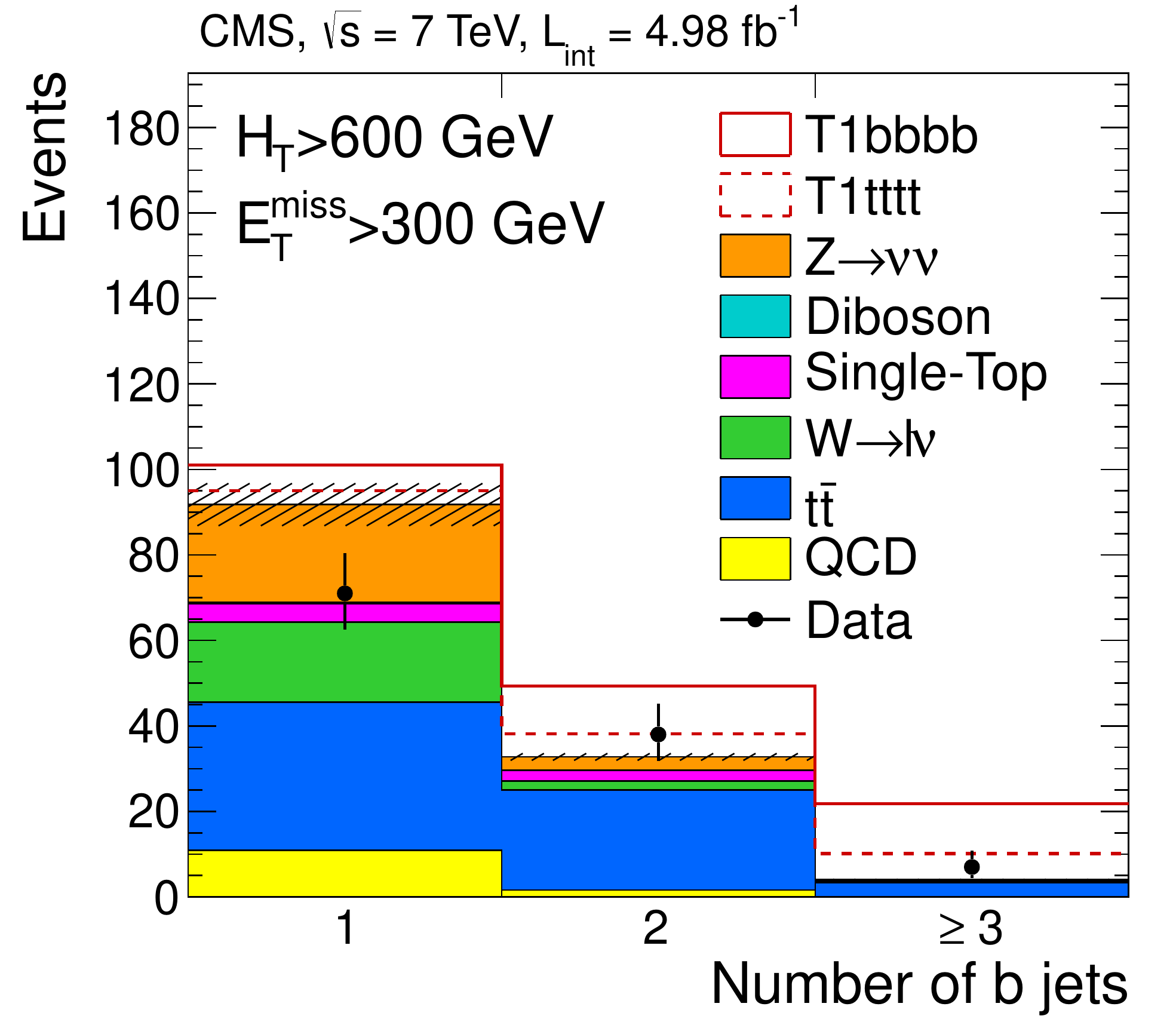}
\end{center}
\caption{
The distributions of the number of tagged {\cPqb} jets
for event samples selected
with the (a) 1BL, (b) 1BT, and (c)~2BT requirements,
except for the requirement on the number of {\cPqb} jets.
The hatched bands show the statistical uncertainty on the
total SM background prediction from simulation.
The open histograms show the expectations for the \tonebbbb (solid line)
and \tonetttt (dashed line) NP models,
both with $m_{\sGlu} = 925$\gev,
$m_{\mathrm{LSP}}=100$\gev,
and normalization to NLO+NLL.
}
\label{fig-nbjets}
\end{figure}

The distributions of the number of tagged {\cPqb} jets
for the 1BL, 1BT, and 2BT samples
(i.e., for the three different sets of
selection criteria on \HT and \met),
except without the requirement on the number of {\cPqb} jets,
are shown in Fig.~\ref{fig-nbjets}.
The results are presented in comparison with Monte Carlo (MC)
simulations of SM processes.
Results from the benchmark \tonebbbb and \tonetttt NP models
mentioned in the Introduction are also shown.
The simulated \ttbar, {\PW}+jets, and {\Z}+jets events are produced
at the parton level with the {\MADGRAPH}{5.1.1.0}~\cite{bib-madgraph} event generator.
Single-top-quark events are generated with the \POWHEG 301~\cite{bib-powheg} program.
The \PYTHIA 6.4.22 program~\cite{bib-pythia} is used to produce diboson and QCD events.
For all simulated samples,
\PYTHIA 6.4 is used to describe parton showering and hadronization.
All samples are generated using the CTEQ6~\cite{bib-cteq}
parton distribution functions.
The description of the detector response is implemented using
the \GEANTfour~\cite{bib-geant} program.
The \ttbar sample is normalized to the measured
cross section~\cite{bib-cms-ttbar}.
The other simulated samples are normalized
using the most accurate cross section calculations currently available,
which is generally NLO.
The jet energy resolution in the simulation is corrected to account for a small
discrepancy with respect to data~\cite{bib-cms-ptres}.
In addition,
the simulated samples are reweighted to describe the probability distribution
observed in data for overlapping pp collisions within a bunch crossing (``pileup'').

\begin{figure}[thbp]
\begin{center}
\includegraphics[width=\cmsFigWidthThree]{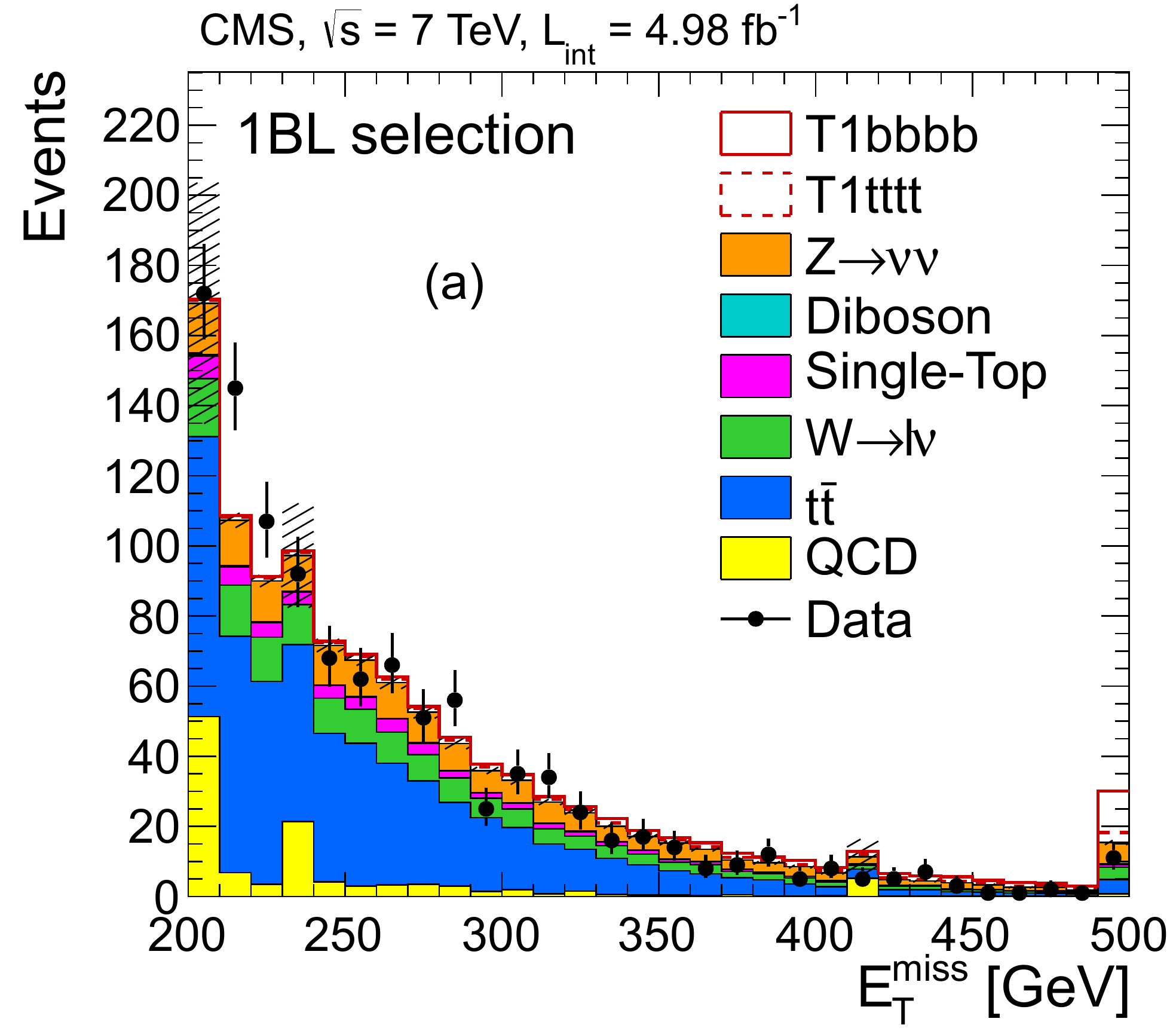}
\includegraphics[width=\cmsFigWidthThree]{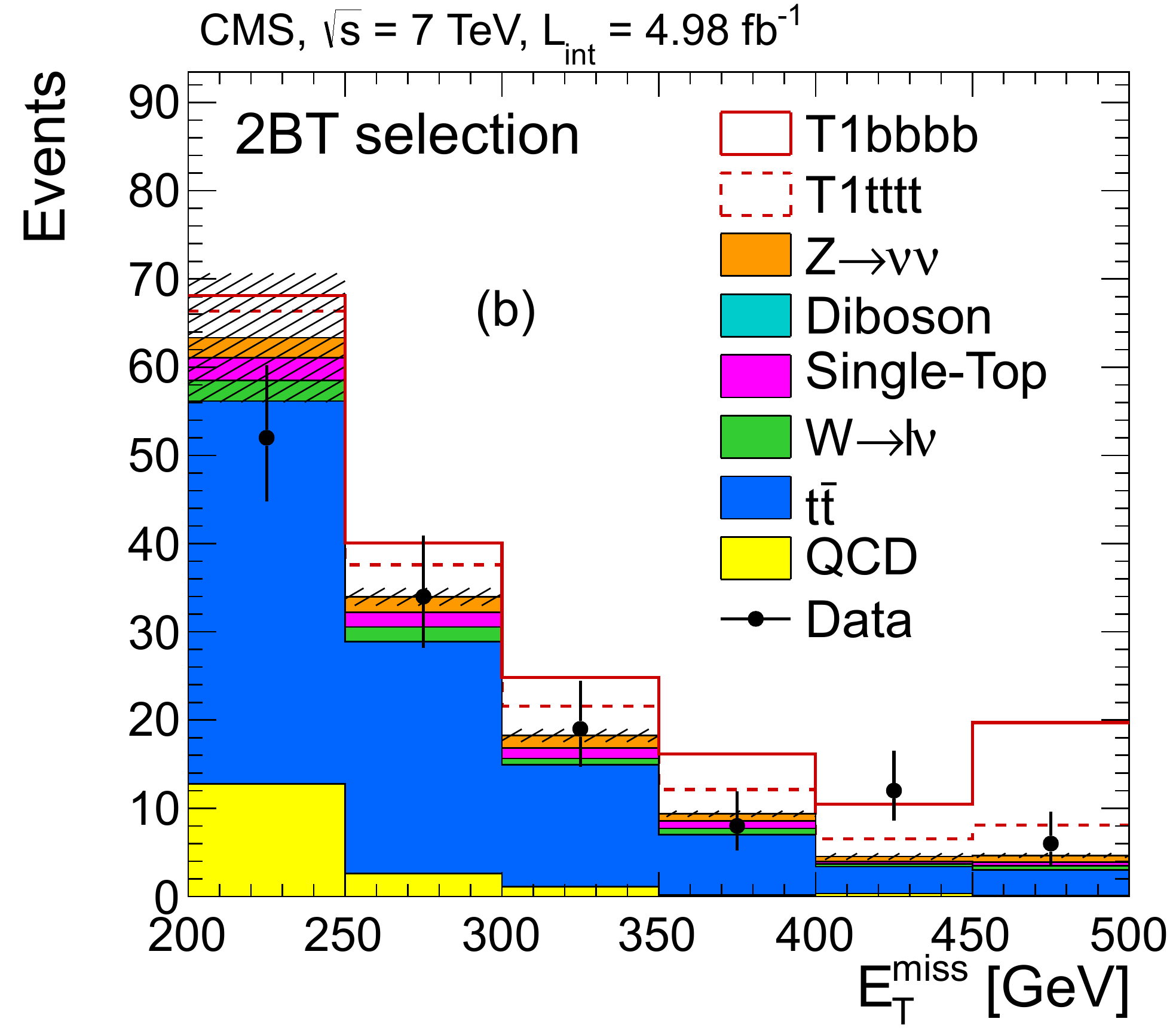}
\includegraphics[width=\cmsFigWidthThree]{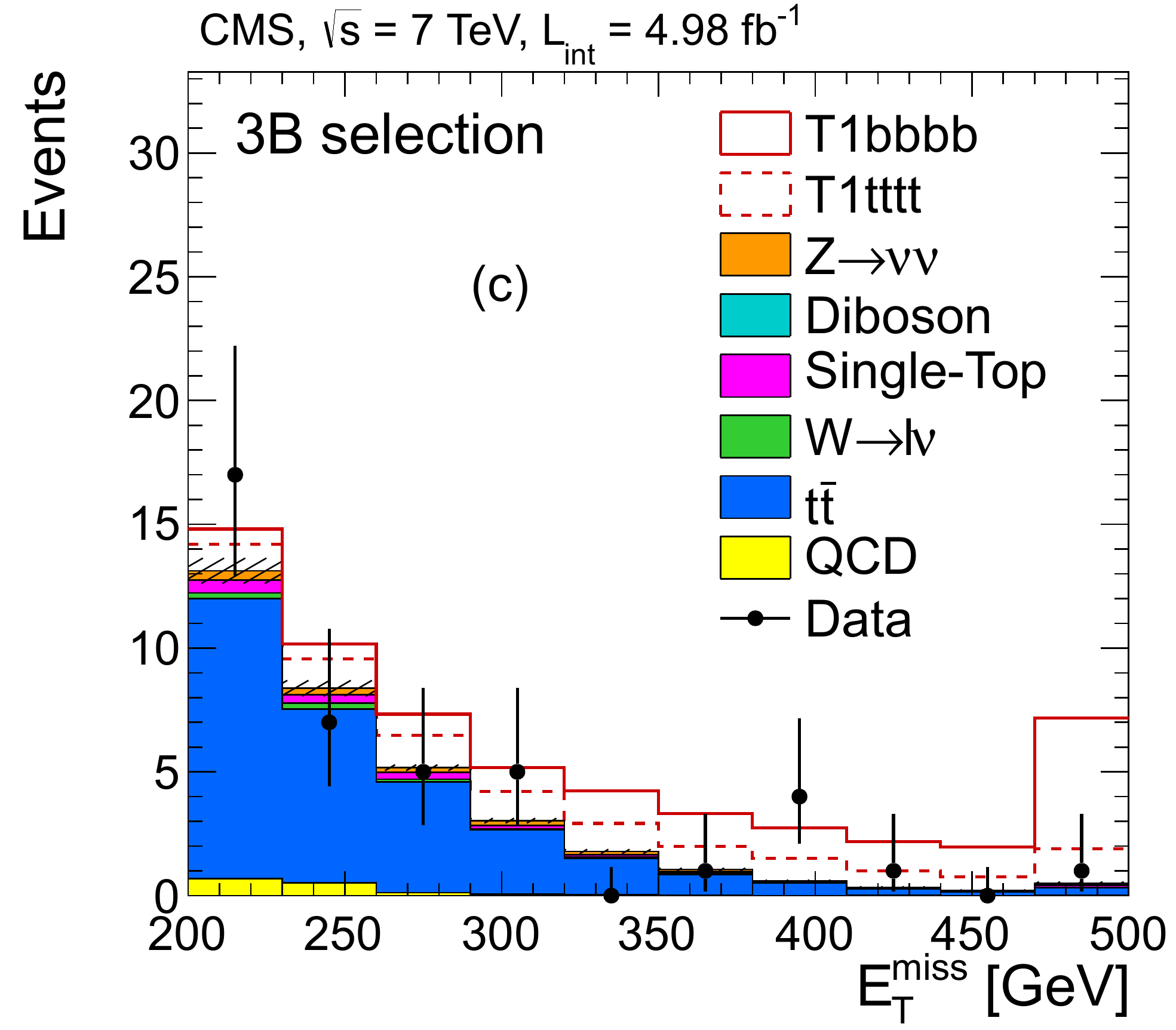}
\end{center}
\caption{
The distributions of \MET for event samples selected
with the (a) 1BL, (b) 2BT, and (c)~3B requirements,
except for the requirement on \MET.
The simulated spectra are normalized as in Fig.~\ref{fig-nbjets}.
The hatched bands show the statistical uncertainty on the
total SM background prediction from simulation.
The rightmost bin in all plots includes event overflow.
The open histograms show the expectations for the \tonebbbb (solid line)
and \tonetttt (dashed line) NP models,
both with $m_{\sGlu} = 925$\gev,
$m_{\mathrm{LSP}}=100$\gev,
and normalization to NLO+NLL.
}
\label{fig-met}
\end{figure}

\begin{table*}[bt]
\topcaption{
The number of data events and corresponding predictions from MC simulation
for the signal regions,
with normalization to 4.98\fbinv.
The uncertainties on the simulated results are statistical.
}
\scotchrule[l|ccccc]
              & 1BL  & 1BT & 2BL & 2BT & 3B \\
\hline
Data          & 478 & 11 & 146 & 45 & 22 \\
Total SM MC   & $496\pm7$   & $13.3\pm0.6$   & $148\pm2$     & $36.8\pm0.9$  & $15.0\pm0.2$ \\
\hline
\ttbar        & $257\pm2$    & $3.6\pm0.2$   & $111\pm1$     & $26.7\pm0.4$  & $12.6\pm0.2$ \\
Single-top quark   & $26.0\pm1.0$ & $0.8\pm0.2$   & $9.1\pm0.5$   & $2.7\pm0.3$   & $0.88\pm0.09$ \\
{\PW}+jets        & $80.0\pm1.0$ & $2.8\pm0.2$   & $7.7\pm0.3$   & $2.2\pm0.2$   & $0.38\pm0.05$ \\
\Zinvisible   & $104\pm2$ & $5.3\pm0.4$  & $13.8\pm0.7$  & $3.5\pm0.3$   & $0.80\pm0.10$ \\
Diboson       & $1.8\pm0.1$  & $0.10\pm0.02$ & $0.27\pm0.04$ & $0.05\pm0.02$ & $0.02\pm0.01$ \\
QCD           & $28.0\pm6.0$ & $0.70\pm0.20$ & $6.0\pm1.0$   & $1.7\pm0.6$   & $0.29\pm0.07$ \\
\donescotchrule
\label{tab-event-count}
\end{table*}

As examples illustrating the characteristics of events with at least one,
two, or three tagged {\b} jets,
the \MET distributions of events in the 1BL, 2BT, and 3B samples
are shown in Fig.~\ref{fig-met}.
The numbers of events in the different signal regions
are listed in Table~\ref{tab-event-count} for data and simulation.
The simulated results are for guidance only and are not used in the analysis.

\section{The \texorpdfstring{\dphin}{Delta Phi[min]} variable}
\label{sec-dphin}

\begin{figure}[thbp]
\begin{center}
\includegraphics[width=\cmsFigWidth]{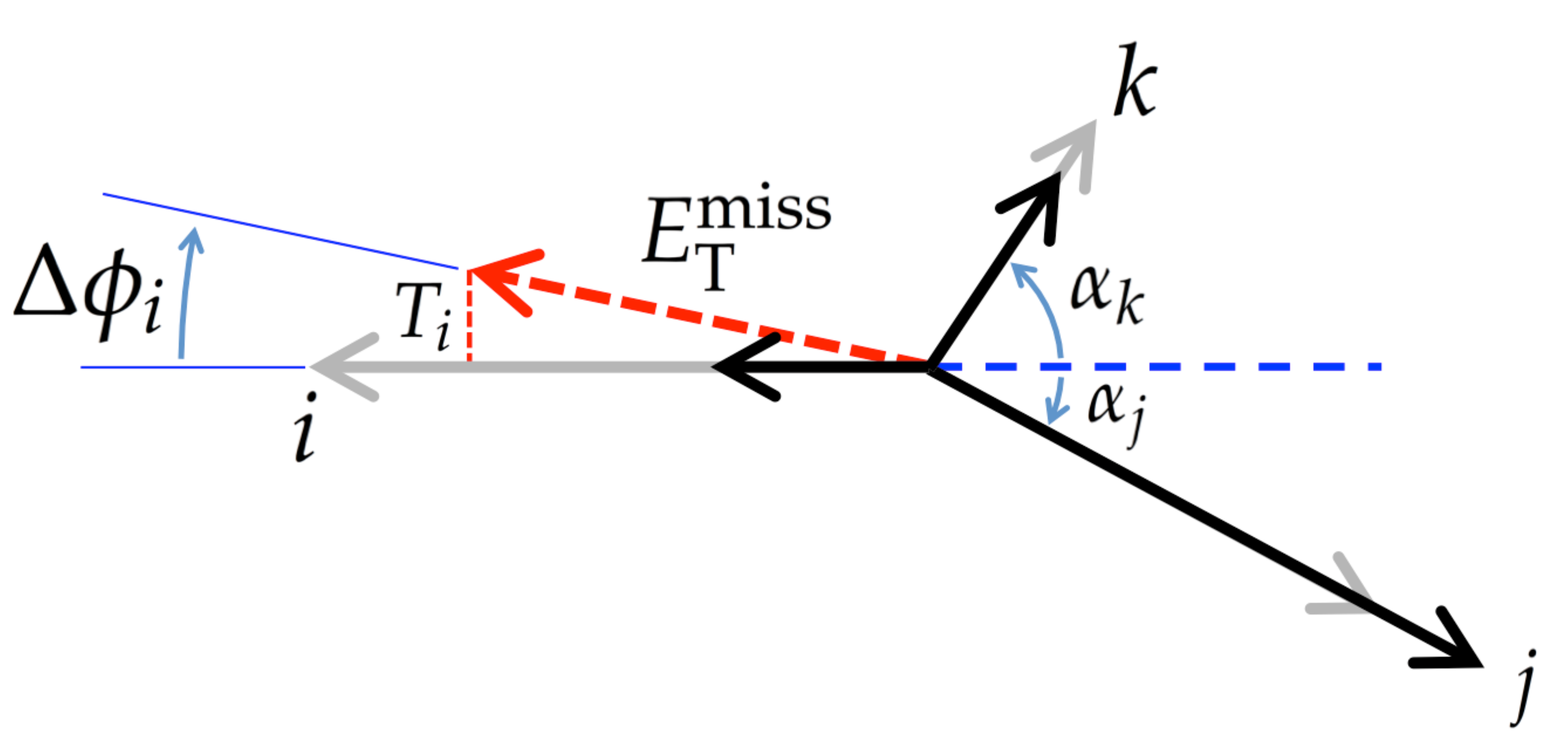}
\end{center}
\caption{
Illustration of variables used to calculate \dphin
for the case of an event with exactly three jets with $\pt>30$\gev.
The light-shaded (light gray) solid arrows show the true \pt values
of the three jets $i$, $j$, and $k$.
The dark-shaded (black) solid arrows show the reconstructed jet \pt values.
The angles of jets $j$ and $k$
with respect to the direction opposite to jet~$i$ are denoted $\alpha_j$ and $\alpha_k$.
The \MET for the event is shown by the dotted (red) arrow.
The component of \MET perpendicular to jet~$i$,
denoted $T_i$,
is shown by the dotted (red) line.
\metti is the uncertainty on~$T_i$.
$\Delta\phi_i$ is the angle between \MET and jet~$i$.
}
\label{fig-dphin-diagram}
\end{figure}

Our method to evaluate the QCD background is based on the \dphin variable.
This method presumes that most \MET in a QCD event
arises from the \pt mismeasurement of a single jet.

The \dphin variable is a modified version of the commonly used quantity
$\dphimin\equiv\min(\dphii)$ ($i=1,2,3$),
the minimum azimuthal opening angle
between the \MET vector and each of the three highest-\pt jets in an event.
Misreconstruction of a jet primarily affects the modulus of its transverse momentum
but not its direction.
Thus QCD background events are characterized by small values of~\dphimin.
The \dphimin variable is strongly correlated with \MET,
as discussed below.
This correlation undermines its utility for
the evaluation of the QCD background from data.
To reduce this correlation,
we divide the \dphii by their estimated resolutions \dphinresi
to obtain
$\dphin\equiv\min(\dphii/\dphinresi)$.

The resolution \dphinresi for jet~$i$ is evaluated
by considering the \pt resolution \ptres of
the other jets in the event.
The uncertainty \metti on the component of the \MET vector perpendicular
to jet~$i$ is found using
$\metti^2\equiv\sum_{n}(\sigma_{\pt,n}\sin\alpha_n)^2$,
where the sum is over all other jets in the event
with $\pt>30$~\gev
and $\alpha_n$ is the angle between jet $n$ and the direction opposite jet~$i$.
The situation is depicted in Fig.~\ref{fig-dphin-diagram} for
an event with exactly three jets with $\pt>30$~\gev.
Our estimate of the $\Delta\phi$ resolution is
$\dphinresi=\arctan(\metti/\MET)$.
[Note: $\arcsin(\metti/\MET)$ is technically more correct in this expression;
we use  $\arctan(\metti/\MET)$
because it is computationally more robust while being equivalent for the
small angles of interest here.]
For the jet \pt resolution,
it suffices to use the simple linear parametrization
$\ptres = 0.10\,\pt$~\cite{bib-cms-ptres}.

\begin{figure}[thbp]
  \begin{center}
   \includegraphics[width=\cmsFigWidthThree]{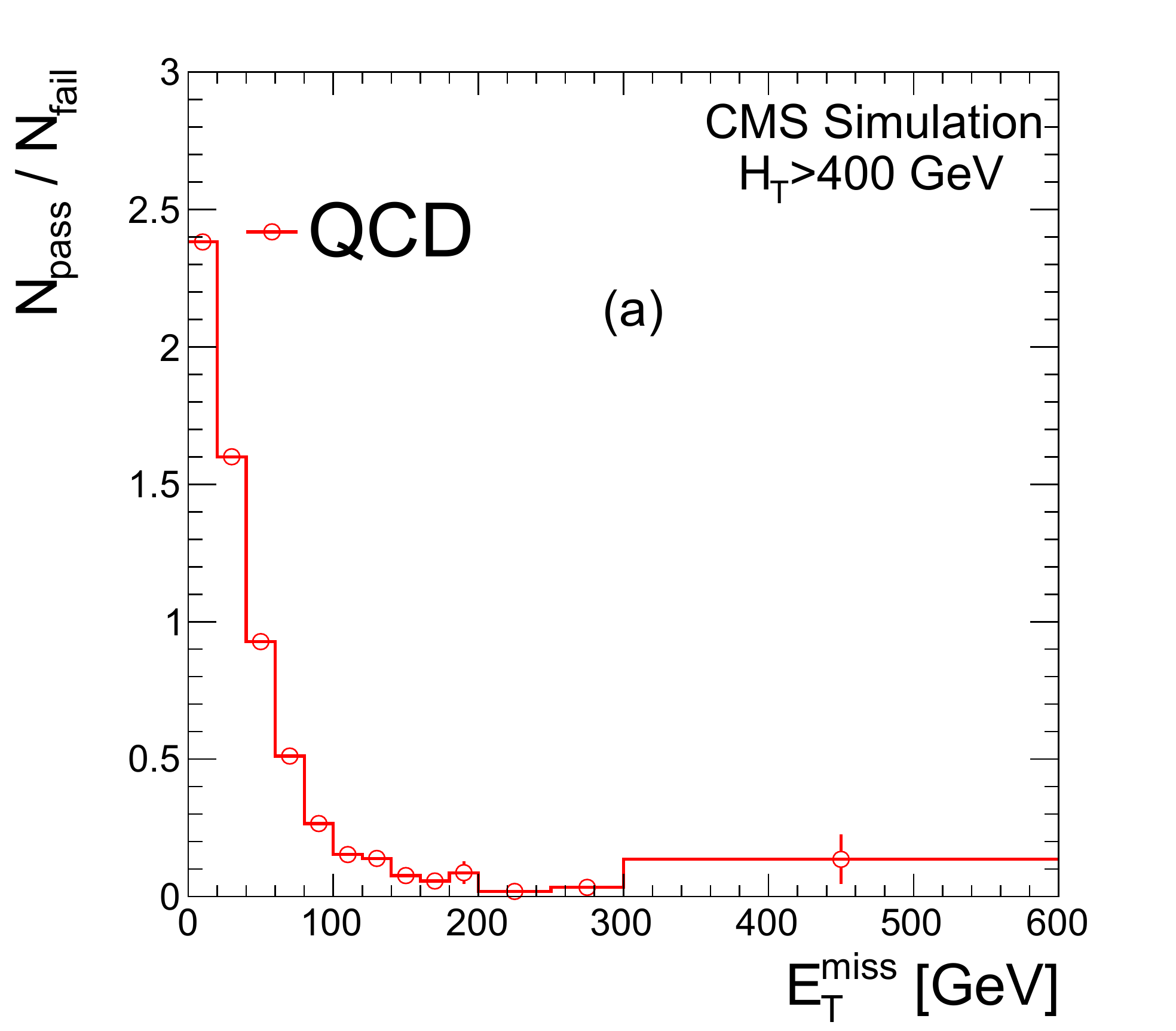}
   \includegraphics[width=\cmsFigWidthThree]{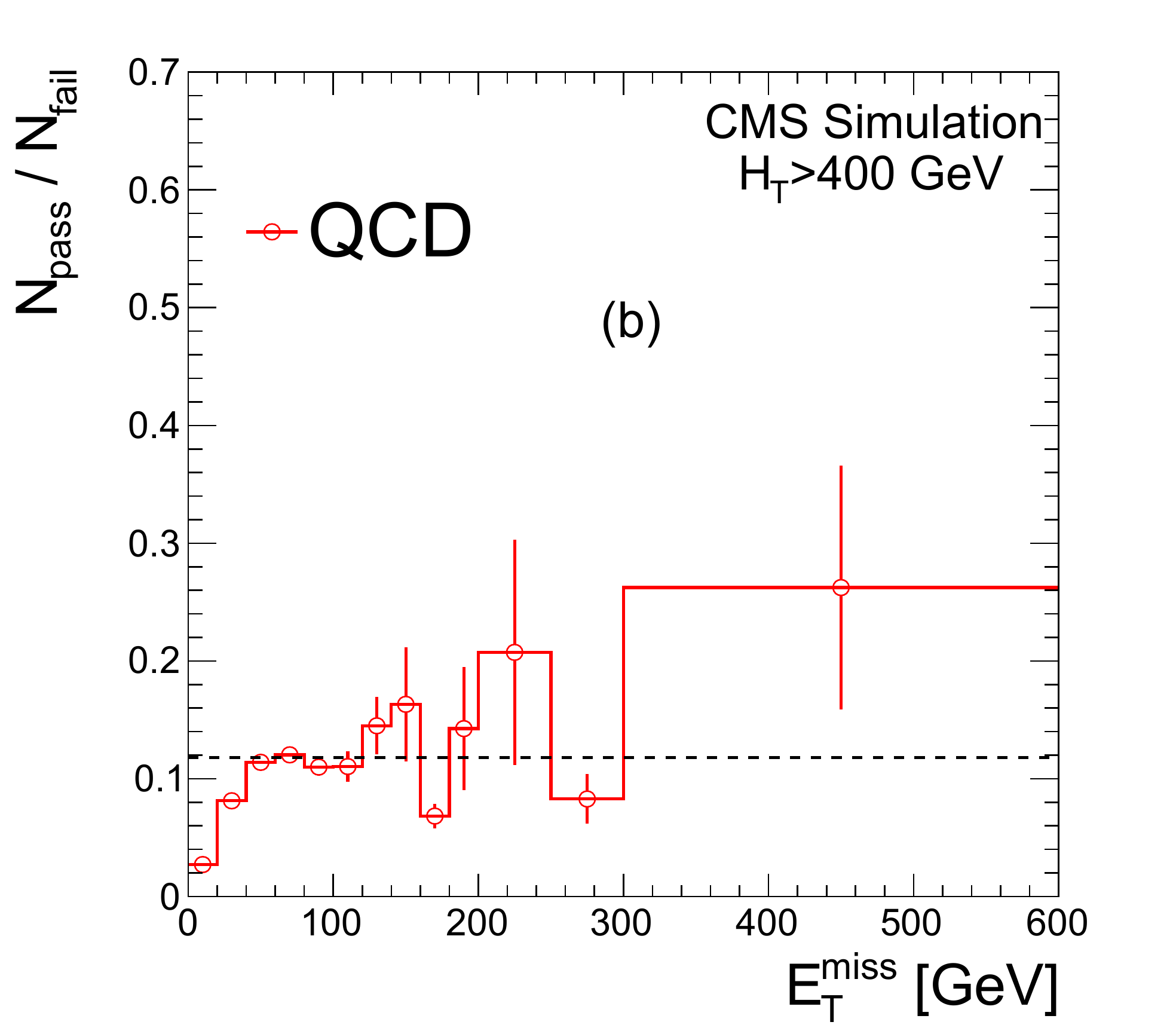}
   \includegraphics[width=\cmsFigWidthThree]{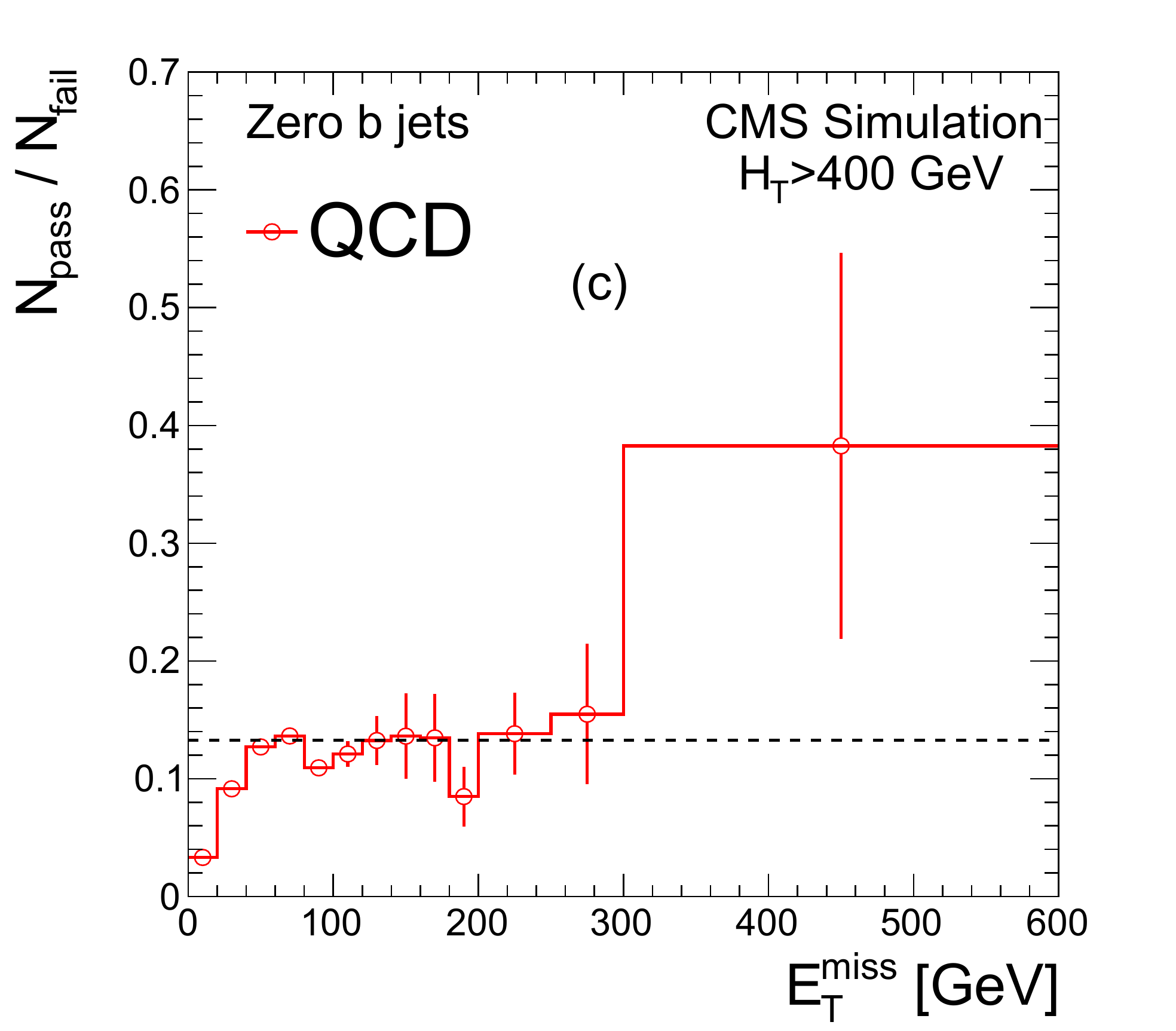}
     \caption{
QCD-simulation results:
(a) the ratio of the number of events that pass
the criterion $\dphimin\geq 0.3$ ($\mathrm{N_{pass}}$)
to the number that fail ($\mathrm{N_{fail}}$)
as a function of~\met,
for events selected with the 1BL requirements
except for those on~\dphin and~\MET;
(b)~The analogous ratio of events with
$\dphin\geq 4.0$ to those with $\dphin<4.0$;
and (c)~the same as (b) for events with zero {\cPqb} jets.
The QCD-background estimate is based on the relative flatness of the
distributions in (b) and (c) for $\met\gsim30\gev$,
as illustrated schematically by the dashed lines.
}
    \label{fig-dphi-comp2}
  \end{center}
\end{figure}

Figure~\ref{fig-dphi-comp2}(a)
shows the ratio of the number of events with $\dphimin>0.3$
to the number with $\dphimin<0.3$ as a function of \met,
for a simulated QCD sample selected with the 1BL requirements
except for those on \dphin and~\MET
($\dphimin>0.3$ or a similar criterion is commonly used to
reject QCD background,
see, \eg,
Refs.~\cite{bib-atlas-susyb,bib-cms-mt2-2011,bib-atlas-susyb-2,bib-atlas-susyb-3}).
The strong correlation between \dphimin and \MET is evident.
The corresponding result based on \dphin is shown in
Fig.~\ref{fig-dphi-comp2}(b).
For the latter figure we choose $\dphin=4.0$
in place of $\dphimin=0.3$,
which yields a similar selection efficiency.
For values of \MET greater than about 30\gev,
the distribution based on \dphin is seen to be
far less dependent on \MET than that based on \dphimin.
Figure~\ref{fig-dphi-comp2}(c) shows the result
corresponding to Fig.~\ref{fig-dphi-comp2}(b)
for events with zero tagged {\cPqb} jets.
Comparing Figs.~\ref{fig-dphi-comp2}(b) and~(c),
it is seen that the ratio \rdphin has an approximately
constant value of about 0.13 (for $\MET>30$\gev)
irrespective of the number of {\cPqb} jets.

\begin{figure}[thbp]
\centering
  \includegraphics[width=\cmsFigWidthThree]{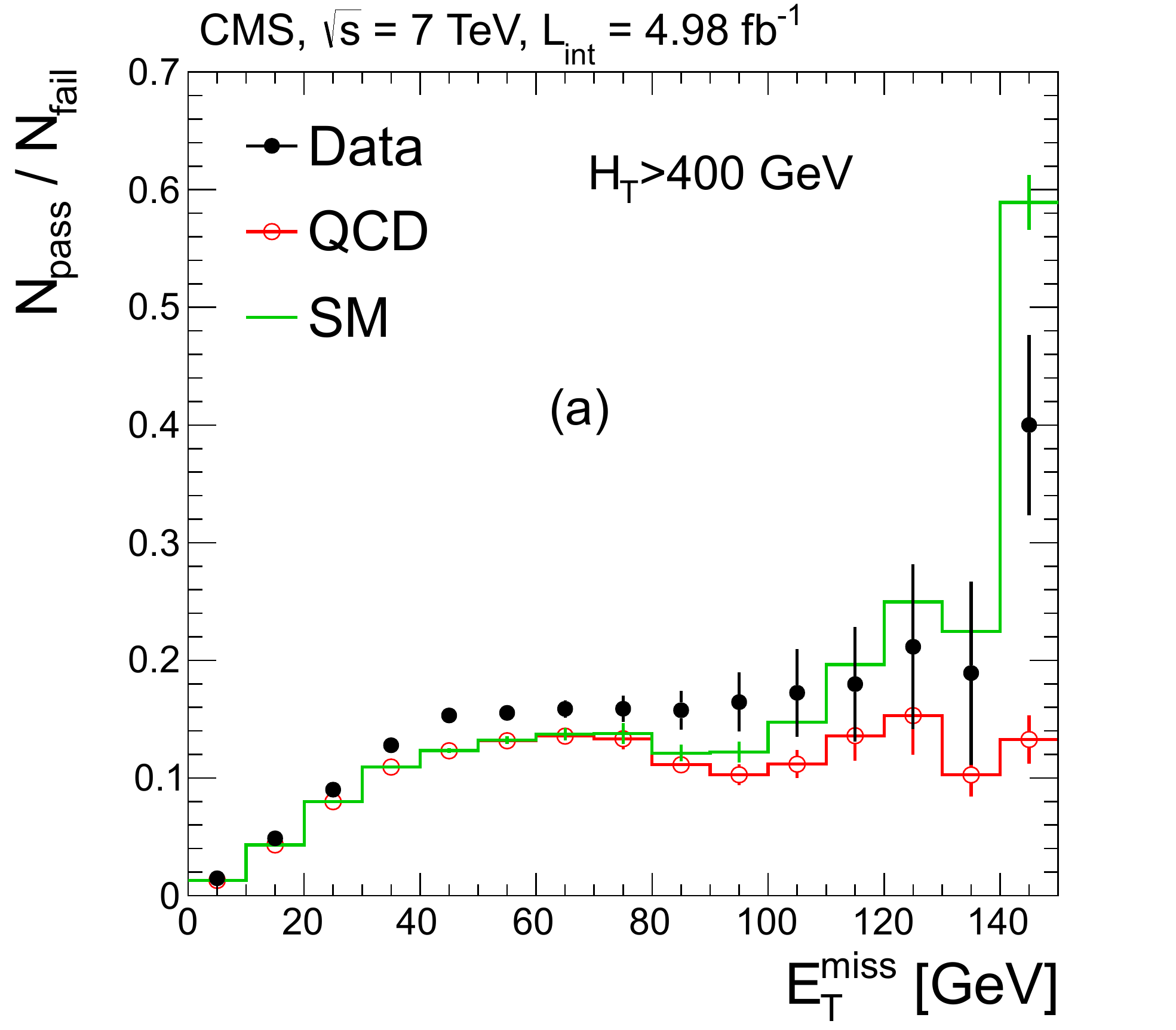}
  \includegraphics[width=\cmsFigWidthThree]{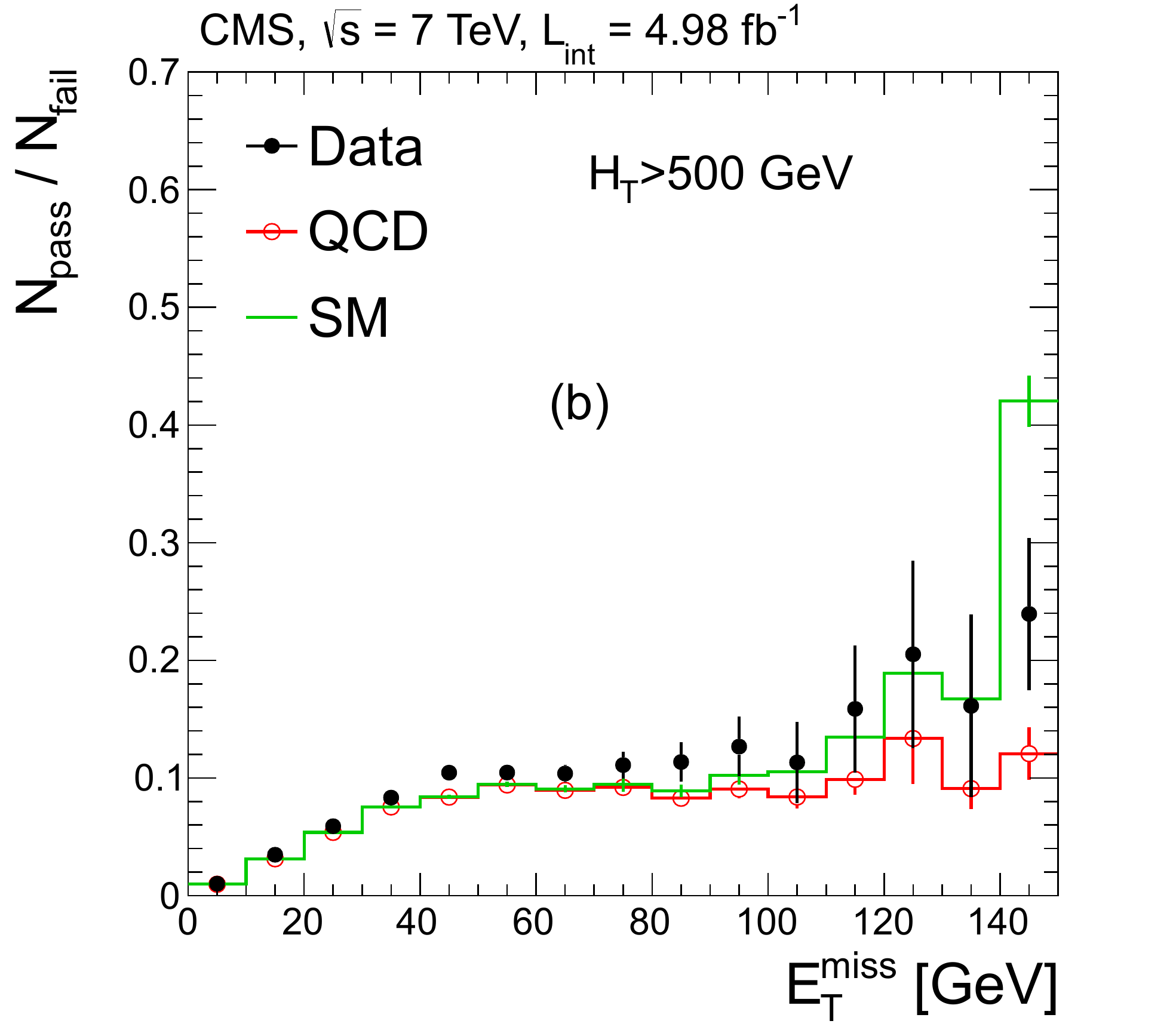}
  \includegraphics[width=\cmsFigWidthThree]{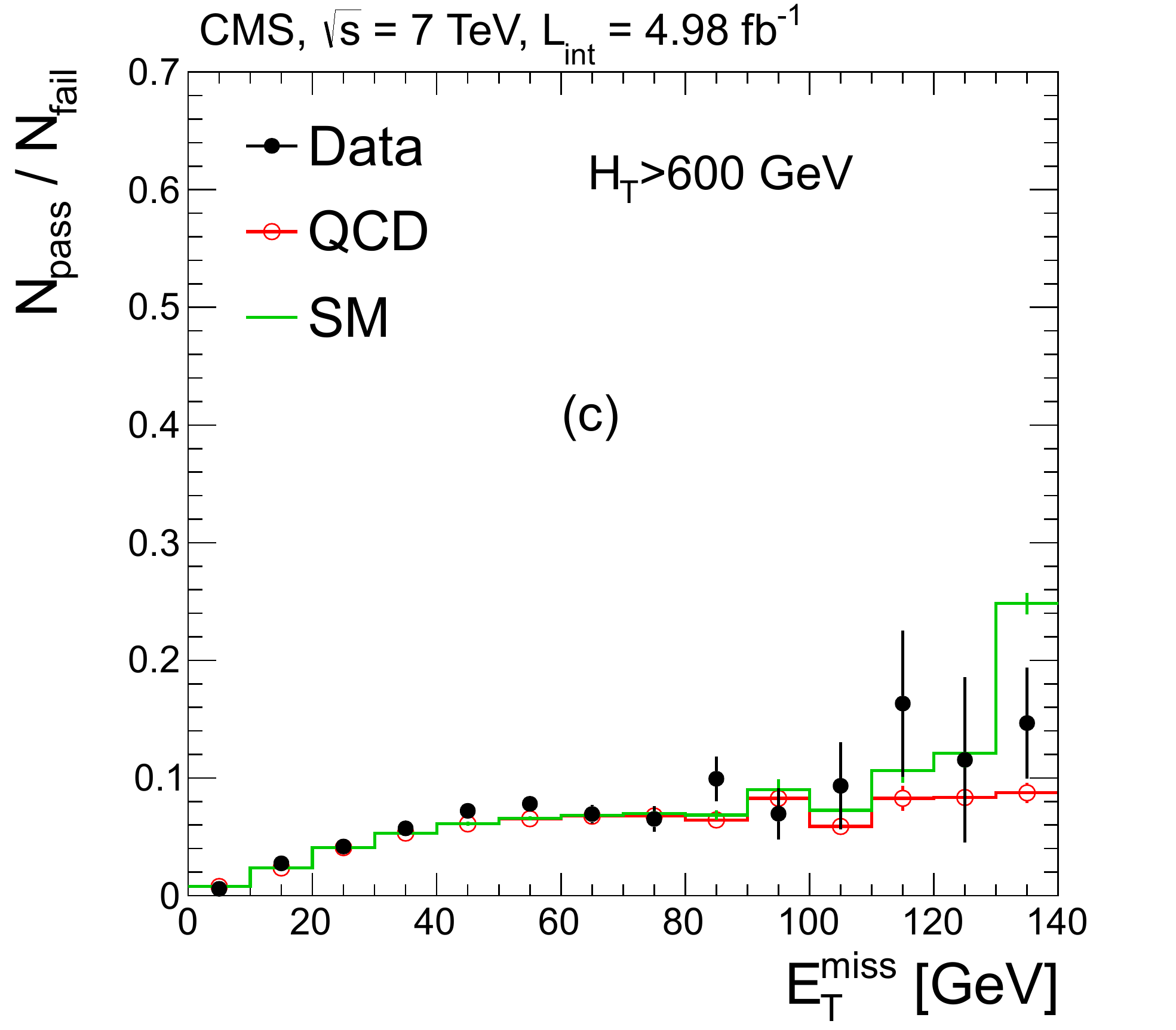}
\caption{
The ratio \rdphin,
denoted $\mathrm{N_{pass}}/\mathrm{N_{fail}}$,
as a function of \met for the zero-{\cPqb}-jet sample,
for events selected with the basic event selection criteria of the analysis
except for the requirements on \met and the number of {\b} jets.
The results are shown for
(a)~$\HT>400$\gev, (b)~$\HT>500$\gev, and (c)~$\HT>600$\gev.
The histograms show simulated predictions for the QCD and total SM background.
}
\label{fig:rMETqcd}
\end{figure}

The measured results for \rdphin with zero {\cPqb} jets,
for events with $\HT>400$\gev, 500\gev, and 600\gev,
are shown in Fig.~\ref{fig:rMETqcd}.
By requiring that there not be a {\cPqb} jet,
we reduce the contribution of top-quark events,
which is helpful for the evaluation of QCD background (Section~\ref{sec-qcd}).
The data in Fig.~\ref{fig:rMETqcd} are collected with a pre-scaled \HT trigger,
allowing events to be selected at low \MET without a trigger bias.
The data in Fig.~\ref{fig:rMETqcd}(a) are seen to somewhat exceed the
simulated predictions.
The trend is visible in Fig.~\ref{fig:rMETqcd}(b) to a lesser extent.
This modest discrepancy arises because the \dphin distribution is
narrower in the simulation than in data.
Since our method to evaluate the QCD background is based on the measured distribution,
this feature of the simulation does not affect our analysis.
The data in Fig.~\ref{fig:rMETqcd} are seen to exhibit the
general behavior expected from the simulation.
The region below around 100\gev is seen to be dominated by the QCD background.

\section{Background evaluation}
\label{sec-background}

In this section we describe our methods to evaluate
the SM background from control samples in data.
Each of the three main backgrounds
--~from QCD, {\Z}+jets, and top-quark and {\PW}+jets events
(where ``top quark'' includes both \ttbar and single-top-quark events)~--
is evaluated separately.
We group top quark and {\PW}+jets events together because they have a similar
experimental signature.
Note that our final results for the total SM background are derived from
a global likelihood procedure that incorporates our
background evaluation procedures into a single fit,
and that also accounts for possible NP contributions to the control regions
in a consistent manner.
The global likelihood procedure
is described in Section~\ref{sec-likelihood}.

QCD background is evaluated using the \dphin variable.
Background from {\Z}+jets events is evaluated by scaling the
measured rates of \zll\ ($\ell =\,${\Pe} or {\Pgm}) events.
To estimate the top-quark and {\PW}+jet background,
we employ two complementary techniques.
One,
which we call the nominal method,
is simple and almost entirely data based,
while the other,
which we call the \MET-reweighting method,
combines results based on data with information from simulation to
examine individual sources of top-quark and {\PW}+jets background in detail.

\subsection{QCD background}
\label{sec-qcd}

The low level of correlation between \dphin and \MET allows us
to employ a simple method to evaluate the QCD background from data.
As discussed in Section~\ref{sec-dphin},
the ratio \rdphin is approximately independent of \met,
and also of the number of {\cPqb} jets,
for QCD events.
Furthermore,
the \met distribution below 100\gev is expected to be dominated by~QCD events,
especially for events with zero {\cPqb} jets (Fig.~\ref{fig:rMETqcd}).
We therefore measure \rdphin in a low \MET region
of the zero-{\cPqb}-jet sample and assume this equals
\num-rdphin \den-rdphin for QCD events at all \MET values,
also for samples with {\cPqb} jets such as our signal samples.

\begin{figure}[thbp]
\centering
\includegraphics[width=\cmsFigWidthFigSix]{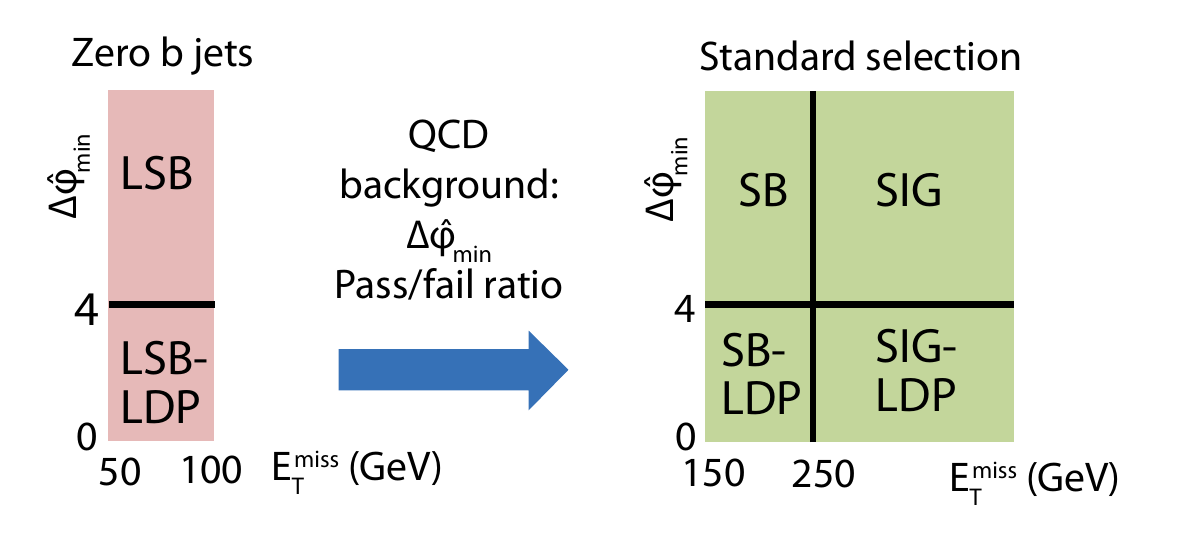}
\caption{
Schematic diagram illustrating the regions used to evaluate
the QCD background.
The low sideband (LSB) and low sideband-low \dphin (LSB-LDP) regions
correspond to $50<\met<100$\gev.
The sideband (SB) and sideband-low \dphin (SB-LDP) regions correspond to
$150<\met<250$\gev.
The signal (SIG) and signal-low \dphin (SIG-LDP) regions
have \met ranges corresponding to those in Table~\ref{tab-sig-regions}.
The designation ``SIG'' generically refers to any of the signal regions
in this table.
The SIG and SIG-LDP regions shown in the diagram explicitly depict
the loose kinematic signal regions 1BL, 2BL, and 3B,
which require $\met>250\gev$,
but implicitly include
the tight kinematic signal regions 1BT and 2BT,
which require $\met>500$\gev and 300\gev, respectively.
For each choice of signal region,
the condition on \HT specified in Table~\ref{tab-sig-regions} for that region
is applied to all six panels of the diagram,
while the condition on the number of {\cPqb} jets is applied
to the four panels denoted ``Standard selection.''
All regions with the low \dphin (LDP) designation require $0.0<\dphin<4.0$,
while the other regions require $\dphin>4.0$.
}
\label{fig:roadmap_qcd}
\end{figure}

\begin{figure}[thbp]
\centering
  \includegraphics[width=\cmsFigWidthThree]{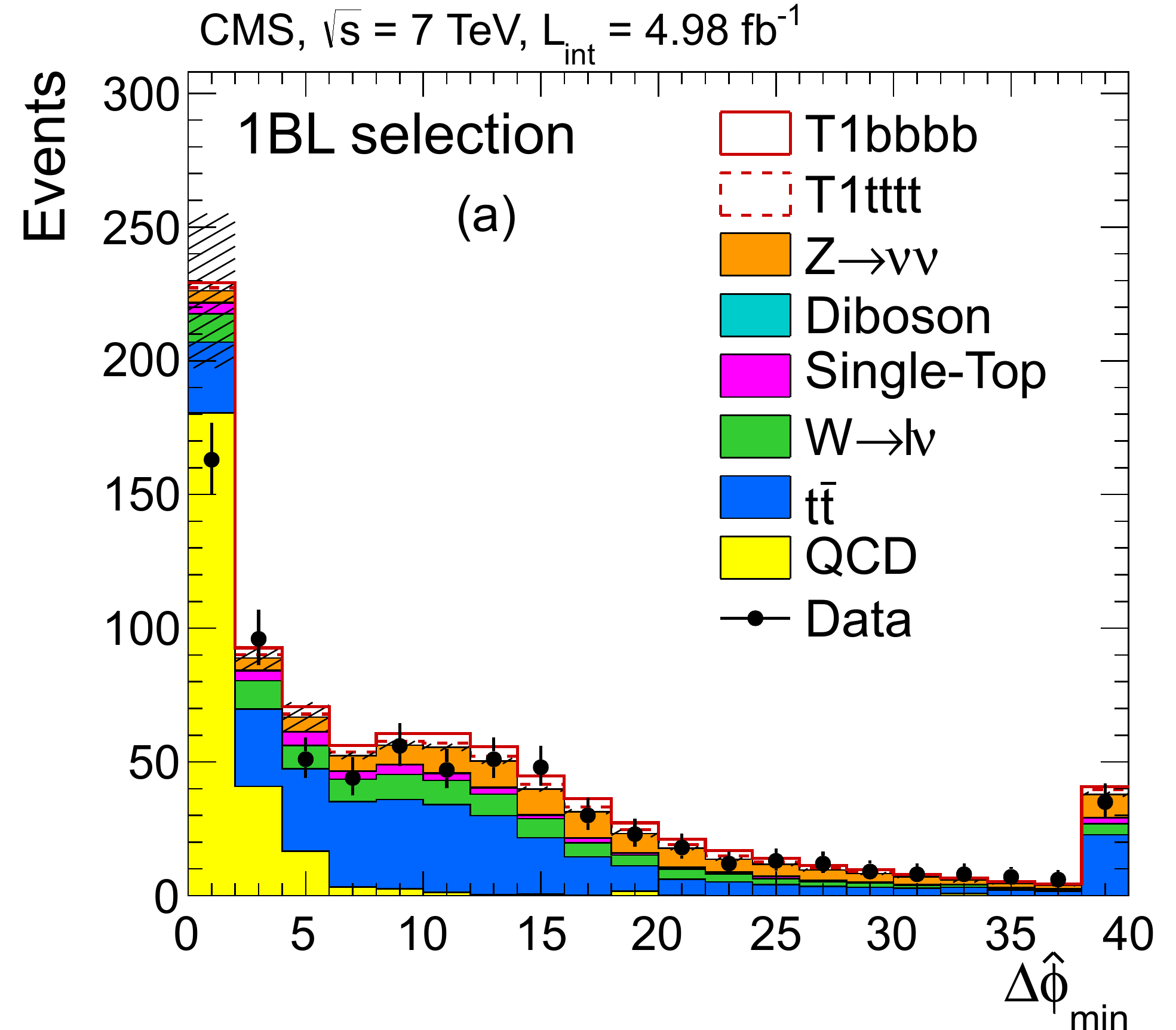}
  \includegraphics[width=\cmsFigWidthThree]{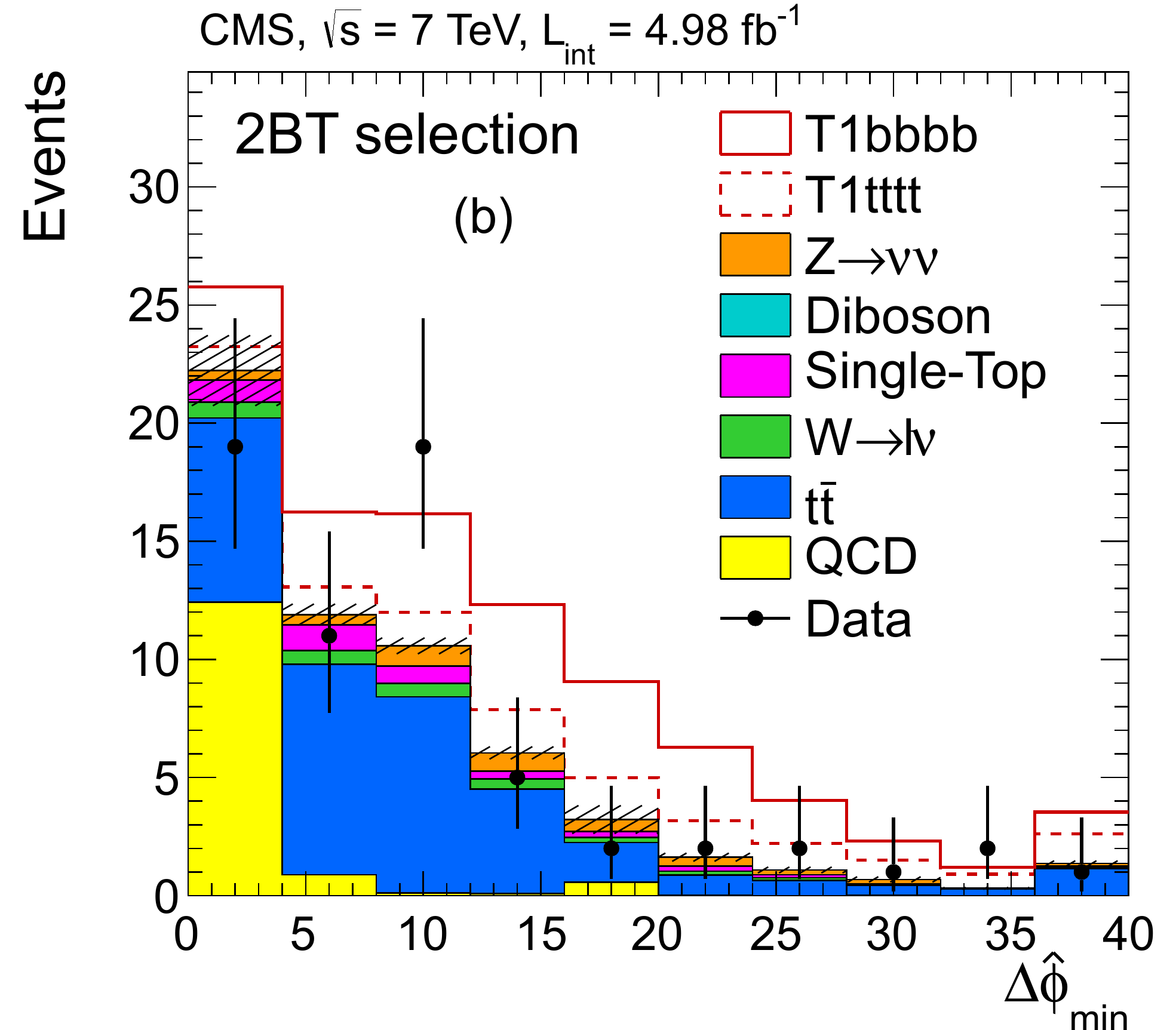}
  \includegraphics[width=\cmsFigWidthThree]{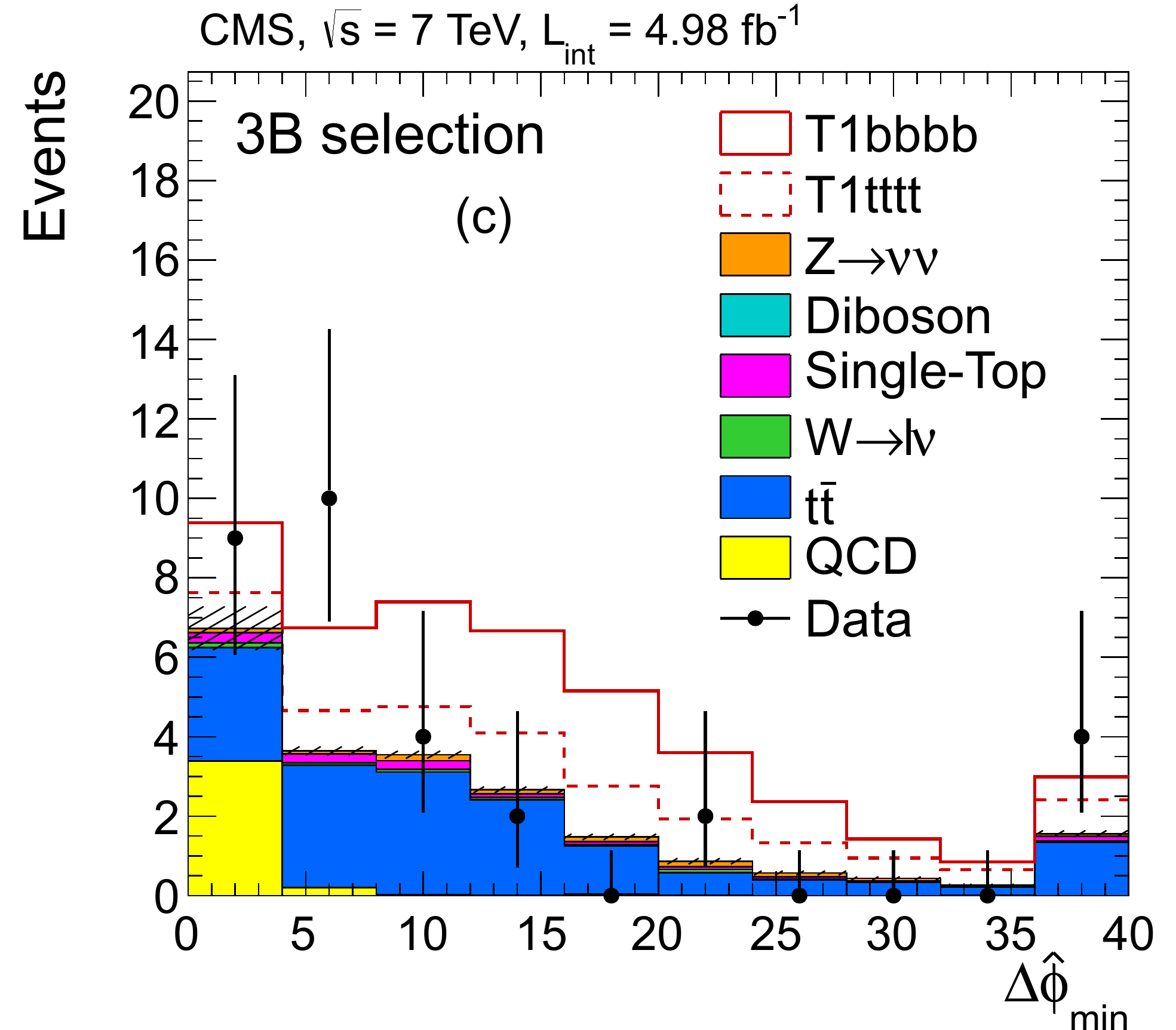}
\caption{
The distributions of \dphin in data and simulation for events selected
with the (a) 1BL, (b) 2BT, and (c)~3B requirements,
except for the requirement on \dphin.
The simulated spectra are normalized as in Fig.~\ref{fig-nbjets}.
The hatched bands show the statistical uncertainty on the
total SM prediction from simulation.
The open histograms show the expectations for the \tonebbbb (solid line)
and \tonetttt (dashed line) NP models,
both with $m_{\sGlu} = 925$\gev, $m_{\mathrm{LSP}}=100$\gev,
and normalization to NLO+NLL.
The SIG-LDP regions correspond to $\dphin<4.0$
and the signal (SIG) regions to $\dphin>4.0$.
}
\label{fig-dphin}
\end{figure}

To perform this measurement,
we divide the data into sideband and signal regions in the
\dphin-\MET plane,
as illustrated schematically in Fig.~\ref{fig:roadmap_qcd}.
We use the low-\MET interval defined by $50<\MET<100$\gev and $\dphin>4.0$.
We call this interval the low sideband (LSB) region.
We also define low \dphin (LDP) intervals $\dphin<4.0$.
We do this not only for the $50<\MET<100$\gev region,
but also for the signal regions (SIG) and for a sideband (SB) region
defined by $150<\met<250$\gev.
We denote these regions LSB-LDP, SIG-LDP, and SB-LDP,
respectively.
The LSB-LDP region is dominated by QCD events.
Similarly,
the SB-LDP and SIG-LDP regions largely consist of QCD events,
as illustrated for the 1BL, 2BT, and 3B SIG-LDP regions in Fig.~\ref{fig-dphin}
(according to simulation,
QCD events comprise between 73\% and 85\%
of the events in the SB-LDP region,
depending on the SIG selection;
the corresponding results for
for the SIG-LDP region lie between 50\% and 70\%).
For higher values of \MET,
contributions to the SB-LDP and SIG-LDP regions
from events with a top quark or a {\PW} or {\Z} boson
become more important.
This contamination is subtracted using simulation.

Applying corrections for the non-QCD components of the SIG-LDP and SB-LDP regions,
our estimates of the QCD yields in the SIG and SB regions are therefore:
\ifthenelse{\boolean{cms@external}}{
\begin{align}
N^{\mathrm{QCD}}_{\mathrm{SIG}} &=  \frac{N_{\mathrm{LSB}}}{N_{\mathrm{LSB-LDP}}}\notag\\& \times
   ( N_{\mathrm{SIG-LDP}}
   - N^{\mathrm{top,MC}}_{\mathrm{SIG-LDP}}- N^{\mathrm{{\PW}\&{\Z},MC}}_{\mathrm{SIG-LDP}} ),
\label{eq:qcdsig}\\
N^{\mathrm{QCD}}_{\mathrm{SB}} &=  \frac{N_{\mathrm{LSB}}}{N_{\mathrm{LSB-LDP}}}\notag\\& \times
   ( N_{\mathrm{SB-LDP}}
   - N^{\mathrm{top,MC}}_{\mathrm{SB-LDP}}- N^{\mathrm{{\PW}\&{\Z},MC}}_{\mathrm{SB-LDP}} ),
\label{eq:qcdsb}
\end{align}
}{
\begin{eqnarray}
N^{\mathrm{QCD}}_{\mathrm{SIG}} &= & \frac{N_{\mathrm{LSB}}}{N_{\mathrm{LSB-LDP}}} \times
   ( N_{\mathrm{SIG-LDP}}
   - N^{\mathrm{top,MC}}_{\mathrm{SIG-LDP}}- N^{\mathrm{{\PW}\&{\Z},MC}}_{\mathrm{SIG-LDP}} ),
\label{eq:qcdsig} \\
N^{\mathrm{QCD}}_{\mathrm{SB}} &= & \frac{N_{\mathrm{LSB}}}{N_{\mathrm{LSB-LDP}}} \times
   ( N_{\mathrm{SB-LDP}}
   - N^{\mathrm{top,MC}}_{\mathrm{SB-LDP}}- N^{\mathrm{{\PW}\&{\Z},MC}}_{\mathrm{SB-LDP}} ),
\label{eq:qcdsb}
\end{eqnarray}
}

where the LSB and LSB-LDP results are derived from the zero-{\cPqb}-jet,
pre-scaled \HT trigger sample mentioned in Section~\ref{sec-dphin}.
The result for $N^{\mathrm{QCD}}_{\mathrm{SB}}$ is used in Section~\ref{sec-ttw}.
The ratio $N_{\mathrm{LSB}}/N_{\mathrm{LSB-LDP}}$ is found to depend on the
number of primary vertices (PV) in the event
and thus on the LHC instantaneous luminosity.
Before evaluating Eqs.~(\ref{eq:qcdsig}) and~(\ref{eq:qcdsb}),
we therefore reweight the events in the pre-scaled sample to
have the same PV distribution as the standard sample.

\begin{table}[!tb]
\topcaption{
The relative systematic uncertainties (\%) for the QCD background estimate in the signal regions.
Because the 1BT QCD background estimate is zero (Section~\ref{sec-background-summary}),
we do not present results for 1BT in this table.
}
\label{tab:qcdsyst}
\scotchrule[l|rrrrr]
                & 1BL & 2BL & 2BT & 3B \\
\hline
MC subtraction  & 23  & 43  & 44  & 24 \\
MC closure      & 37  & 41  & 150 & 45 \\
LSB reweighting & 7.9 & 7.9 & 9.8 & 7.9 \\
\hline
Total           & 44  & 60  & 160 & 52 \\
\donescotchrule
\end{table}

Systematic uncertainties are summarized in Table~\ref{tab:qcdsyst}.
The systematic uncertainty associated with the
subtraction of events with either a top quark or a {\PW} or {\Z} boson
from the SIG-LDP and SB-LDP regions
is determined by varying the subtracted values
by their uncertainties,
evaluated as described in Section~\ref{sec-simplified}.
The systematic uncertainty associated with the assumption that
\met and \dphin are uncorrelated is evaluated with an MC closure test,
namely by determining the ability of the method to predict the correct
yield using simulated samples.
We compute $(N_{\mathrm{true}} - N_{\mathrm{pred}})/N_{\mathrm{pred}}$,
where $N_{\mathrm{pred}}$ is the predicted number of
QCD events in the signal region,
estimated by applying the above procedure to simulated samples
treated as data,
and $N_{\mathrm{true}}$ is the true number.
We assign the result,
added in quadrature with its statistical uncertainty,
as a symmetric systematic uncertainty.
This uncertainty is dominated by statistical uncertainties for $N_{\mathrm{true}}$.
The closure test is performed both for the standard
simulated samples and for simulated samples that are reweighted to account for
discrepancies in the jet multiplicity distributions between data
and simulation;
we take the larger closure discrepancy as the uncertainty.
A third systematic uncertainty is evaluated by
taking $\pm100$\% of the shift in the result
caused by the PV reweighting of $N_{\mathrm{LSB}}/N_{\mathrm{LSB-LDP}}$.
The systematic uncertainty associated with the
trigger efficiency is found to be negligible.

As a cross-check,
we vary the definition of the LSB by raising and
lowering its lower edge by 10\gev,
which alters the number of events in the LSB by more than a factor of two
in each case.
The observed change in the QCD background estimate is negligible.

\subsection{\texorpdfstring{\Z{}+jets}{Z+jets} background}
\label{sec-zjets}
Events with a \Z\ boson and one or more {\cPqb} jets present an
irreducible background when the \Z\ decays to two neutrinos.
We evaluate this background by reconstructing
\zll\ events ($\ell =\,${\Pe} or {\Pgm})
and removing the $\ell^+$ and $\ell^-$.
Fits are performed to determine the \zll\ yields,
which are then corrected for background and efficiency.
The efficiency is
$\epsilon = {\cal A} \cdot \epsilon_{\mathrm{trig}}
  \cdot \epsilon_{\mathrm{\ell\, reco}}^2
  \cdot \epsilon_{\mathrm{\ell\, sel}}^2$,
where the geometrical acceptance ${\cal A}$ is determined from simulation
while the trigger $\epsilon_{\mathrm{trig}}$,
lepton reconstruction $\epsilon_{\mathrm{\ell\, reco}}$,
and lepton selection $\epsilon_{\mathrm{\ell\, sel}}$
efficiencies are determined from data.
The corrected $\zll$ yields are used to estimate the $\znunu$
background through scaling by the ratio of branching fractions,
$\mathrm{BR}\,(\znunu) / \mathrm{BR}\,(\zll) = 5.95\pm0.02$~\cite{pdg},
after accounting for the larger acceptance of \znunu events.

The \zll\ yields are small or zero in the signal regions.
To increase these yields,
we select events with the signal-sample requirements except
with a significantly looser {\cPqb}-tagging definition.
A scale factor derived from a control sample in data is then applied
to estimate the number of \zll\ events in the signal regions.
The control sample is defined with the same loosened {\cPqb}-tagging definition,
but without requiring the presence of a {\Z} boson,
and also by reversing the \dphin requirement,
i.e., we require $\dphin<4.0$,
which yields a control sample with
a {\cPqb}-jet content similar to that in the \zll\ and \znunu\ events.
All other selection criteria are the same as for the corresponding signal sample.
The scale factors are given by the fraction of events in the control
sample that passes the nominal {\cPqb}-tagging requirements.
The scale factors have values around 0.30, 0.07, and 0.01
for the samples with $\geq1$, $\geq2$, and $\ge3$ {\cPqb} jets, respectively.
We verify that the output of the {\cPqb}-tagging algorithm is
independent of the presence of a~\Z.

We validate our method with a consistency test,
applying the above procedure to data samples with loosened
restrictions on \HT\ and \MET.
We find the number of predicted and observed \zll\ events
to be in close agreement.

\begin{table*}[!t]
\caption{
The relative systematic uncertainties (\%) for the \znunu\ background estimate
in the signal regions,
determined for {\zee} (\zmumu) events.
}
\label{tab:ZinvSyst}
\scotchrule[l|cccccc]
  & 1BL & 1BT & 2BL & 2BT & 3B \\
\hline
Scale factors                 & 17 (20) &  17 (20) & 49 (61) & 49 (61) & 140 (110) \\
Non-resonant $\ell^{+}\ell^{-}$  background
                              &  10 (8) &  10  (8) & 10  (8) & 10  (8) &  10   (8)\\
Acceptance                    &  3  (3) &  6   (8) &  3  (3) &  4  (4) &   3   (3)\\
Lepton selection efficiency   &  5  (4) &  5   (4) &  5  (4) &  5  (4) &   5   (4) \\
Trigger efficiency            &  5  (5) &  5   (5) &  5  (5) &  5  (5) &   5   (5) \\
MC closure                    & 11 (19) & 11  (19) & 11 (19) & 11 (19) &  11  (19) \\
\hline
Total                         & 24 (30) &  25 (30) & 52 (65) & 52 (65) & 150 (110) \\
\donescotchrule
\end{table*}

Systematic uncertainties are summarized in Table~\ref{tab:ZinvSyst}.
We evaluate a systematic uncertainty on the scale factors
by loosening and tightening the {\cPqb}-tagging criterion of
the control sample and taking half the difference between the two results
as an uncertainty.
The size of the control sample changes by about $\pm30$\% in these variations.
In addition,
we use $\dphin>4.0$ rather than $\dphin<4.0$ to define the control sample
and calculate the difference with respect to the nominal results.
Finally, we evaluate the percentage difference between the number of predicted
and observed events found with the consistency test described above.
The three terms are added in quadrature to define the systematic uncertainty
of the scale factors.
We evaluate a systematic uncertainty associated with the
non-resonant $\ell^+\ell^-$ background to {\zll} events
by comparing the fraction of fitted events in the \zll\ peak from the
nominal fit with those found
using either a loosened \HT or a loosened \MET restriction.
The RMS of the three results is added in quadrature with the
statistical uncertainty from the nominal fit to define the
systematic uncertainty.
The 1BL selection is used to determine this uncertainty for all signal regions.
A systematic uncertainty for the acceptance is defined by
recalculating the acceptance after varying the \pt and $\eta$
ranges of the $\ell^+$ and $\ell^-$.
The largest difference with respect to the nominal result
is added in quadrature with the statistical uncertainty of the acceptance.
A systematic uncertainty is defined for the lepton selection efficiency,
and analogously for the trigger efficiency,
by recalculating the respective efficiency after varying the requirements
on \HT, \MET, \dphin, the number of jets, and the number of {\cPqb} jets
(the number of jets is found using all jets with $\pt>50$\gev and $|\eta|<2.4$).
We also use alternative signal and background shapes in the fits used to
extract the \zll\ event yields.
The maximum variations from each case are added in quadrature
with the statistical uncertainty from the nominal method to define
the systematic uncertainties.
Finally,
we evaluate a systematic uncertainty based on an MC closure test
in the manner described in Section~\ref{sec-qcd}.
We use the SB region to determine this uncertainty.

An analogous procedure to that described above
is used to evaluate the number of \znunu\
events $N^{\Zinvisible}_{\mathrm{SB}}$
in the SB regions ($150<\met<250$\gev),
along with the corresponding uncertainty.

\subsection{Top-quark and \texorpdfstring{\PW}{W}+jets background (nominal)}
\label{sec-ttw}

\begin{figure}[thbp]
\begin{center}
 \includegraphics[width=\cmsFigWidthFigEight]{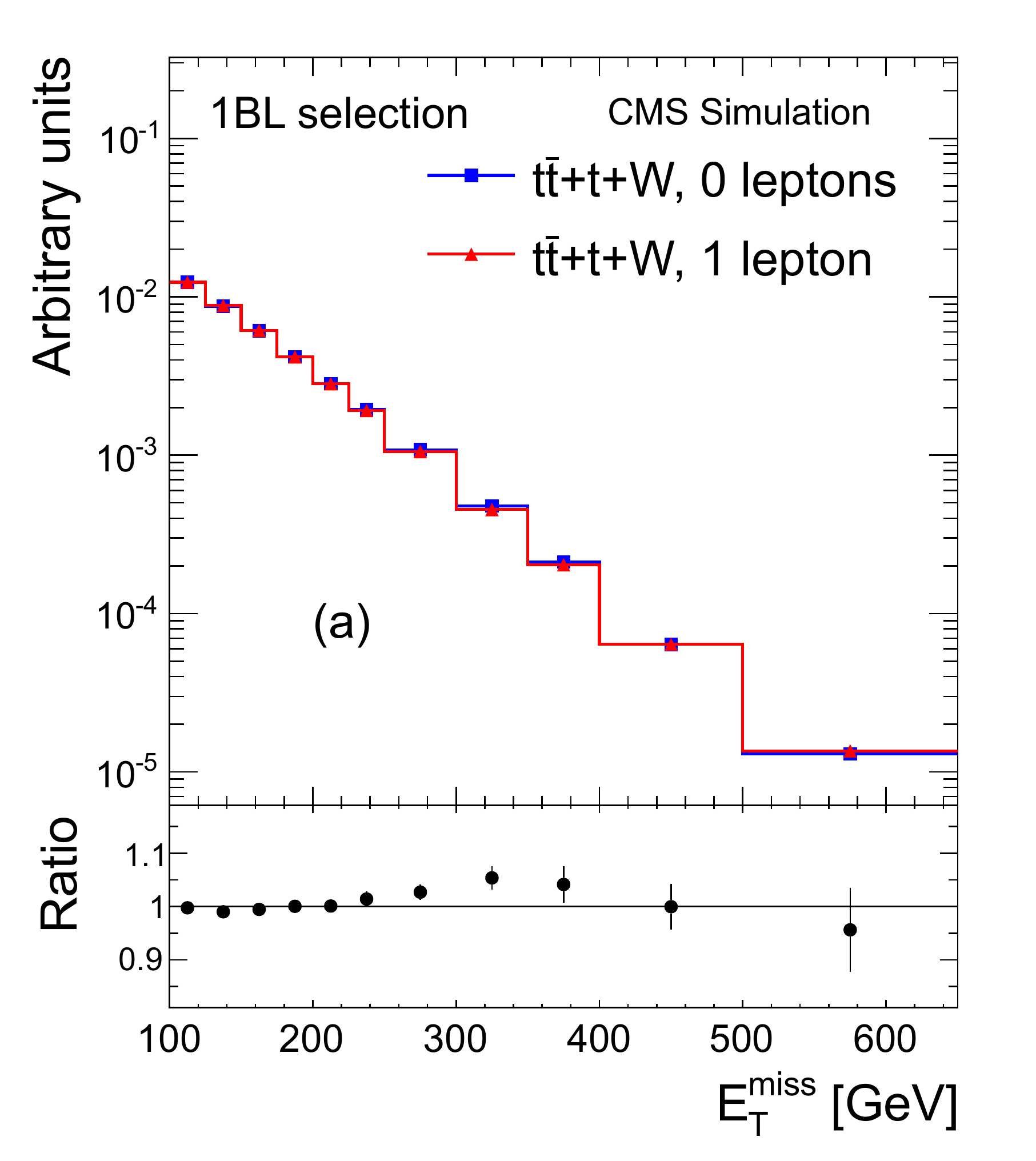}
 \includegraphics[width=\cmsFigWidthFigEight]{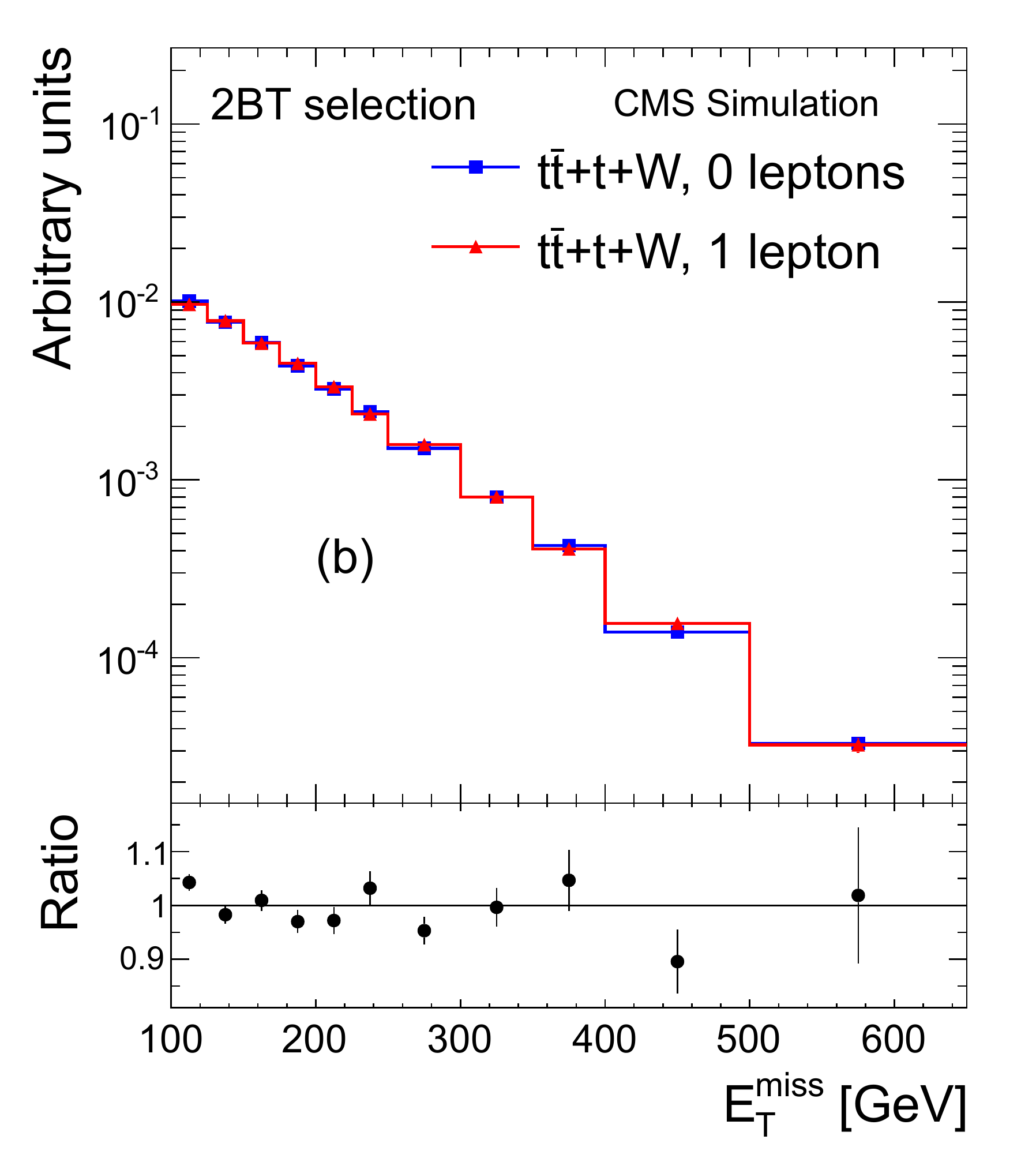}
 \includegraphics[width=\cmsFigWidthFigEight]{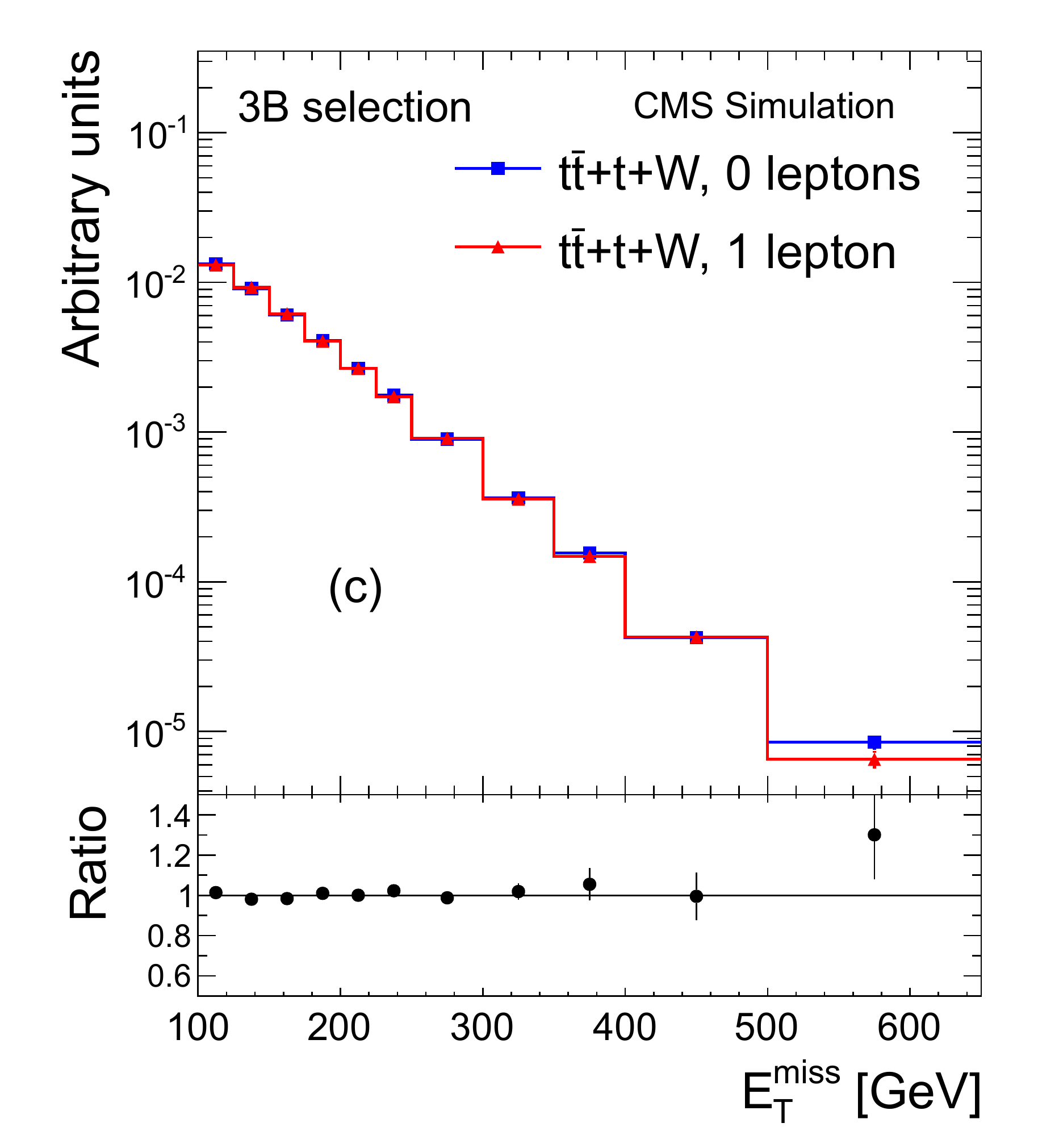}
\end{center}
\caption{
The distributions of \met in simulated events selected
with the (a)~1BL, (b)~2BT, and (c)~3B requirements,
except for the requirement on \MET.
The square (triangle) symbols show the results for signal
(single-lepton SL control) sample events.
The small plots below the main figures show the ratio
of the signal to SL sample curves.
The event samples include \ttbar, {\PW}+jet, and single-top-quark events.
}
\label{fig:ttbarmetshape}
\end{figure}

For most signal regions,
\ttbar events are expected to be the dominant background
(Table~\ref{tab-event-count}).
Backgrounds from single-top-quark and {\PW}+jets events are expected to be
smaller but to have a similar signature.
Almost all top-quark and {\PW}+jets background in our analysis
arises either because a {\PW} boson decays leptonically to an \Pe\ or a {\Pgm},
with the \Pe\ or {\Pgm} unidentified, not isolated,
or outside the acceptance of the analysis,
or because a {\PW} boson decays to a hadronically decaying {\Pgt} lepton.
We find empirically, through studies with simulation,
that the shape of the \MET distribution is
similar for all top-quark and {\PW}+jets background categories
that enter the signal (Table~\ref{tab-sig-regions})
or sideband ($150<\met<250$\gev) regions,
regardless of whether the {\PW} boson decays to {\Pe}, {\Pgm}, or {\Pgt},
or whether a {\Pgt} lepton decays hadronically or leptonically:
the decay of the {\PW} boson in {\PW}+jets events
generates an \MET spectrum (from the neutrino) that is similar to
the \MET spectrum generated by the {\PW} boson produced directly in
the decay of a top quark in top quark events.
Additional, softer neutrinos in events with a $\tau$ lepton
do not much alter this spectrum.
We also find that this shape is well-modeled by the \MET
distribution of a single-lepton (SL) control sample
formed by inverting the lepton veto,
i.e.,  by requiring that exactly one \Pe\ or one {\Pgm} be present
using the lepton identification criteria of Section~\ref{sec-selection},
in a sample whose selection is otherwise the same as the
corresponding signal sample,
except to reduce the potential contribution of NP to the SL samples,
we impose an additional restriction $\transmass<100$\gev
on the SL samples (only),
where \transmass is the transverse {\PW}-boson mass formed from the
charged lepton and \MET momentum vectors.
As an illustration,
Fig.~\ref{fig:ttbarmetshape} shows a comparison based on simulation
of the \met distributions in the signal and SL samples,
for events selected with the 1BL, 2BT, and 3B criteria.

\begin{figure}[thbp]
\begin{center}
 \includegraphics[width=\cmsFigWidthThree]{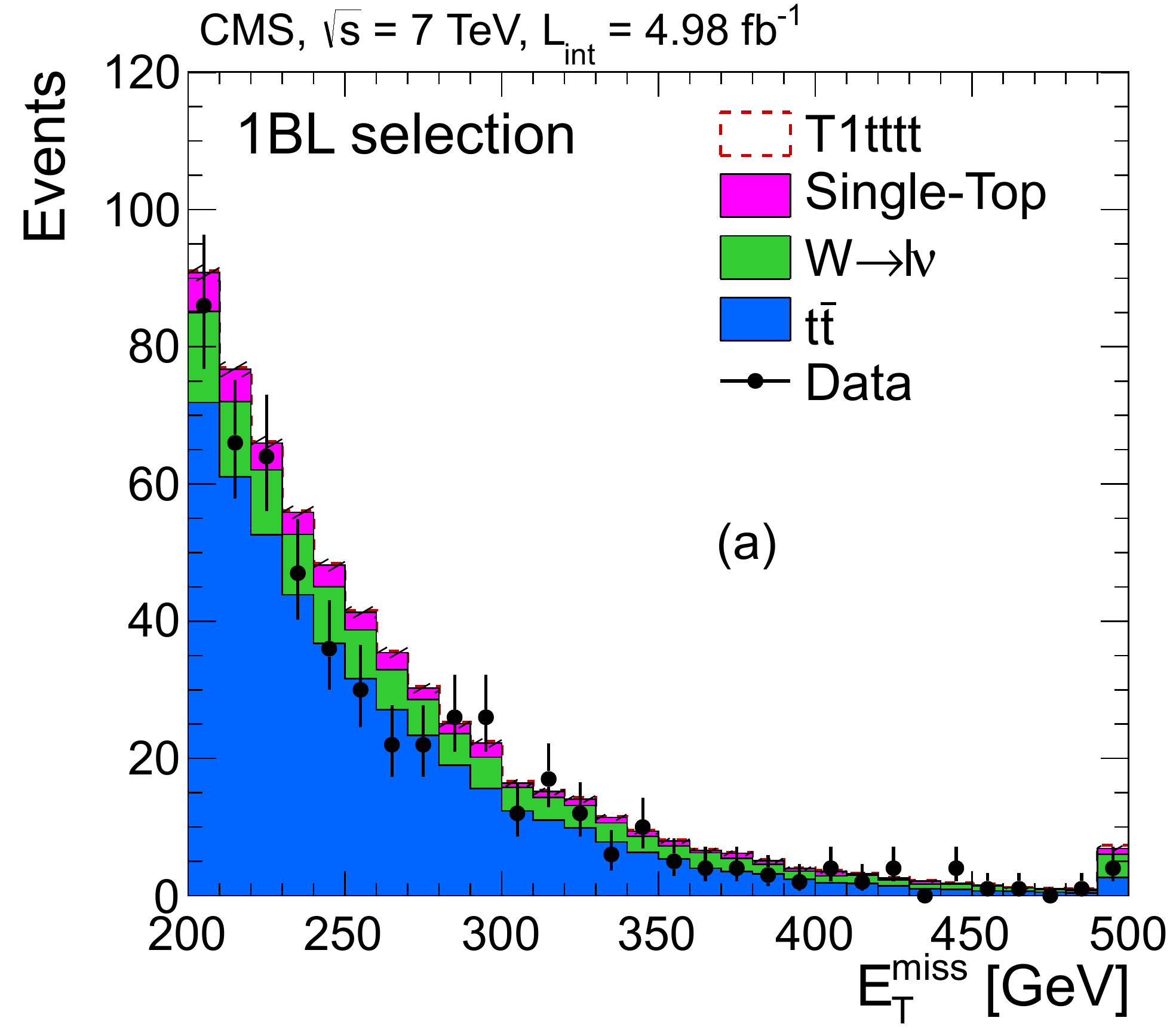} \includegraphics[width=\cmsFigWidthThree]{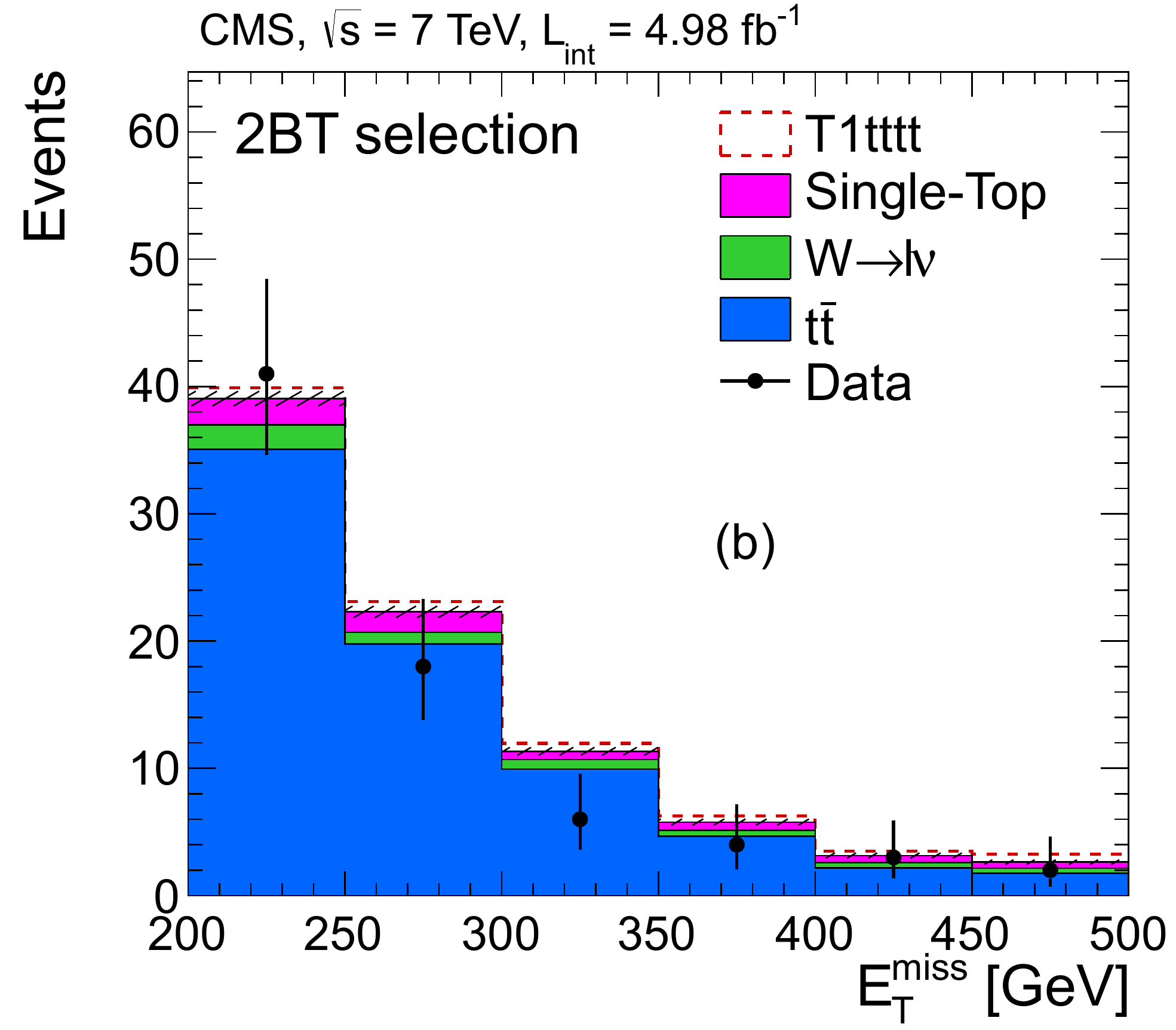}
 \includegraphics[width=\cmsFigWidthThree]{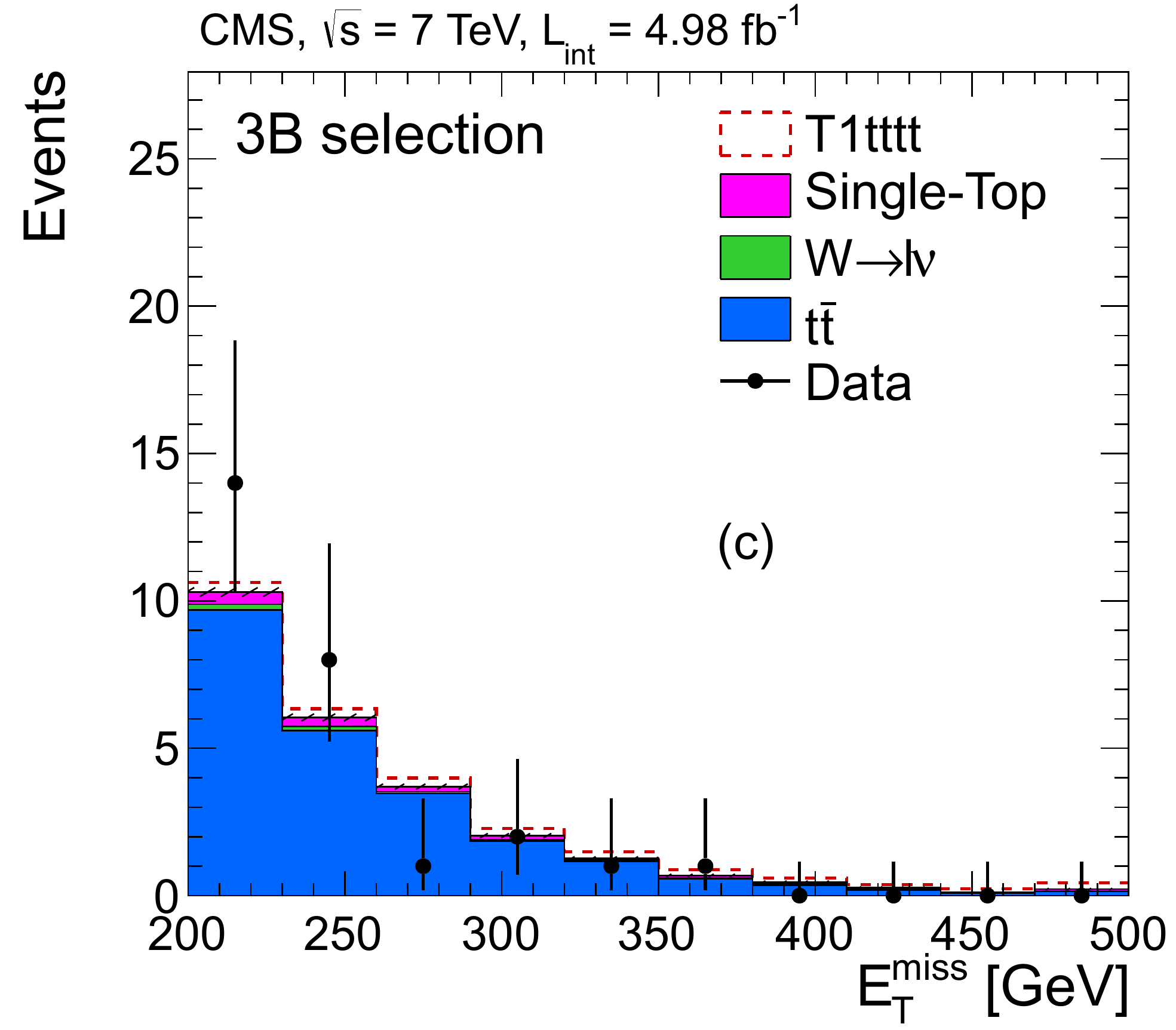}
\end{center}
\caption{
The distributions of \met for the SL
control sample for events selected
with the (a)~1BL, (b)~2BT, and (c)~3B requirements,
except for the requirement on \MET.
The simulated spectra are normalized as in Fig.~\ref{fig-nbjets}.
The hatched bands show the statistical uncertainty on the
total SM prediction from simulation.
The open dashed histogram
shows the expectations for the \tonetttt NP model
with $m_{\sGlu} = 925$\gev, $m_{\mathrm{LSP}}=100$\gev,
and normalization to NLO+NLL
(the corresponding contributions from the \tonebbbb model
are negligible and are not shown).
}
\label{fig:SL_MET}
\end{figure}

The \met distributions of events in the SL samples
with the 1BL, 2BT, and 3B requirements are shown in Fig.~\ref{fig:SL_MET}.
The distributions are seen to be overwhelmingly composed of~\ttbar events
(for example,
according to simulation,
top and {\PW}+jets events comprise over 98\% of the events in
the SB-SL samples for all SIG selections).
The expected contributions of the benchmark \tonebbbb NP scenario are found to be negligible,
while those of the benchmark \tonetttt scenario are seen to be small in Fig.~\ref{fig:SL_MET}
compared to Fig.~\ref{fig-met}.

\begin{figure}[thbp]
\centering
\includegraphics[width=0.8\cmsFigWidth]{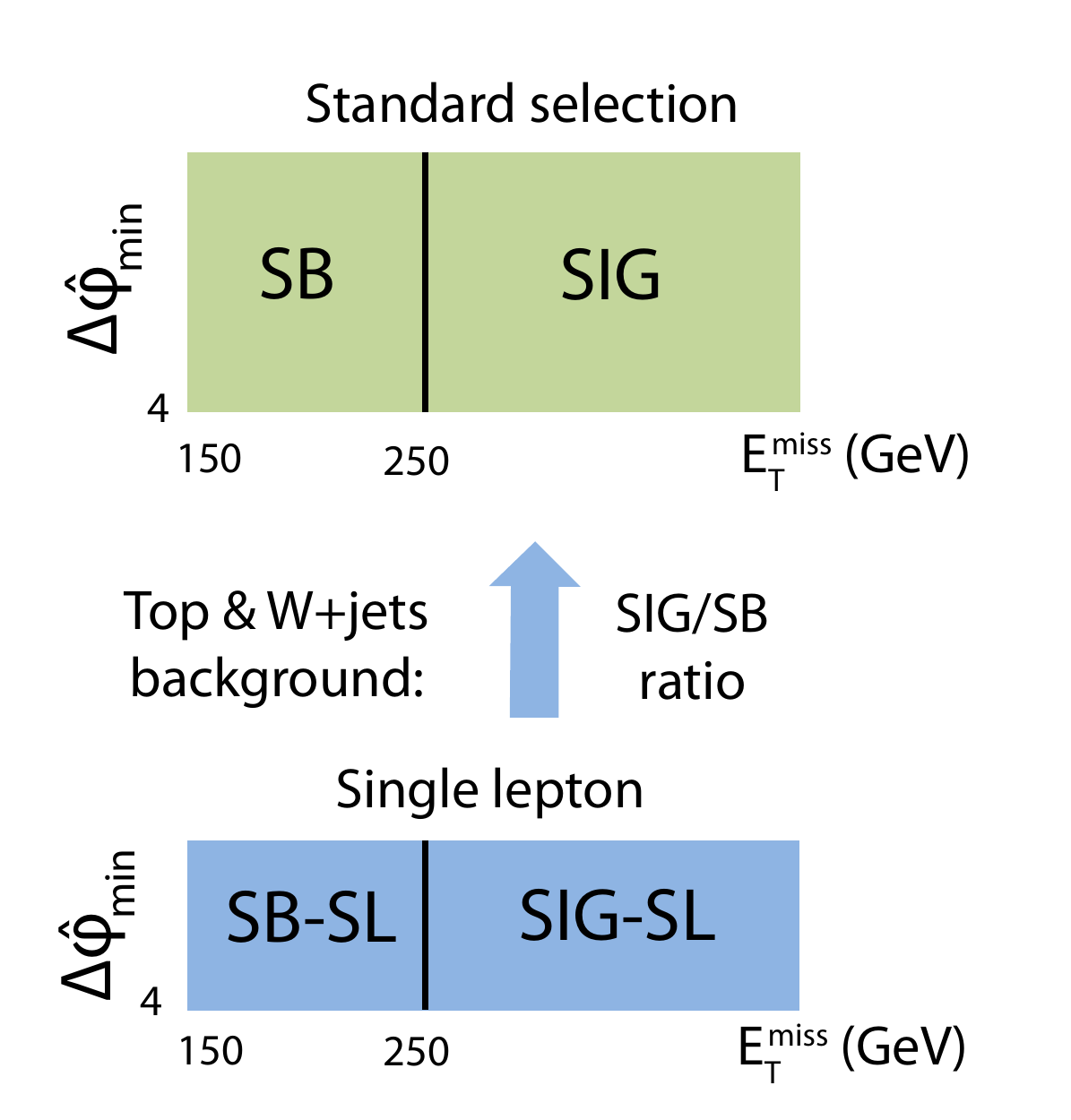}
\caption{
Schematic diagram illustrating the regions used to evaluate
the top and {\PW}+jets background with the nominal method.
The sideband (SB) and signal (SIG) regions
are described in the caption to Fig.~\ref{fig:roadmap_qcd}.
The sideband-single-lepton (SB-SL) and signal-single-lepton (SIG-SL)
regions correspond to the SB and SIG regions,
respectively,
except an electron or muon is required to be present and a requirement
is placed on the transverse {\PW} boson mass
$\transmass<100$\gev.
}
\label{fig:roadmap-ttbar-nominal}
\end{figure}

Based on these observations,
we implement a template method in which the shape
of the \met distribution in an SL sample
is used to describe the shape of the \met distribution in the
corresponding signal sample of Table~\ref{tab-sig-regions},
for all top-quark and {\PW}+jets categories.
An uncertainty for our presumption of the similarity
of the \MET spectra between different top and {\PW}+jet categories
is evaluated through the closure test described below.
We split each SL sample into
a sideband \met region SB-SL defined by
$150<\met<250$\gev,
and a signal \met region SIG-SL
given by the corresponding \met requirement in Table~\ref{tab-sig-regions}.
The templates are normalized based on the number of top-quark plus
{\PW}+jets events observed in the SB regions ($150<\MET<250$\gev)
of samples selected with the requirements
of Table~\ref{tab-sig-regions}
except for that on~\MET.
A schematic diagram of the different regions
used to evaluate the top and {\PW}+jets background
with the nominal method is presented in
Fig.~\ref{fig:roadmap-ttbar-nominal}.
Contributions to the SB region from QCD and \Zinvisible
events are taken from the data-based estimates
of Sections~\ref{sec-qcd} and~\ref{sec-zjets}.
Small, residual contributions from other
backgrounds such as diboson events
are subtracted using simulation.

Our estimate of the top-quark and {\PW}+jets background in the SIG
region is therefore:
\begin{equation}
N^{\mathrm{top + {\PW}}}_{\mathrm{SIG}} =
  \frac{ N_{\mathrm{SIG-SL}} }{ N_{\mathrm{SB-SL}} } \times
  (N_{\mathrm{SB}}
  - N^{\Zinvisible}_{\mathrm{SB}}- N^{\mathrm{QCD}}_{\mathrm{SB}}
       - N^{\mathrm{other,MC}}_{\mathrm{SB}} ).
\label{eq:ttwsig}
\end{equation}
Contamination of the SB region in the benchmark \tonebbbb (\tonetttt)
NP scenario is predicted to
be around 1\% (1\%)  for the 1BL, 1BT, and 2BL selections,
4\% (3\%) for the 2BT selection,
and 7\% (5\%) for the 3B selection.
The likelihood procedure described in Section~\ref{sec-likelihood}
accounts for NP contributions to all control regions
in a coherent manner.

\begin{table}[!t]
\topcaption{
The relative systematic uncertainties (\%) for the nominal top-quark and {\PW}+jets
background estimate in the signal regions.
}
\label{tab:ttwsyst}
\scotchrule[l|rrrrr]
                             & 1BL  & 1BT  & 2BL  & 2BT  & 3B \\
\hline
MC closure                   & 4.6  & 15   & 5.4  & 4.6  & 2.8  \\
Subtraction of QCD           & 13   & 19   & 8.2  & 20   & 8.0  \\
Subtraction of \Zinvisible\  & 3.4  & 3.9  & 5.4  & 5.9  & 15   \\
MC subtraction               & 0.6  & 0.6  & 0.2  & 0.4  & 0.1  \\
Trigger efficiency           &  13  &  14  &  11  &  11  & 10 \\
\hline
Total                        &  19  &  28  &  15  &  24  & 20 \\
\donescotchrule
\end{table}

Systematic uncertainties are summarized in Table~\ref{tab:ttwsyst}.
We consider the systematic uncertainty associated with MC closure,
evaluated as described in Section~\ref{sec-qcd}.
The closure is evaluated separately for the nominal combined
top-quark and {\PW}+jets simulated sample,
with the {\PW}+jets cross section increased by 50\% and the
single-top-quark cross section by 100\%,
and  with the {\PW}+jets cross section decreased by 50\% and the
single-top-quark cross section by 100\%
(these variations account for uncertainties on the relative cross sections;
they are based on the uncertainties of the NLO calculations and on
comparisons between data and simulation).
We take the largest closure discrepancy as the uncertainty.
We also consider the systematic uncertainty associated with
subtraction of the QCD- and \Zinvisible-background estimates in the SB region,
evaluated by varying these estimates by their uncertainties.
The systematic uncertainty associated with other backgrounds is
evaluated by varying the MC-based background estimates in the SB region
by their uncertainties,
which we assume to be $\pm 100\%$ for these small terms.
A final systematic uncertainty accounts for the uncertainty
on the trigger efficiency.

\subsection{Top-quark and \texorpdfstring{\PW}{W}+jets background (\texorpdfstring{\MET}{MET}-reweighting)}
\label{sec-dthetat}

We perform a second, complementary evaluation
of the top-quark and {\PW}+jets background,
which we refer to as the \MET-reweighting method.
The \MET distribution is determined separately
for each of the three principal top-quark and {\PW}+jets background categories:
\begin{enumerate}
\item top-quark or {\PW}+jets events in which exactly one {\PW} boson decays into
  an \Pe\ or {\Pgm},
  or into a {\Pgt} that decays into an \Pe\ or {\Pgm},
  while the other {\PW} boson (if any) decays hadronically;
\item top-quark or {\PW}+jets events in which exactly one {\PW} boson decays into a
  hadronically decaying {\Pgt},
  while the other {\PW} boson (if any) decays hadronically;
\item \ttbar events in which both {\PW} bosons decay
  into an {\Pe}, {\Pgm} or {\Pgt},
  with the {\Pgt} decaying either leptonically or hadronically.
\end{enumerate}
For the 1BL selection,
these three categories represent, respectively,
approximately 44\%, 49\%, and 7\% of the total
expected background from top-quark and {\PW}+jets events,
as determined from simulation.

\subsubsection{Single \texorpdfstring{\Pe\ or {\Pgm}}{electron or muon} events: category~1}
\label{ssec:XcheckCat1}

Category~1 top-quark and {\PW}+jets background is evaluated
with the SL data control sample
introduced in Section~\ref{sec-ttw}.
To relate event yields in the SL and SIG samples,
we use constraints derived from knowledge
of the {\PW}-boson polarization.
The polarization of the {\PW} boson governs the angular distribution
of leptons in the {\PW} boson rest frame.
Because forward-going leptons are boosted to higher momentum,
and backward-going leptons to lower momentum,
the {\PW}-boson polarization is directly related to the lepton
momentum spectrum in the laboratory frame.
{\PW}-boson polarization is predicted to high precision in the SM,
with calculations carried out to the next-to-next-to-leading order
for \ttbar events ~\cite{bib-wpolar-1} and to NLO
for {\PW}+jets events~\cite{bib-wpolar-2}.
The results of these calculations are consistent with
measurements~\cite{bib-CDF_topDecayWPol,bib-D0_topDecayWPol,bib-wpolar-3,Aad:2012ky}.

To construct a distribution sensitive to the {\PW}-boson polarization in
{\PW}$\,\rightarrow\ell\overline{\nu}$ ($\ell=\Pe$, {\Pgm}) events
(we include
{\PW}$\,\rightarrow\Pgt\overline{\nu} \rightarrow \ell\overline{\nu}\nu\overline{\nu}$
events in this category),
we calculate the angle \dthetat between the direction of the {\PW} boson in
the laboratory frame and the direction of the \Pe\ or {\Pgm}
in the {\PW} boson rest frame,
all defined in the transverse plane.
The \pt of the {\PW} boson is given by the vector sum of the \MET
and charged lepton \pt vectors.
When \dthetat is small,
the charged lepton is produced along the \pt direction of the {\PW} boson,
typically resulting in a high-\pt charged
lepton and a low-\pt neutrino (and therefore low \MET)
in the laboratory frame.
Such events usually appear in the SL sample.
Conversely, when \dthetat is large,
the charged lepton (neutrino) has lower (higher) \pt,
typically leading to larger \MET,
a charged lepton that fails our \Pe\ or {\Pgm} identification criteria,
and an event that appears as background in the signal samples.

\begin{figure}[thbp]
\begin{center}
   \includegraphics[width=0.9\cmsFigWidthThree]{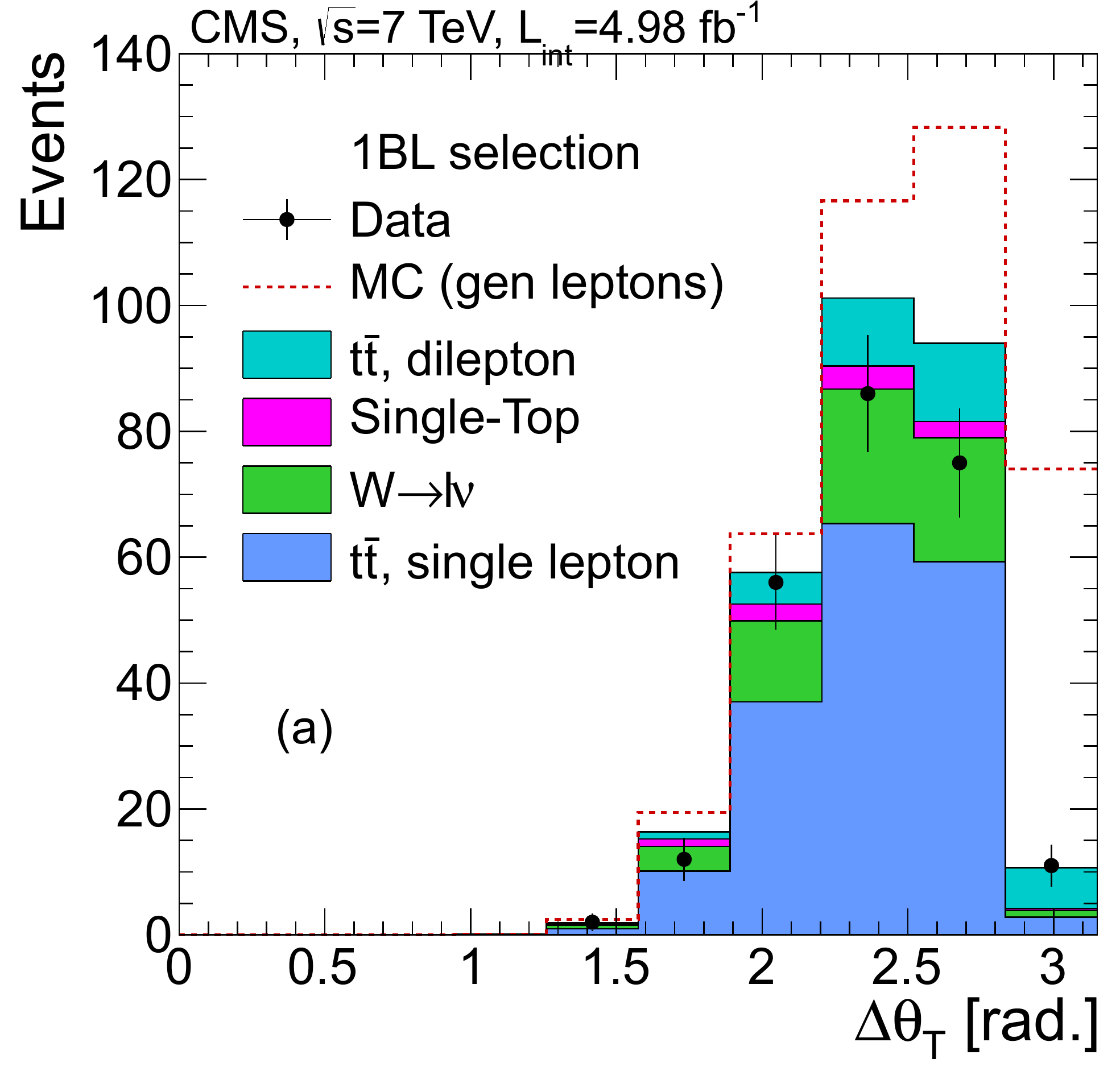}
   \includegraphics[width=0.9\cmsFigWidthThree]{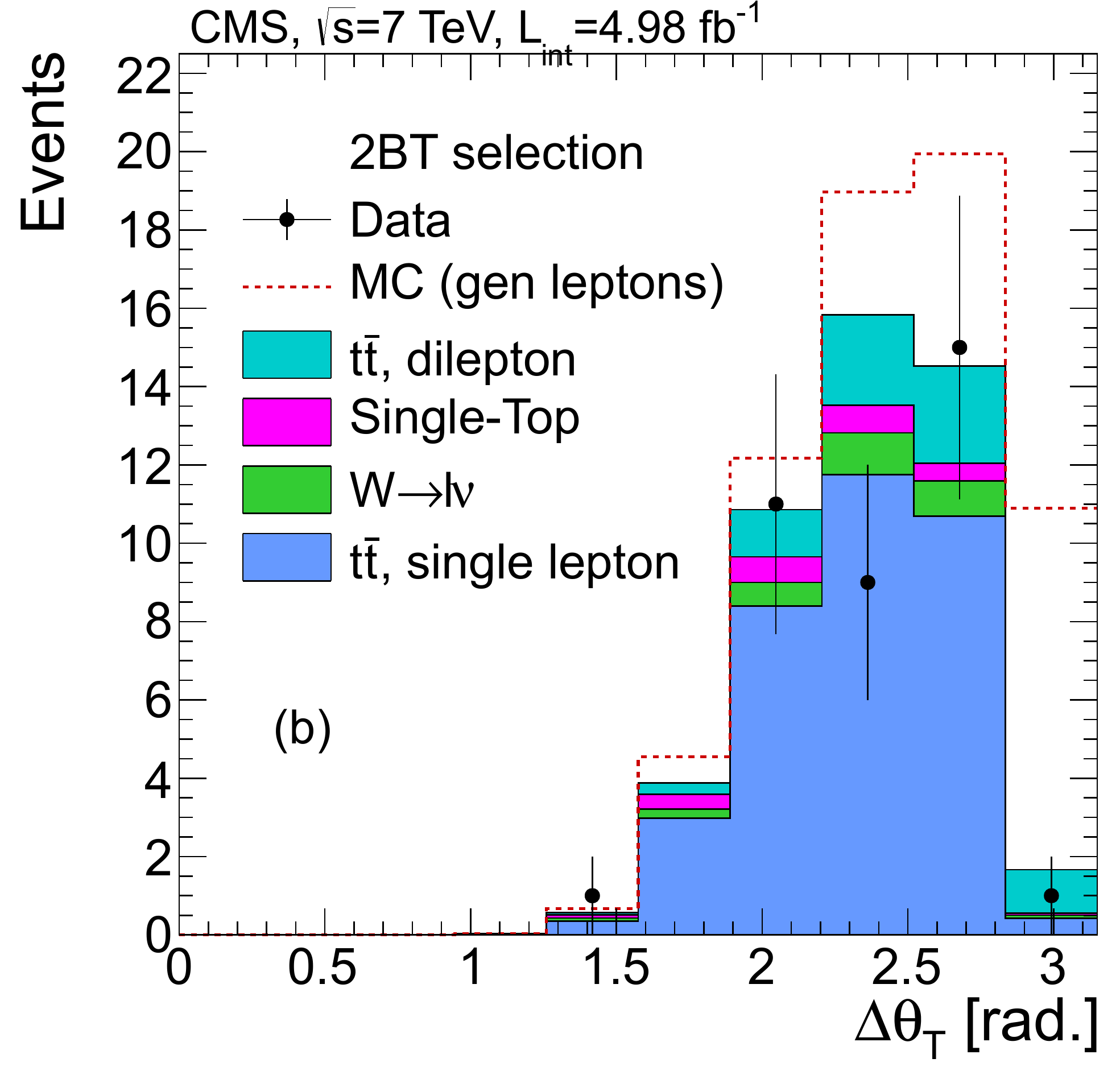}
   \includegraphics[width=0.9\cmsFigWidthThree]{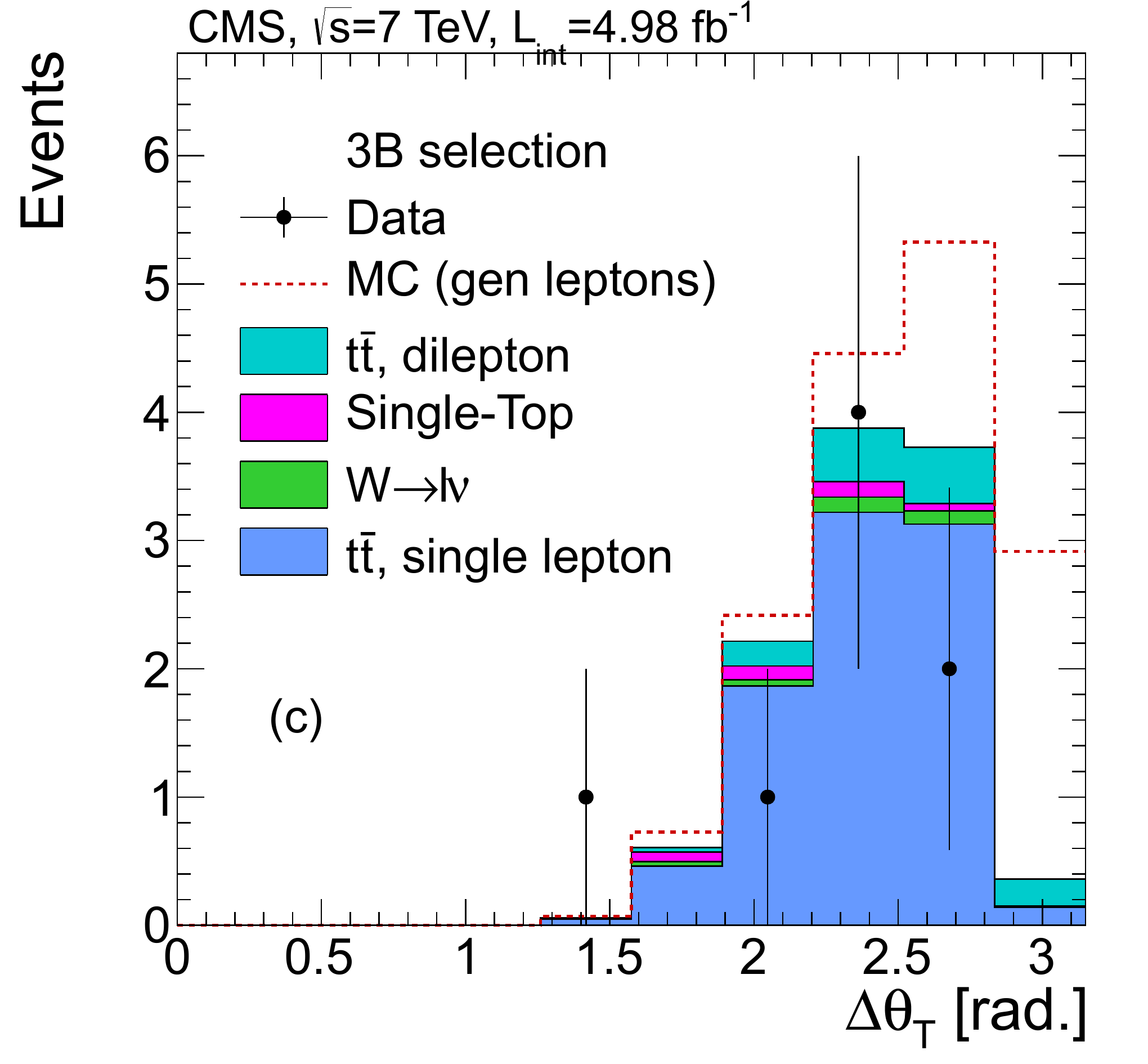}
\caption{
The distributions of \dthetat for events with a single \Pe\ or {\Pgm}
for the (a) 1BL, (b)~2BT, and (c)~3B selection criteria
except with a loosened \met restriction for (b)
as described in the text.
The stacked, filled histograms show simulated predictions for events
in the SL sample.
The dashed histogram shows the corresponding simulated prediction
in the limit of perfect charged lepton reconstruction.
The simulated results are normalized as in Fig.~\ref{fig-nbjets}.
}
\label{fig:dtheta_method}
\end{center}
\end{figure}

Figure~\ref{fig:dtheta_method} shows the distribution of \dthetat
in data and simulation
for SL events selected with the 1BL, 2BT, and 3B criteria,
except a looser $\MET$ requirement ($\met>250\gev$) is used
for the 2BT region to reduce statistical fluctuations.
These results can be compared to those expected
in the limit of perfect charged lepton reconstruction,
indicated by the dashed histograms in Fig.~\ref{fig:dtheta_method},
which show
the corresponding simulated predictions,
including simulation of the detector,
for top-quark and {\PW}+jets events
with a single {\PW}$\,\rightarrow\ell\nu$ decay,
where the {\Pe} or {\Pgm} needs only to be present at the generator level.
The difference between the dashed histogram and
the sum of the histograms with exactly one true
\Pe\ or {\Pgm} found in the event
represents {\PW}$\,\rightarrow\ell\nu$ events in which the \Pe\ or {\Pgm}
is either not reconstructed or does not meet the selection criteria
of Section~\ref{sec-selection}.

To estimate the \met distribution of category~1 events,
we measure the \met distribution of SL events in bins of~\dthetat.
The \MET distribution for each bin is then multiplied by an MC
scale factor, determined as follows.
The numerator equals the difference between the total yield from
single-lepton processes (the dashed histograms in Fig.~\ref{fig:dtheta_method})
and the subset of those events that enter the SL sample,
both determined for that bin.
The denominator equals the corresponding number of events that appear
in the SL sample from all sources.
The definition of the denominator therefore corresponds to the
SL observable in data.
The normalization of the \met distribution in each \dthetat bin
is thus given by the corresponding measured yield,
corrected by a scale factor that accounts for
the \Pe\ or {\Pgm} acceptance and reconstruction efficiency.
The corrected \met spectra from the different \dthetat bins are summed
to provide the total \met distribution for category~1 events.

\begin{table*}[!bt]
\topcaption{
The relative systematic uncertainties (\%) for the \MET-reweighting estimate of
the top-quark and {\PW}+jets background,
for category~1 (category~2) events.
}
\label{tab:cc-sys}
\scotchrule[l|rrrrr]
  & 1BL & 1BT & 2BL & 2BT & 3B \\
\hline
$\sigma(\WJets)$/$\sigma(\ttbar)$ ratio
                       & 0.1 (0.7) & 3.3 (3.9) & 0.1 (0.3) & 0.2 (0.3) & 0.6 (0.3) \\
Lepton efficiency      & 2.0 (2.0) & 2.0 (2.9) & 2.0 (2.0) & 2.0 (2.2) & 2.0 (2.0) \\
Top-quark \pt spectrum       & 0.1 (2.2) & 6.8 (0.7) & 0.6 (3.2) & 1.6 (0.7) & 1.6 (2.7)  \\
Jet energy scale       & 1.6 (3.0) & 5.0 (5.2) & 1.7 (2.1) & 1.2 (4.9) & 1.1 (4.1)  \\
Jet energy resolution  & 0.2 (0)   & 0.4 (1.6) & 0.2 (0.2) & 0.5 (0.4) & 0.3 (0.2) \\
{\cPqb}-tagging efficiency   & 0.2 (0.4) & 1.0 (2.8) & 0.4 (0.6) & 0.5 (0.5) & 0.3 (0.4) \\
MC closure             & 10 (4.7)  & 55 (29)   & 12 (5.1)  & 17 (16)   & 21 (6.6) \\
{\Pgt} visible energy  & --- (1.5) & --- (3.1) & --- (1.9) & --- (2.0) & --- (2.1) \\
\hline
Total                  & 10 (6.5)  & 56 (30)    & 12 (7.0)   & 17 (17)   & 21 (8.7) \\
\donescotchrule
\end{table*}

Systematic uncertainties
are summarized in Table~\ref{tab:cc-sys}.
To evaluate a systematic uncertainty associated with the
relative \ttbar and {\PW}+jets cross sections,
we vary the {\PW}+jets cross section by $\pm50$\%.
From studies of \zll\ events,
the systematic uncertainty associated with the lepton reconstruction
efficiency is determined to be~2\%.
A systematic uncertainty associated with the top-quark \pt spectrum
is evaluated by varying the {\PW}-boson \pt distribution
in the simulated \ttbar sample.
In these variations,
the number of events in the upper 10\% of the distribution changes by
two standard deviations of the corresponding result in data.
The systematic uncertainties associated with the jet energy scale,
jet energy resolution, and {\cPqb}-tagging efficiency are evaluated
as described in Section~\ref{sec-simplified}.
A systematic uncertainty to account for MC closure
is evaluated as described in Section~\ref{sec-ttw}.

\subsubsection{\texorpdfstring{$\Pgt\to\text{hadrons}$}{tau to hadrons}: category~2}
\label{ssec:XcheckCat2}

Category~2 top-quark and {\PW}+jets background is evaluated using
a single-muon data control sample.
The muon in the event is replaced with a simulated
hadronically decaying {\Pgt} (a {\Pgt} jet) of the same momentum.
To account for the addition of the {\Pgt} jet,
the initial selection criteria are
less restrictive than those of the nominal analysis.
We require two or more jets, $\MET>100$\gev,
and do not place restrictions on \HT or \dphin.
To ensure compatibility with the triggers used to define this
single-muon control sample,
the minimum muon \pt is set to 25\gev,
and the muon isolation requirement is also more stringent than the
nominal criterion of Section~\ref{sec-selection}.

The visible energy fraction of the {\Pgt} jet,
namely its visible energy divided by its \pt value,
is determined by sampling \pt-dependent MC distributions
(``response templates'')
of the {\Pgt} visible energy distribution,
for a given underlying value of {\Pgt} lepton~\pt.
The {\Pgt} jet visible energy is added to the event.
The modified event is then subjected to our standard signal region
selection criteria.
A normalization factor derived from simulation
accounts for the relative rates
of category 2 and single-muon control sample events.

The same systematic uncertainties are considered as for category~1 events.
In addition,
we evaluate an uncertainty for the {\Pgt} jet visible energy
by varying the {\Pgt} energy scale by $\pm3$\%~\cite{bib-cms-tau}.
Systematic uncertainties are summarized in Table~\ref{tab:cc-sys}.

\subsubsection{\texorpdfstring{\ttbar}{t t-bar} dilepton events: category~3}
\label{XcheckCat3}

The contribution of category 3 top-quark and {\PW}+jets background events
is determined using dilepton data control samples.
When both leptons are electrons or both are muons,
or when one is an electron and the other a muon
(where the \Pe\ or {\Pgm} can either be from a {\PW} boson or {\Pgt} decay),
we use simulated predictions to describe the shape of
the \MET distribution.
The normalization is derived from data,
by measuring the number of dilepton events that satisfy
loosened selection criteria for each class of events
({\Pe\Pe}, {\Pgm\Pgm}, or {\Pe\Pgm}) individually.
The measured value is multiplied by an MC scale factor,
defined by the number of corresponding \ttbar dilepton events that
satisfy the final selection criteria
divided by the number that satisfy the loosened criteria.

When one or both of the leptons is a hadronically decaying {\Pgt},
we apply a procedure similar to that described for category~2 events.
Data control samples of {\Pe\Pgm}+jets and {\Pgm\Pgm}+jets events are selected
with the loosened criteria of Section~\ref{ssec:XcheckCat2}.
One or both muons is replaced by a {\Pgt}-jet using MC response templates.
The signal sample selection criteria are applied to the modified events,
and the resulting \MET distributions normalized
by scaling the number of events in the respective control samples
with factors derived from MC simulation.

The \MET distributions of all six dilepton categories are summed to
provide the total category~3 prediction.
A systematic uncertainty is evaluated based on MC closure
in the manner described in Section~\ref{sec-qcd}.

\subsection{Summary of the data-based background estimates}
\label{sec-background-summary}

\begin{table*}[!t]
\centering
\topcaption{
The SM background estimates from the procedures of
Sections~\ref{sec-qcd}-\ref{sec-dthetat}
in comparison with the observed number of events in data.
The first uncertainties are statistical and the second systematic.
For the total SM estimates,
we give the results based both on the nominal and \MET-reweighting
methods to evaluate the top-quark and {\PW}+jets background.
}
\label{tab:results}
{\footnotesize
\scotchrule[l|ccccc]
  & 1BL & 1BT & 2BL & 2BT & 3B \\
\hline
QCD            & $28\pm3\pm12$   & $0.0\pm0.2\pm0.3$ & $4.7\pm1.3\pm2.8$ & $0.8\pm0.4\pm1.2$  & $1.0\pm0.5\pm0.5$ \\
\Zinvisible    & $154\pm20\pm32$ & $2.4\pm1.9\pm0.5$ & $32\pm5\pm20$     & $6.2\pm2.0\pm3.9$  & $4.7\pm1.3\pm6.5$ \\
top quark \& {\PW}+jets: & & & & & \\
\hspace*{2mm}
   nominal     & $337\pm30\pm63$ & $6.5\pm3.3\pm1.8$ & $123\pm17\pm19$   & $22.8\pm6.9\pm5.5$ & $8.8\pm4.0\pm1.8$ \\
\hspace*{2mm}
   \MET-reweighting
               & $295\pm16\pm17$ & $4.0\pm1.2\pm1.5$ & $116\pm8\pm8$    & $19.8\pm2.5\pm2.2$ & $13.6\pm3.2\pm1.2$ \\
\hline
Total SM:  & & & & & \\
\hspace*{2mm}
   nominal  & $519\pm36\pm72$ & $8.9\pm3.8\pm1.9$ & $159\pm18\pm28$   & $29.8\pm7.2\pm6.8$ & $14.4\pm4.2\pm6.8$ \\
\hspace*{2mm}
   \MET-reweighting
            & $477\pm26\pm38$ & $6.4\pm2.3\pm1.6$ & $153\pm10\pm22$   & $26.8\pm3.2\pm4.6$ & $19.3\pm3.5 \pm6.6$ \\
\hline
Data           & 478  & 11 & 146  & 45  & 22 \\
\donescotchrule
} % small
\end{table*}

A summary of the background estimates
is given in Table~\ref{tab:results}.
The results from the three categories of Section~\ref{sec-dthetat}
are summed to provide the total \MET-reweighting top-quark and {\PW}+jets prediction.
The estimates from the \MET-reweighting method
are seen to be consistent with those from the nominal method and
to yield smaller uncertainties.
Note that there are statistical correlations between the
nominal and \met-reweighting methods because
they both make use of the SIG-SL region
of Fig.~\ref{fig:roadmap-ttbar-nominal}.
However,
the nominal method relies on the SB
and SB-SL regions of
Fig.~\ref{fig:roadmap-ttbar-nominal},
while the \met-reweighting method does not.
The \met-reweighting method makes use of MC scale factors
and data selected with lepton-based triggers
(for category~2 and~3 events),
while the nominal method does not.
Furthermore, the systematic uncertainties of the two
methods are largely uncorrelated
(compare Tables~\ref{tab:ttwsyst} and~\ref{tab:cc-sys}).

The data are generally in good agreement with the SM expectations.
However,
for 2BT,
the data lie 1.1 and 2.2 standard deviations ($\sigma$) above the predictions
(including systematic uncertainties)
for the nominal and \MET-reweighting methods, respectively.
For 3B,
the corresponding deviations are 1.2$\sigma$ and 0.7$\sigma$.
Since these deviations are not significant,
we do not consider them further.

\begin{figure}[thbp]
\begin{center}
  \includegraphics[width=\cmsFigWidthThree]{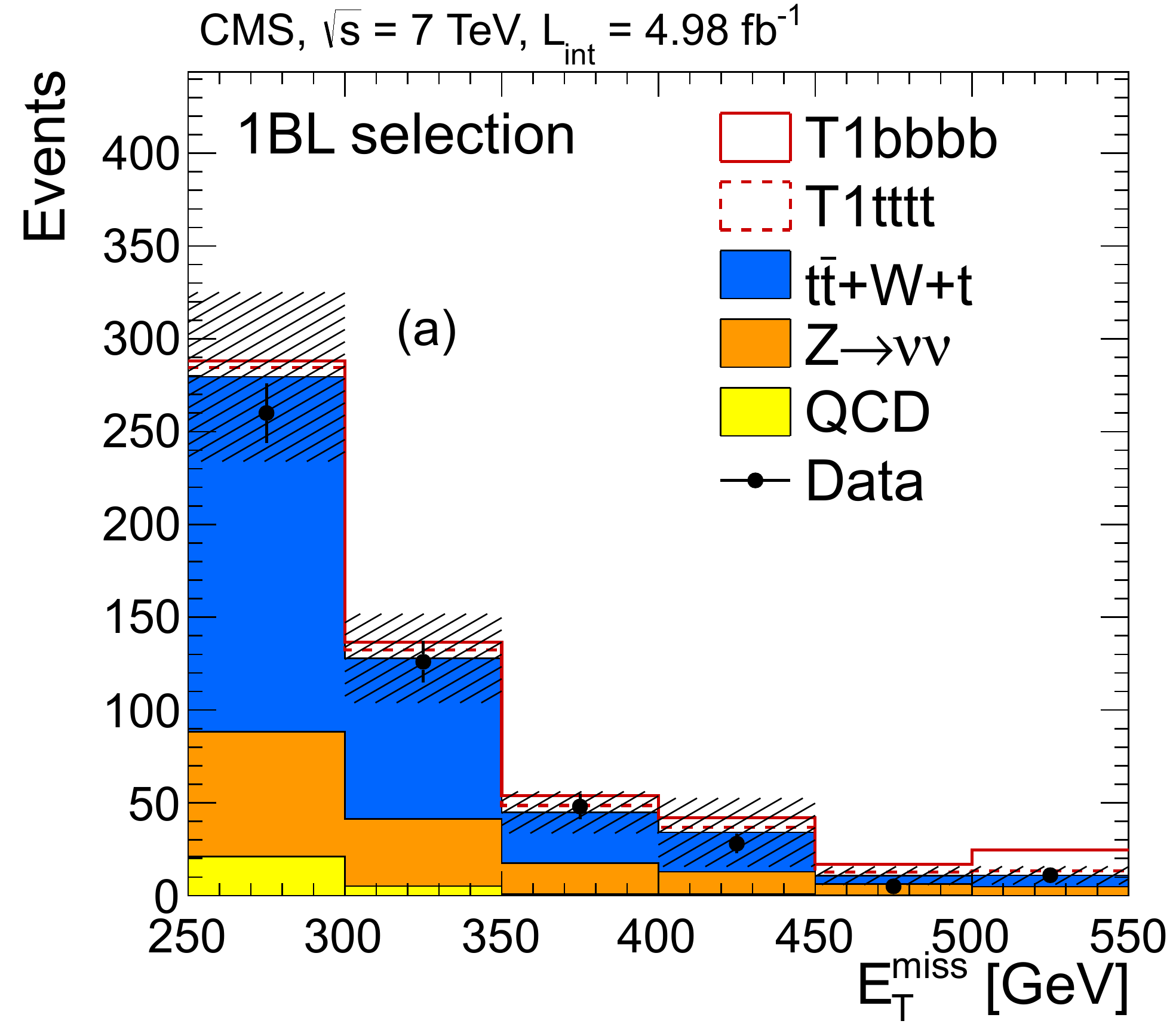}
  \includegraphics[width=\cmsFigWidthThree]{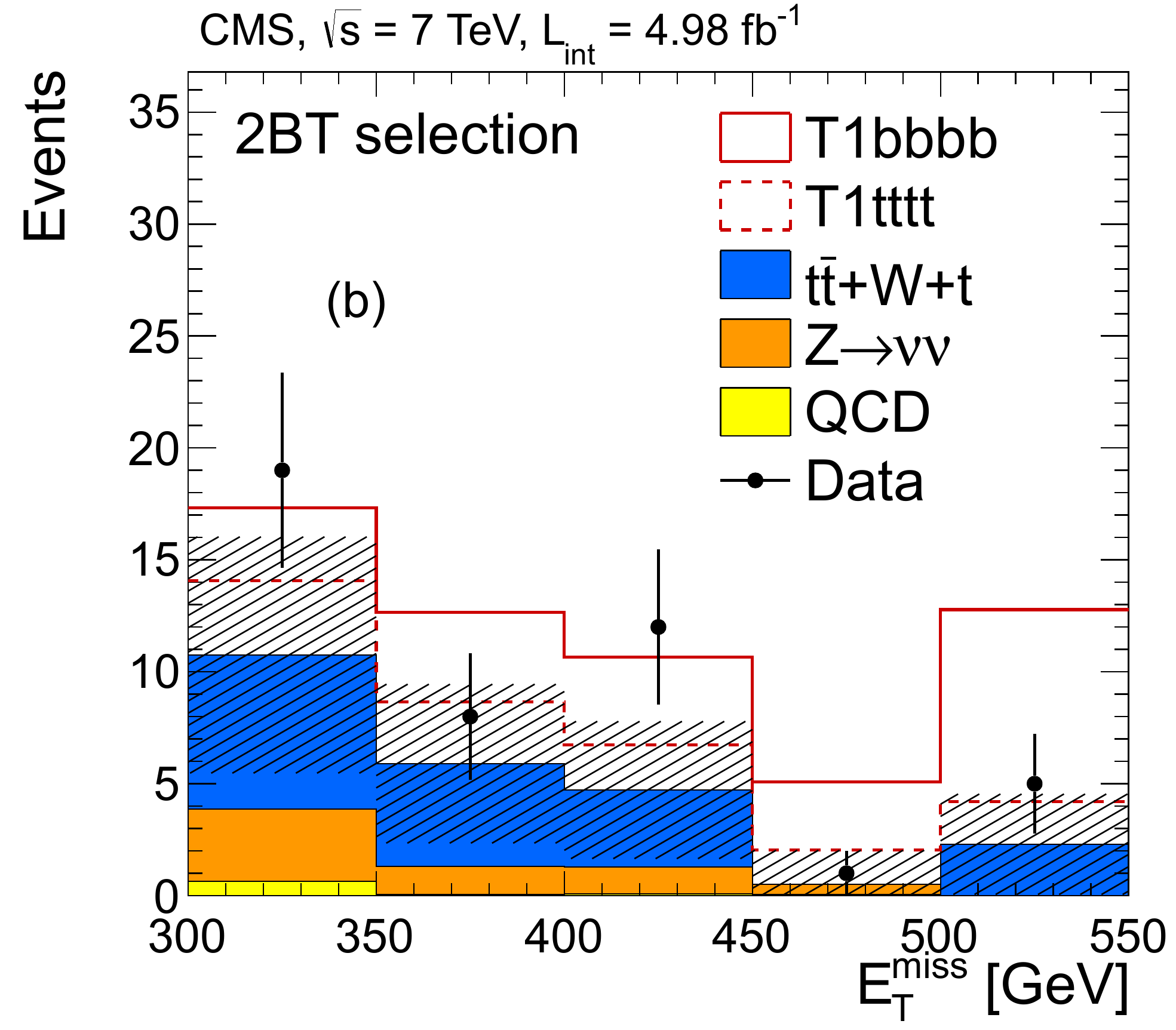}
  \includegraphics[width=\cmsFigWidthThree]{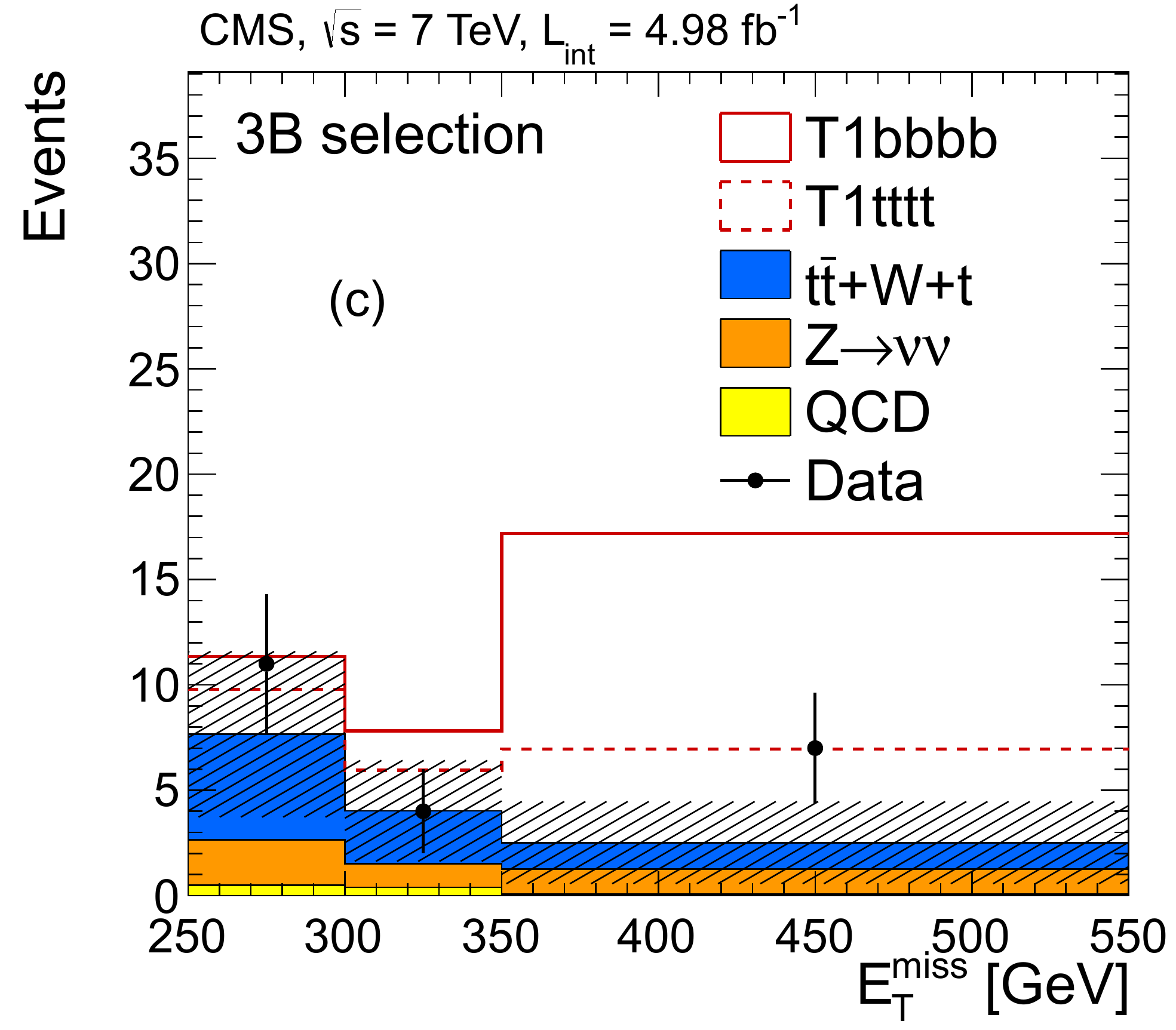}
\caption{
The data-based SM background predictions for \MET in the
(a)~1BL, (b)~2BT, and (c)~3B signal regions in comparison to data.
The top-quark and {\PW}+jets estimate is based on the nominal method.
The hatched bands show the total uncertainty on the prediction,
including systematic uncertainties.
The uncertainties are correlated between bins.
The open histograms show the expectations for the \tonebbbb (solid line)
and \tonetttt (dashed line) NP models,
both with $m_{\sGlu} = 925$\gev,
$m_{\mathrm{LSP}}=100$\gev,
and normalization to NLO+NLL.
}
\label{fig-met-predictions}
\end{center}
\end{figure}

As an illustration,
Fig.~\ref{fig-met-predictions}
presents the background predictions
in comparison to data for the 1BL, 2BT, and 3B selections.
These results are based on
the nominal top-quark and {\PW}+jets background estimate.

\section{Likelihood analysis}
\label{sec-likelihood}

We perform a global likelihood fit that simultaneously
determines the SM background and yield of a NP model,
using the background estimation techniques
of Section~\ref{sec-background}.
The likelihood analysis allows us to treat the SM backgrounds in a more
unified manner than is possible through
the collection of individual results in Table~\ref{tab:results}.
Furthermore,
it allows us to account for NP contributions to the control regions
(``signal contamination''),
as well as to the signal region,
in a comprehensive and consistent manner.

It is difficult to account for signal contamination
using the \met-reweight\-ing method,
in contrast to the nominal method.
Therefore, signal contamination is evaluated for the nominal method only.
Of the two NP scenarios we consider,
one of them,
the \tonetttt model,
exhibits non-negligible contamination of the SL samples,
while the other,
the \tonebbbb model,
does not.
Since the \tonebbbb model does not exhibit significant signal contamination,
we employ both the nominal- and \MET-reweighting-based likelihood fits for this model.
For the \tonetttt model,
we employ only the likelihood fit based on the nominal method.

\begin{table}[!tb]
\topcaption{
The observables (number of data events)
of the likelihood analysis for the nominal method,
representing the signal region and ten control regions.
The seven observables listed in the upper portion of the table are
subject to contributions from the signal model in our analysis.
The \met SB region corresponds to $150<\met<250$\gev
while the SIG regions correspond to the \met regions
listed in Table~\ref{tab-sig-regions}.
The low \dphin region corresponds to $\dphin<4.0$.
}
\label{tab:lhbins-notation}
\scotchrule[l|l]
      SIG &  Standard selection, \met SIG region  \\
      SB  &  Standard selection, \met SB region \\
      SIG--LDP &  Standard selection, \met SIG/low \dphin region\\
      SB--LDP  &  Standard selection, \met SB/low \dphin region \\
      SIG--SL  &  Single-lepton selection, \met SIG region \\
      SB--S\Pe\  &  Single-electron selection, \met SB region \\
      SB--S{\Pgm}   &  Single-muon selection, \met SB region \\
      \hline
      SIG--\Pe\Pe\ &  \zee\ selection, \met SIG region \\
      SB--\Pe\Pe\  &  \zee\ selection, \met SB region  \\
      SIG--{\Pgm\Pgm} & \zmumu\ selection, \met SIG region \\
      SB--{\Pgm\Pgm} & \zmumu\ selection, \met SB region  \\
\donescotchrule
\end{table}

For the nominal method,
the data are divided into 11 mutually exclusive bins,
corresponding to the 11 observables
listed in Table~\ref{tab:lhbins-notation},
where each ``observable'' corresponds to the number of data events
recorded for that bin.
Note that the SB-SL events of Fig.~\ref{fig:roadmap-ttbar-nominal}
are divided into two components,
one for electrons (denoted SB--S{\Pe})
and the other for muons (denoted SB--S{\Pgm}),
because their trigger efficiencies and uncertainties differ.
Similarly,
the reconstruction efficiencies of
{\zee} and {\zmumu} events differ,
so we divide the {\zll} events of Section~\ref{sec-zjets}
according to the lepton flavor.
We further divide the {\zll} events according to whether they
appear in the sideband ($150<\met<250$\gev)
or signal regions (Table~\ref{tab-sig-regions}) of \met.
The four {\zll} samples are denoted
{SIG-\Pe\Pe} and SIG-{\Pgm\Pgm}
for events in the signal regions,
and SB-{\Pe\Pe} and SB-{\Pgm\Pgm}
for events in the sideband region.

The likelihood model provides a prediction for the mean expected value of each
observable in terms of the parameters of the
signal and background components.
The likelihood function is the product of 11 Poisson
probability density functions,
one for each observable,
$\beta$ distributions~\cite{bib-johnson-1995}
that parametrize efficiencies and acceptances,
and $\beta^\prime$ distributions~\cite{bib-johnson-1995}
that account for systematic uncertainties
and uncertainties on external parameters.
(External parameters include such quantities
as the acceptance ${\cal A}$
and scale factors between the samples with
loose and nominal b-tagging requirements
discussed in Section~\ref{sec-zjets}.)
The new physics scenarios considered here can contribute significantly to the seven
observables listed
in the upper portion of Table~\ref{tab:lhbins-notation}.
In our model,
the relative contributions of NP to these seven observables are
taken from the NP model under consideration.
The NP yield in the SIG bin is a free parameter.
The NP contributions to the other six bins thus depend on
the NP yield in the SIG bin.

Analogous procedures are used to define the likelihood
function for the \met-reweighting method,
with simplifications since there is no SB
region in this case.

The likelihood function is used to set limits on NP models.
Upper limits at 95\% confidence level (CL) are evaluated
taking into account the effects of variation of the
external parameters and their correlations.
All upper limits are determined
using a modified frequentist technique (\cls)~\cite{junk1999,bib-cls}.

\section{Limits on the \texorpdfstring{\tonebbbb and \tonetttt models}{T1bbbb and T1tttt}}
\label{sec-simplified}

Simulated \tonebbbb and \tonetttt event samples are
generated for a range of gluino and LSP masses using \PYTHIA,
with $m_{\mathrm{LSP}}<m_{\sGlu}$.
For increased efficiency when performing scans
over the SMS parameter space (see below),
we base simulation of the CMS detector response
on the fast simulation program~\cite{bib-cms-fastsim},
accounting for modest differences observed with respect
to the \GEANTfour simulation.

\begin{table}[t]
\topcaption{
The relative systematic uncertainties (\%) for the signal efficiency
of the T1bbbb SMS model
with $m_{\sGlu} = 925$\gev and $m_{\mathrm{LSP}}=100$\gev.
}
\scotchrule[l|rrrrr]
                              & 1BL  & 1BT & 2BL  & 2BT  & 3B \\
\hline
Jet energy scale              & 2.1  & 11  & 2.1  & 3.5  & 1.9  \\
Unclustered energy            & 0.2  & 0.8 & 0.2  & 0.2  & 0.2  \\
Jet energy resolution         & 1.0  & 2.0 & 1.0  & 1.0  & 1.0  \\
Pileup                        & 1.0  & 1.0 & 1.0  & 1.0  & 1.0  \\
{\cPqb}-jet tagging efficiency    & 0.8  & 0.9 & 3.8  & 3.9  & 9.0  \\
Trigger efficiency            & 3.6  & 3.6 & 3.6  & 3.6  & 3.6  \\
Parton distribution functions & 0.4  & 1.6 & 0.4  & 0.7  & 0.5  \\
Anomalous \MET                & 1.0  & 1.0 & 1.0  & 1.0  & 1.0  \\
Lepton veto                   & 3.0  & 3.0 & 3.0  & 3.0  & 3.0  \\
Luminosity                    & 2.2  & 2.2 & 2.2  & 2.2  & 2.2  \\
\hline
Total uncertainty             & 5.9  & 12  & 7.0  & 7.6  & 11 \\
\donescotchrule
\label{tab:lm9syst}
\end{table}

Systematic uncertainties on signal efficiency
are summarized in Table~\ref{tab:lm9syst},
using the \tonebbbb benchmark model as an example.
A systematic uncertainty associated with the jet energy scale
is evaluated by varying this scale by its {\pt}- and
$\eta$-dependent uncertainties.
A systematic uncertainty associated with unclustered energy
is evaluated by varying  the transverse energy in an event
that is not clustered into a physics object by $\pm 10\%$.
The systematic uncertainties associated with the correction
to the jet energy resolution,
the pileup
reweighting method mentioned in Section~\ref{sec-selection},
the {\cPqb}-jet tagging efficiency scale factor,
and the trigger efficiency,
are evaluated by varying the respective quantities by their uncertainties.
The uncertainty for the trigger efficiency
includes a 2.5\% uncertainty for the plateau efficiency.
Systematic uncertainties associated with the parton distribution
functions are evaluated following the recommendations
of Ref.~\cite{PDF4LHC}.
The systematic uncertainty associated with anomalous \met values,
caused by beam background and reconstruction effects,
is~1\%.
The systematic uncertainty associated with the lepton veto
is determined from studies of {\zll} events in data to be 3.0\%.
The uncertainty in the luminosity determination
is 2.2\%~\cite{CMS-PAS-SMP-12-008}.

We determine 95\% CL upper limits on the SMS cross sections
as a function of the gluino and LSP masses.
Using the NLO+NLL cross section as a reference,
we also evaluate 95\% CL exclusion curves.
The jet energy scale, unclustered energy,
parton distribution function,
and {\cPqb}-jet tagging efficiency uncertainties are evaluated for each scan point.
Other uncertainties are fixed to the values in Table~\ref{tab:lm9syst}.
For each choice of gluino and LSP mass,
we use the combination of
the top-quark and {\PW}+jets background estimation method,
and the signal selection
(Table~\ref{tab-sig-regions}),
that provides the best expected limit.
We do not include results for points near the
$m_{\sGlu} = m_{\mathrm{LSP}}$ diagonal
because of neglected uncertainties from initial-state radiation (ISR),
which are large in this region.
Specifically,
we remove from consideration any point for which the
signal efficiency changes by more than 50\% when the
ISR radiation in \PYTHIA is (effectively) turned off.

For the \tonebbbb model,
the \met-reweighting method is always found to provide the best expected result:
we therefore use this method to determine the \tonebbbb limits.
The \met-reweighting method incorporates an additional constraint
compared to the nominal method,
namely the normalization of the SM
prediction for the \met distribution from the SIG-SL sample
(Fig.~\ref{fig:roadmap-ttbar-nominal}),
and not merely the \met distribution shape.
As a consequence,
it has greater discrimination power against NP scenarios.

\begin{figure}[tbhp]
\begin{center}
    \includegraphics[width=\cmsFigWidthTwo]{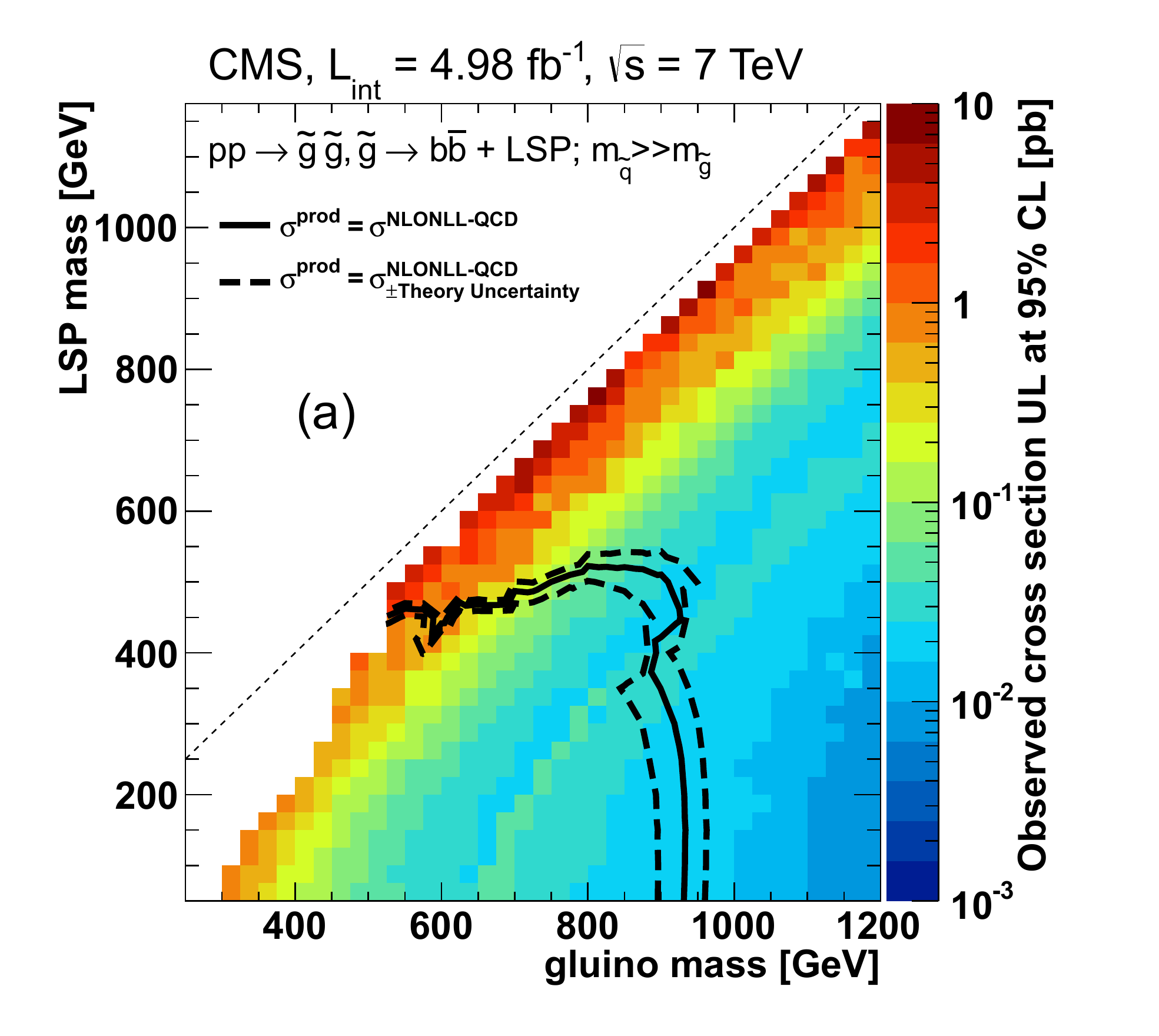}
    \includegraphics[width=\cmsFigWidthTwo]{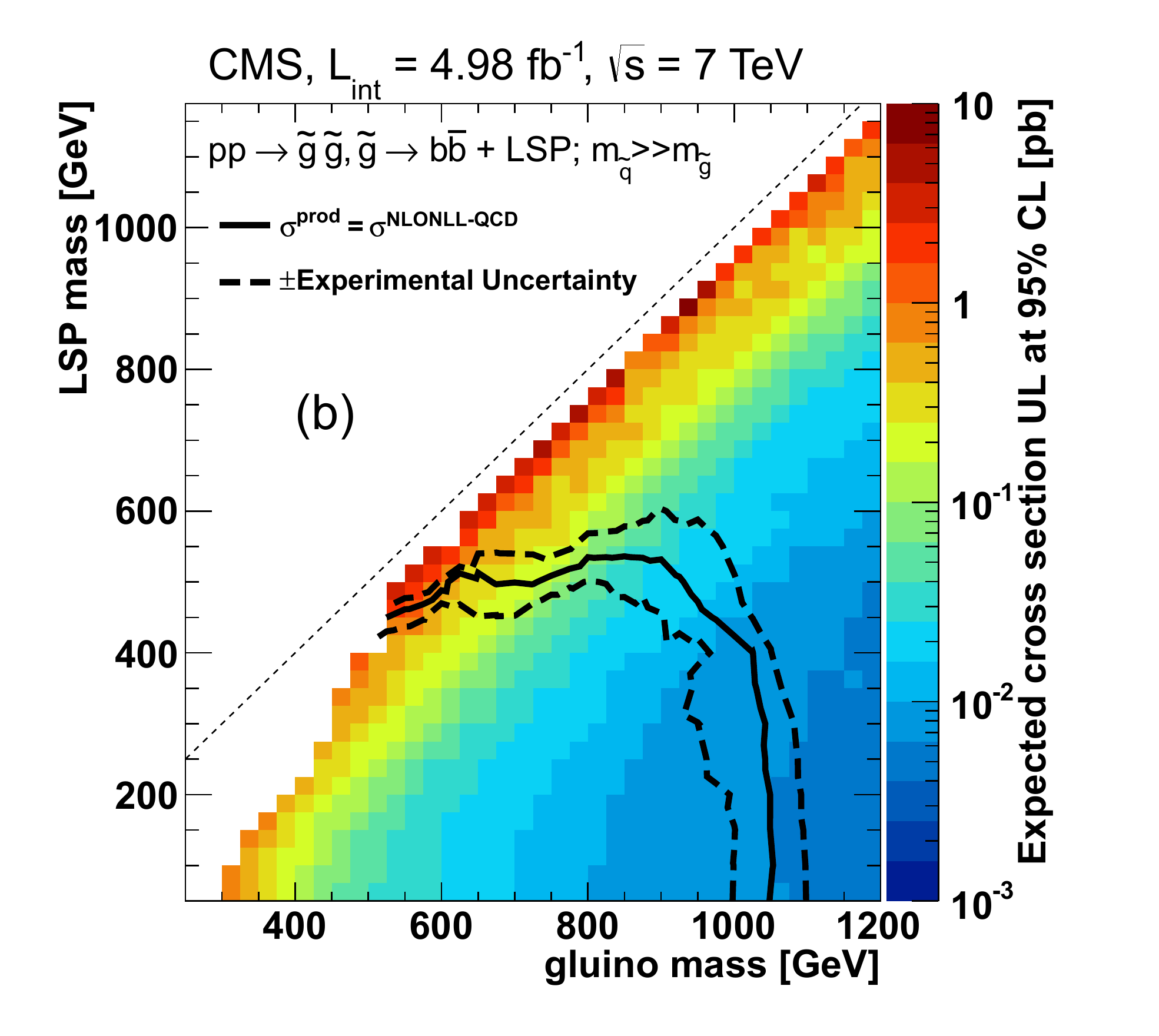}
\caption{
(a) The 95\% CL observed cross section upper limits (UL) for the \tonebbbb SMS model,
based on the \MET-reweighting method to evaluate the top-quark and {\PW}+jets background.
For each point, the selection that provides the best expected
cross section limit is used.
The solid contour shows the 95\%~CL exclusion limits on the gluino and LSP masses
using the NLO+NLL cross section for new physics.
The dashed contours represent the theory uncertainties.
(b)~The corresponding expected limits.
The dashed contours represent the uncertainties on
the SM background estimates.
}
\label{fig-t14b-kristen}
\end{center}
\end{figure}

\begin{figure}[tbhp]
\begin{center}
    \includegraphics[width=\cmsFigWidthTwo]{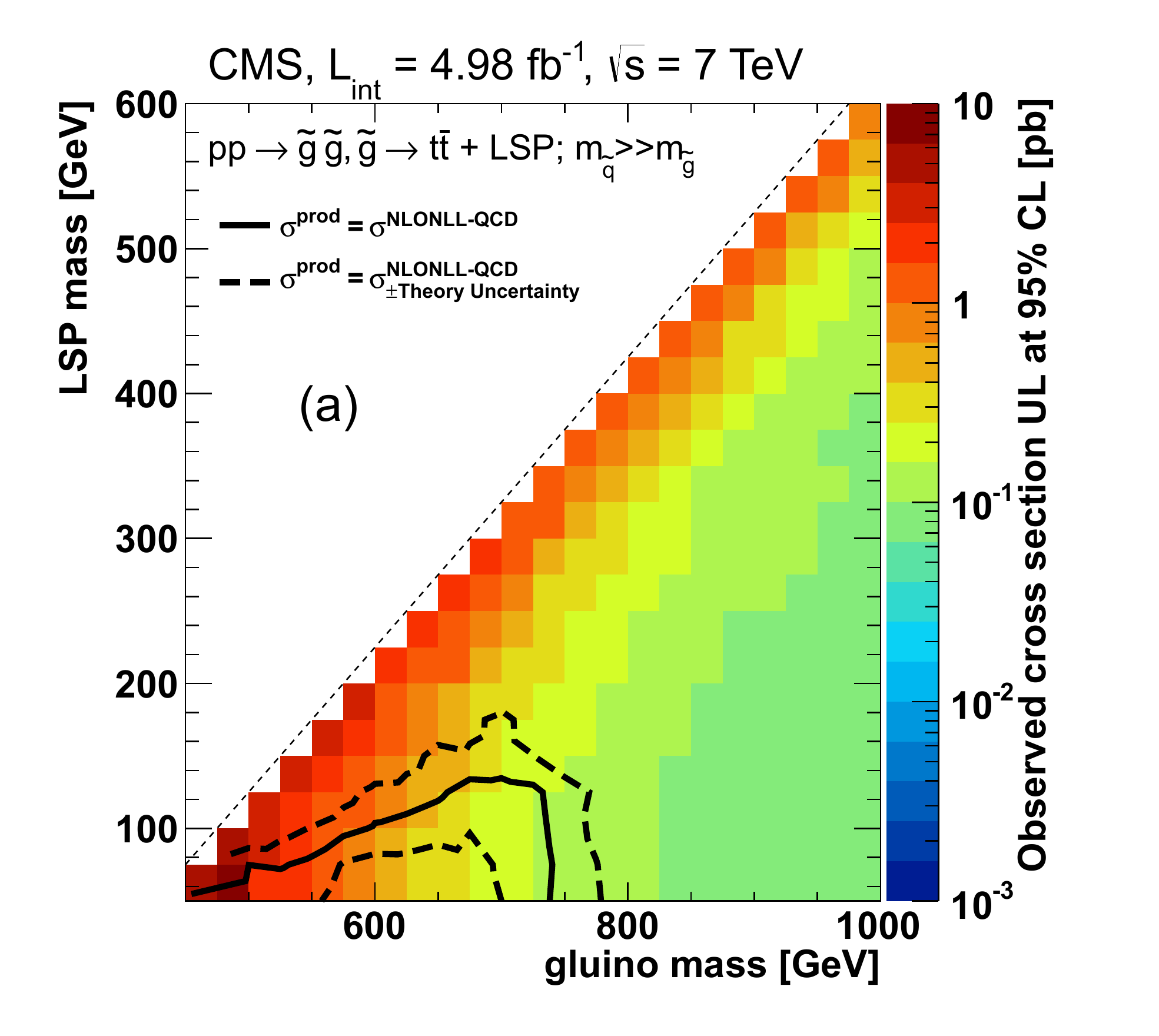}
    \includegraphics[width=\cmsFigWidthTwo]{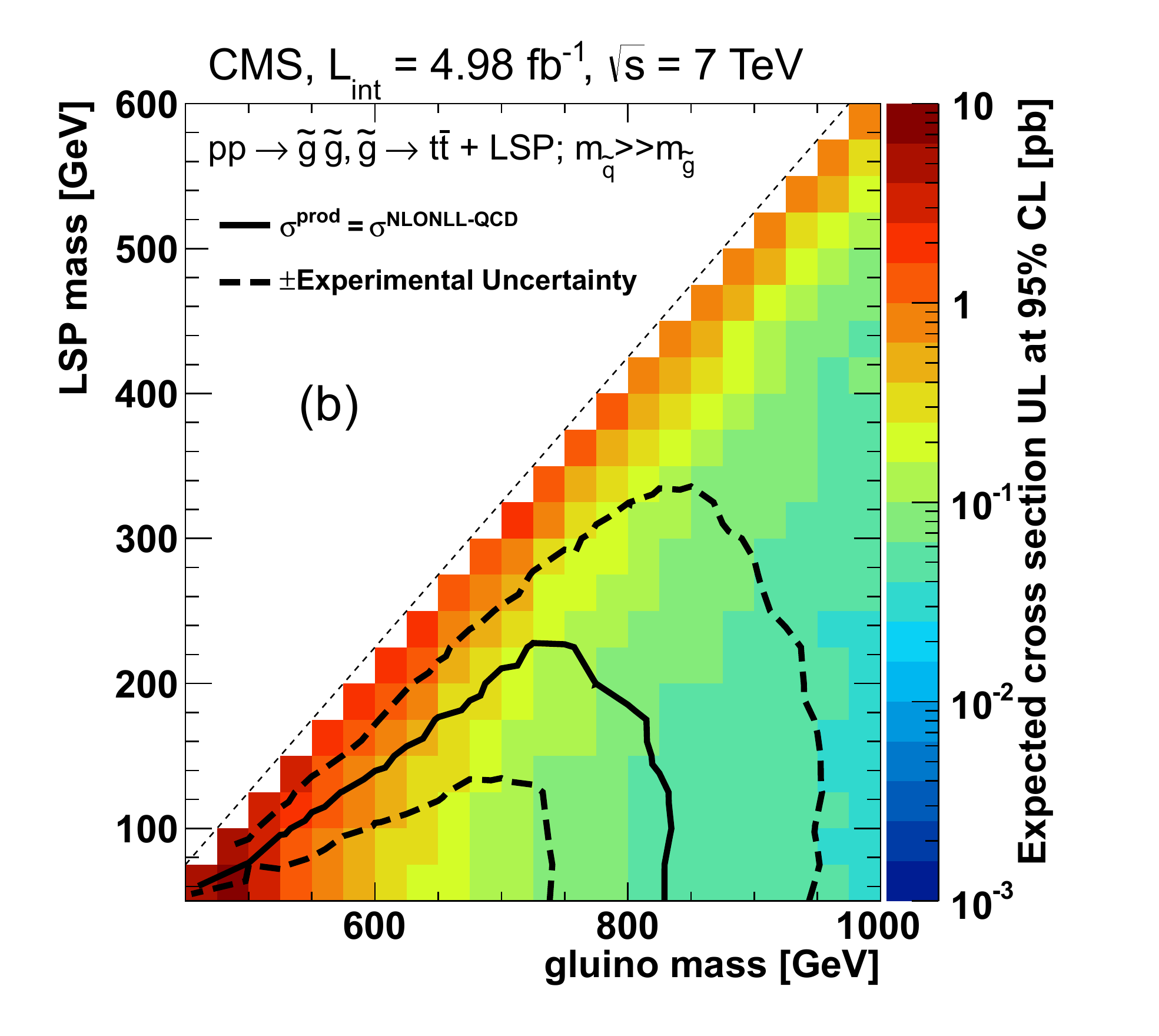}
\caption{
(a) The 95\% CL observed cross section upper limits (UL) for the \tonetttt SMS model,
based on the nominal method to evaluate the top-quark and {\PW}+jets background.
For each point, the selection that provides the best expected
cross section limit is used
(the best selection is virtually always 3B).
The solid contour shows the 95\%~CL exclusion limits on the gluino and LSP masses
using the NLO+NLL cross section for new physics.
The dashed contours represent the theory uncertainties.
(b)~The corresponding expected limits.
The dashed contours represent the uncertainties on
the SM background estimates.
}
\label{fig-t14t}
\end{center}
\end{figure}

The results for \tonebbbb are shown
in Fig.~\ref{fig-t14b-kristen}(a).
The 1BT selection is found to provide the best expected result
in the bottom right corner of the distribution,
corresponding to the region of large gluino-LSP mass splitting.
The 2BT selection is best
for the swath roughly parallel to the diagonal
defined by gluino masses between around 650 and 900\gev
along the bottom edge of the plot.
The 3B selection is generally best elsewhere.
The solid contour shows the 95\% CL exclusion curve for
the reference cross section.
The zigzagging structure around
$m_{\sGlu} = 900$\gev, $m_{\mathrm{LSP}}=450$\gev
is due to the transition from the region where 3B is the
best expected selection to that where 2BT is best,
in conjunction with the slight excess observed in data
for the 2BT selection in comparison with the SM prediction
for the \met-reweighting method
(Section~\ref{sec-background-summary}).
The dashed contours
represent the results when the reference cross section
is varied by the theory uncertainty~\cite{bib-NLO-NLL}.
Our results improve those of Ref.~\cite{bib-cms-mt2-2011}
for large LSP mass values.
For example,
for gluino masses around 800\gev,
we extend the exclusion of the reference cross section
from an LSP mass of about 400\gev~\cite{bib-cms-mt2-2011}
to about 500\gev,
where these numerical values are
given by the observed results minus the one standard deviation
theory uncertainties.

Fig.~\ref{fig-t14b-kristen}(b) shows the best expected results
for the \tonebbbb model.
In this case the dashed contours represent the results when the
SM background estimates (Table~\ref{tab:results})
are varied by their uncertainties.

Even with the selection requirement against ISR events described above,
the effect of ISR events can be significant for \tonebbbb scenarios
that lie within three or four cell widths of the diagonal,
where the size of a cell is indicated by
the small boxes visible near the diagonal in Fig.~\ref{fig-t14b-kristen}.
We do not account for the effect of ISR for these scenarios.
However,
for gluino masses less than around 500\gev,
and for points more than three to four cell widths from the diagonal,
the effect of ISR is negligible for the points that are retained.
This includes the entire exclusion region
for the NLO+NLL reference cross section.

The corresponding results for the \tonetttt model
are presented in Fig.~\ref{fig-t14t}.
Our \tonetttt results are based on the nominal top-quark and {\PW}+jets background
estimation method
for the reason stated in Section~\ref{sec-likelihood}.
In this case,
the best expected selection is essentially always~3B.
Note that the observed limits for \tonetttt,
shown in Fig~\ref{fig-t14t}(a),
are not as stringent as the expected limits,
shown in Fig~\ref{fig-t14t}(b),
because of the slight excess of data events in the 3B
sample for the nominal method,
compared to the SM expectation (Table~\ref{tab:results}).

\section{Summary}
\label{sec-summary}

In this paper,
we present a search for an anomalous rate of events with
three or more jets,
at least one, two, or three tagged bottom-quark jets,
no identified, isolated leptons,
and large missing transverse energy~\MET.
The study is based on a sample of proton-proton collision data
collected at $\sqrt{s}=7$\TeV
with the CMS detector at the LHC during 2011,
corresponding to an integrated luminosity of 4.98\fbinv.
The principal standard model backgrounds,
arising from top-quark, {\PW}+jets, {\Z}+jets, and QCD-multijet events,
are evaluated from data.
We introduce a variable \dphin
that allows us to address the QCD-multijet background with a simple approach.
The top-quark and {\PW}+jets background is evaluated with two complementary methods,
which yield consistent results.
In the \met-reweighting method to evaluate the top-quark and {\PW}+jets background,
we introduce a technique based on the {\PW} polarization in
\ttbar and {\PW}+jets events.
Our analysis is performed in a likelihood framework
in order to account for backgrounds,
and for new physics contamination of the control regions,
in a unified and consistent manner.

We find no evidence for a significant excess of events beyond the expectations
of the standard model and
set limits on new physics in the context of the {\cPqb}-jet-rich
\tonebbbb and \tonetttt simplified model spectra,
in which new strongly interacting particles decay to two {\cPqb}-quark jets,
or two t-quark jets,
plus an undetected particle.
For the \tonebbbb scenario,
our results improve on those of Ref.~\cite{bib-cms-mt2-2011} for large LSP masses.

\section*{Acknowledgements}

\hyphenation{Bundes-ministerium Forschungs-gemeinschaft Forschungs-zentren} We congratulate our colleagues in the CERN accelerator departments for the excellent performance of the LHC machine. We thank the technical and administrative staff at CERN and other CMS institutes. This work was supported by the Austrian Federal Ministry of Science and Research; the Belgian Fonds de la Recherche Scientifique, and Fonds voor Wetenschappelijk Onderzoek; the Brazilian Funding Agencies (CNPq, CAPES, FAPERJ, and FAPESP); the Bulgarian Ministry of Education and Science; CERN; the Chinese Academy of Sciences, Ministry of Science and Technology, and National Natural Science Foundation of China; the Colombian Funding Agency (COLCIENCIAS); the Croatian Ministry of Science, Education and Sport; the Research Promotion Foundation, Cyprus; the Ministry of Education and Research, Recurrent financing contract SF0690030s09 and European Regional Development Fund, Estonia; the Academy of Finland, Finnish Ministry of Education and Culture, and Helsinki Institute of Physics; the Institut National de Physique Nucl\'eaire et de Physique des Particules~/~CNRS, and Commissariat \`a l'\'Energie Atomique et aux \'Energies Alternatives~/~CEA, France; the Bundesministerium f\"ur Bildung und Forschung, Deutsche Forschungsgemeinschaft, and Helmholtz-Gemeinschaft Deutscher Forschungszentren, Germany; the General Secretariat for Research and Technology, Greece; the National Scientific Research Foundation, and National Office for Research and Technology, Hungary; the Department of Atomic Energy and the Department of Science and Technology, India; the Institute for Studies in Theoretical Physics and Mathematics, Iran; the Science Foundation, Ireland; the Istituto Nazionale di Fisica Nucleare, Italy; the Korean Ministry of Education, Science and Technology and the World Class University program of NRF, Korea; the Lithuanian Academy of Sciences; the Mexican Funding Agencies (CINVESTAV, CONACYT, SEP, and UASLP-FAI); the Ministry of Science and Innovation, New Zealand; the Pakistan Atomic Energy Commission; the Ministry of Science and Higher Education and the National Science Centre, Poland; the Funda\c{c}\~ao para a Ci\^encia e a Tecnologia, Portugal; JINR (Armenia, Belarus, Georgia, Ukraine, Uzbekistan); the Ministry of Education and Science of the Russian Federation, the Federal Agency of Atomic Energy of the Russian Federation, Russian Academy of Sciences, and the Russian Foundation for Basic Research; the Ministry of Science and Technological Development of Serbia; the Secretar\'{\i}a de Estado de Investigaci\'on, Desarrollo e Innovaci\'on and Programa Consolider-Ingenio 2010, Spain; the Swiss Funding Agencies (ETH Board, ETH Zurich, PSI, SNF, UniZH, Canton Zurich, and SER); the National Science Council, Taipei; the Scientific and Technical Research Council of Turkey, and Turkish Atomic Energy Authority; the Science and Technology Facilities Council, UK; the US Department of Energy, and the US National Science Foundation.

Individuals have received support from the Marie-Curie programme and the European Research Council (European Union); the Leventis Foundation; the A. P. Sloan Foundation; the Alexander von Humboldt Foundation; the Belgian Federal Science Policy Office; the Fonds pour la Formation \`a la Recherche dans l'Industrie et dans l'Agriculture (FRIA-Belgium); the Agentschap voor Innovatie door Wetenschap en Technologie (IWT-Belgium); the Council of Science and Industrial Research, India; the Compagnia di San Paolo (Torino); and the HOMING PLUS programme of Foundation for Polish Science, cofinanced from European Union, Regional Development Fund.

\bibliography{auto_generated}   % will be created by the tdr script.

\providecommand{\href}[2]{#2}\begingroup\raggedright\begin{thebibliography}{10}%
\makeatletter
\providecommand{\hrefCMSnoop }[0]{\@secondoftwo}%
\makeatother
\providecommand{\doi}{\texttt{doi:}\begingroup \urlstyle{tt}\Url}

\bibitem{bib-rparity}
\hrefCMSnoop {} {G.~R. Farrar and P.~Fayet, ``Phenomenology of the production,
  decay, and detection of new hadronic states associated with supersymmetry'',}
  \textit{ Phys. Lett. B} \textbf{ 76} (1978) 575,
  \href{http://dx.doi.org/10.1016/0370-2693(78)90858-4}{\doi{10.1016/0370-2693(78)90858-4}}.

\bibitem{bib-susy}
\hrefCMSnoop {} {J.~Wess and B.~Zumino, ``Supergauge transformations in four
  dimensions'',} \textit{ Nucl. Phys. B} \textbf{ 70} (1974) 39,
  \href{http://dx.doi.org/10.1016/0550-3213(74)90355-1}{\doi{10.1016/0550-3213(74)90355-1}}.

\bibitem{Martin:1997ns}
\hrefCMSnoop {} {S.~P. Martin, ``{A supersymmetry primer}'',} (1997).
\href{http://www.arXiv.org/abs/hep-ph/9709356}{\texttt{ arXiv:hep-ph/9709356}}.
%%CITATION = HEP-PH/9709356;%%.

\bibitem{bib-cms-ra1b}
\hrefCMSnoop {} {{ CMS} Collaboration, ``Search for supersymmetry in events
  with b jets and missing transverse energy at the {LHC}'',} \textit{ J. High
  Energy Phys.} \textbf{ 07} (2011) 113,
  \href{http://dx.doi.org/10.1007/JHEP07(2011)113}{\doi{10.1007/JHEP07(2011)113}}.

\bibitem{bib-atlas-susyb}
\hrefCMSnoop {} {{ ATLAS} Collaboration, ``Search for supersymmetry in pp
  collisions at {$\sqrt{s}=7\TeV$} in final states with missing transverse
  momentum and b jets'',} \textit{ Phys. Lett. B} \textbf{ 701} (2011) 398,
  \href{http://dx.doi.org/10.1016/j.physletb.2011.06.015}{\doi{10.1016/j.physletb.2011.06.015}}.

\bibitem{bib-atlas-susyb-2}
\hrefCMSnoop {} {{ ATLAS} Collaboration, ``Search for supersymmetry in pp
  collisions at $\sqrt{s}=7$~TeV in final states with missing transverse
  momentum and b-jets with the {ATLAS} Detector'',} \textit{ Phys. Rev. D}
  \textbf{ 85} (2012) 112006,
  \href{http://dx.doi.org/10.1103/PhysRevD.85.112006}{\doi{10.1103/PhysRevD.85.112006}}.

\bibitem{bib-cms-mt2-2011}
\hrefCMSnoop {} {{ CMS} Collaboration, ``Search for supersymmetry in hadronic
  final states using {$\mathrm{M_{T2}}$} in pp collisions at
  {$\sqrt{s}=7$\TeV}'',} (2012).
  \href{http://www.arXiv.org/abs/1207.1798}{\texttt{ arXiv:1207.1798}}.
In press in J. High Energy Phys.
%%CITATION = HEP-EX 1207.1798;%%.

\bibitem{bib-atlas-susyb-3}
\hrefCMSnoop {} {{ ATLAS} Collaboration, ``Search for top and bottom squarks
  from gluino pair production in final states with missing transverse energy
  and at least three {\b}-jets with the {ATLAS} detector'',} (2012).
  \href{http://www.arXiv.org/abs/1207.4686}{\texttt{ arXiv:1207.4686}}.
Submitted to Eur.\ Phys.\ Journal C.
%%CITATION = ARXIV:1207.4686;%%.

\bibitem{bib-sms-1}
N.~Arkani-Hamed\hrefCMSnoop {} { {et~al.}, ``{{MARMOSET}: The path from {LHC}
  data to the new standard model via on-shell effective theories}'',} (2007).
\href{http://www.arXiv.org/abs/hep-ph/0703088}{\texttt{ arXiv:hep-ph/0703088}}.
%%CITATION = HEP-PH/0703088;%%.

\bibitem{bib-sms-2}
\hrefCMSnoop {} {J.~Alwall, P.~C. Schuster, and N.~Toro, ``Simplified models
  for a first characterization of new physics at the {LHC}'',} \textit{ Phys.
  Rev. D} \textbf{ 79} (2009) 075020,
  \href{http://dx.doi.org/10.1103/PhysRevD.79.075020}{\doi{10.1103/PhysRevD.79.075020}}.

\bibitem{bib-sms-3}
J.~Alwall\hrefCMSnoop {} { {et~al.}, ``{Model-independent jets plus missing
  energy searches}'',} \textit{ Phys. Rev. D} \textbf{ 79} (2009) 015005,
  \href{http://dx.doi.org/10.1103/PhysRevD.79.015005}{\doi{10.1103/PhysRevD.79.015005}}.

\bibitem{bib-sms-4}
\hrefCMSnoop {} {D.~Alves {et~al.}, ``Simplified models for {LHC} new physics
  searches'',} (2011).
\href{http://www.arXiv.org/abs/1105.2838}{\texttt{ arXiv:1105.2838}}.
%%CITATION = arXiv:1105.2838;%%.

\bibitem{bib-nlo-nll-01}
W.~Beenakker\hrefCMSnoop {} { {et~al.}, ``Squark and gluino production at
  hadron colliders'',} \textit{ Nucl. Phys. B} \textbf{ 492} (1997) 51,
\href{http://dx.doi.org/10.1016/S0550-3213(97)00084-9}{\doi{10.1016/S0550-3213(97)00084-9}}.
%%CITATION = HEP-PH/9610490;%%.

\bibitem{bib-nlo-nll-02}
\hrefCMSnoop {} {A.~Kulesza and L.~Motyka, ``{Threshold resummation for
  squark-antisquark and gluino-pair production at the {LHC}}'',} \textit{ Phys.
  Rev. Lett.} \textbf{ 102} (2009) 111802,
\href{http://dx.doi.org/10.1103/PhysRevLett.102.111802}{\doi{10.1103/PhysRevLett.102.111802}}.
%%CITATION = ARXIV:0807.2405;%%.

\bibitem{bib-nlo-nll-03}
\hrefCMSnoop {} {A.~Kulesza and L.~Motyka, ``{Soft gluon resummation for the
  production of gluino-gluino and squark-antisquark pairs at the {LHC}}'',}
  \textit{ Phys. Rev. D} \textbf{ 80} (2009) 095004,
\href{http://dx.doi.org/10.1103/PhysRevD.80.095004}{\doi{10.1103/PhysRevD.80.095004}}.
%%CITATION = ARXIV:0905.4749;%%.

\bibitem{bib-nlo-nll-04}
W.~Beenakker\hrefCMSnoop {} { {et~al.}, ``{Soft-gluon resummation for squark
  and gluino hadroproduction}'',} \textit{ J. High Energy Phys.} \textbf{ 12}
  (2009) 041,
\href{http://dx.doi.org/10.1088/1126-6708/2009/12/041}{\doi{10.1088/1126-6708/2009/12/041}}.
%%CITATION = ARXIV:0909.4418;%%.

\bibitem{bib-nlo-nll-05}
W.~Beenakker\hrefCMSnoop {} { {et~al.}, ``{Squark and gluino
  hadroproduction}'',} \textit{ Int. J. Mod. Phys. A} \textbf{ 26} (2011) 2637,
\href{http://dx.doi.org/10.1142/S0217751X11053560}{\doi{10.1142/S0217751X11053560}}.
%%CITATION = ARXIV:1105.1110;%%.

\bibitem{bib-cms-detector}
\hrefCMSnoop {} {{ CMS} Collaboration, ``The {CMS} experiment at the {CERN}
  {LHC}'',} \textit{ Journal of Instrum.} \textbf{ 03} (2008) S08004,
  \href{http://dx.doi.org/10.1088/1748-0221/3/08/S08004}{\doi{10.1088/1748-0221/3/08/S08004}}.

\bibitem{bib-cms-pf}
\href {http://cdsweb.cern.ch/record/1194487} {{ CMS} Collaboration, ``Particle
  flow event reconstruction in {CMS} and performance for jets, taus
  and~\met'',} CMS Physics Analysis Summary CMS-PAS-PFT-09-001, (2009).

\bibitem{bib-antikt}
\hrefCMSnoop {} {M.~Cacciari, G.~P. Salam, and G.~Soyez, ``The anti-$k_t$ jet
  clustering algorithm'',} \textit{ J. High Energy Phys.} \textbf{ 04} (2008)
  063,
  \href{http://dx.doi.org/10.1088/1126-6708/2008/04/063}{\doi{10.1088/1126-6708/2008/04/063}}.

\bibitem{bib-cms-ptres}
\hrefCMSnoop {} {{ CMS} Collaboration, ``Determination of jet energy
  calibration and transverse momentum resolution in {CMS}'',} \textit{ Journal
  of Instrum.} \textbf{ 06} (2011) 11002,
  \href{http://dx.doi.org/10.1088/1748-0221/6/11/P11002}{\doi{10.1088/1748-0221/6/11/P11002}}.

\bibitem{CMS-PAS-TRK-10-005}
\href {http://cdsweb.cern.ch/record/1279383} {{ CMS} Collaboration, ``Tracking
  and primary vertex results in first 7~{TeV} collisions'',} CMS Physics
  Analysis Summary CMS-PAS-TRK-10-005, (2010).

\bibitem{CMS-PAS-EGM-10-004}
\href {http://cdsweb.cern.ch/record/1299116} {{ CMS} Collaboration, ``Electron
  reconstruction and identification at $\sqrt{s}=7$\TeV'',} CMS Physics
  Analysis Summary CMS-PAS-EGM-10-004, (2010).

\bibitem{bib-cms-muon}
\hrefCMSnoop {} {{ CMS} Collaboration, ``Performance of CMS muon reconstruction
  in pp collision events at {$\sqrt{s}=7$\TeV}'',} (2012).
  \href{http://www.arXiv.org/abs/1206.4071}{\texttt{ arXiv:1206.4071}}.
Submitted to Journal of Instrum.
%%CITATION = ARXIV:1206.4071;%%.

\bibitem{bib-cms-btagging}
\href {http://cdsweb.cern.ch/record/1427247} {{ CMS} Collaboration, ``b-jet
  identification in the {CMS} experiment'',} CMS Physics Analysis Summary
  CMS-PAS-BTV-11-004, (2012).

\bibitem{bib-madgraph}
J.~Alwall\hrefCMSnoop {} { {et~al.}, ``{M}ad{G}raph5: going beyond'',} \textit{
  J. High Energy Phys.} \textbf{ 06} (2011) 128,
\href{http://dx.doi.org/10.1007/JHEP06(2011)128}{\doi{10.1007/JHEP06(2011)128}}.
%%CITATION = ARXIV:1106.0522;%%.

\bibitem{bib-powheg}
\hrefCMSnoop {} {S.~Frixione, P.~Nason, and C.~Oleari, ``Matching {NLO} {QCD}
  computations with parton shower simulations: the {POWHEG} method'',} \textit{
  J. High Energy Phys.} \textbf{ 11} (2007) 070,
  \href{http://dx.doi.org/10.1088/1126-6708/2007/11/070}{\doi{10.1088/1126-6708/2007/11/070}}.

\bibitem{bib-pythia}
\hrefCMSnoop {} {T.~Sj{\"o}strand, S.~Mrenna, and P.~Skands, ``{PYTHIA} 6.4
  physics and manual'',} \textit{ J. High Energy Phys.} \textbf{ 05} (2006)
  026,
  \href{http://dx.doi.org/10.1088/1126-6708/2006/05/026}{\doi{10.1088/1126-6708/2006/05/026}}.

\bibitem{bib-cteq}
J.~Pumplin\hrefCMSnoop {} { {et~al.}, ``New generation of parton distributions
  with uncertainties from global {QCD} analysis'',} \textit{ J. High Energy
  Phys.} \textbf{ 07} (2002) 12,
\href{http://dx.doi.org/10.1088/1126-6708/2002/07/012}{\doi{10.1088/1126-6708/2002/07/012}}.
%%CITATION = HEP-PH 0201195;%%.

\bibitem{bib-geant}
\hrefCMSnoop {} {S.~Agostinelli {et~al.}, ``{GEANT4}---a simulation toolkit'',}
  \textit{ Nucl. Instr. and Meth. A} \textbf{ 506} (2003) 250,
  \href{http://dx.doi.org/10.1016/S0168-9002(03)01368-8}{\doi{10.1016/S0168-9002(03)01368-8}}.

\bibitem{bib-cms-ttbar}
\hrefCMSnoop {} {{ CMS} Collaboration, ``Measurement of the \ttbar production
  cross section in pp collisions at 7~TeV in lepton + jets events using b-quark
  jet identification'',} \textit{ Phys. Rev. D} \textbf{ 84} (2011) 092004,
  \href{http://dx.doi.org/10.1103/PhysRevD.84.092004}{\doi{10.1103/PhysRevD.84.092004}}.

\bibitem{pdg}
\hrefCMSnoop {} {K.~Nakamura {et~al.}, ``Review of particle physics'',}
  \textit{ J. Phys. G} \textbf{ 37} (2010) 075021,
  \href{http://dx.doi.org/10.1088/0954-3899/37/7A/075021}{\doi{10.1088/0954-3899/37/7A/075021}}.

\bibitem{bib-wpolar-1}
\hrefCMSnoop {} {A.~Czarnecki, J.~G. K\"{o}rner, and J.~H. Piclum, ``Helicity
  fractions of {W} bosons from top quark decays at {NNLO} in {QCD}'',} \textit{
  Phys. Rev. D} \textbf{ 81} (2010) 111503,
  \href{http://dx.doi.org/10.1103/PhysRevD.81.111503}{\doi{10.1103/PhysRevD.81.111503}}.

\bibitem{bib-wpolar-2}
Z.~Bern\hrefCMSnoop {} { {et~al.}, ``Left-handed {W} bosons at the {LHC}'',}
  \textit{ Phys. Rev. D} \textbf{ 84} (2011) 034008,
\href{http://dx.doi.org/10.1103/PhysRevD.84.034008}{\doi{10.1103/PhysRevD.84.034008}}.
%%CITATION = ARXIV:1103.5445;%%.

\bibitem{bib-CDF_topDecayWPol}
\hrefCMSnoop {} {{ CDF} Collaboration, ``Measurement of {W} boson polarization
  in top quark decay in p$\mathrm{\overline{p}}$ collisions at $\sqrt
  s=1.96$~TeV'',} \textit{ Phys. Rev. Lett.} \textbf{ 105} (2010) 042002,
  \href{http://dx.doi.org/10.1103/PhysRevLett.105.042002}{\doi{10.1103/PhysRevLett.105.042002}}.

\bibitem{bib-D0_topDecayWPol}
\hrefCMSnoop {} {{ D0} Collaboration, ``Measurement of the {W} boson helicity
  in top quark decays using 5.4\fbinv of p$\mathrm{\overline{p}}$ collision
  data'',} \textit{ Phys. Rev. D} \textbf{ 83} (2011) 032009,
  \href{http://dx.doi.org/10.1103/PhysRevD.83.032009}{\doi{10.1103/PhysRevD.83.032009}}.

\bibitem{bib-wpolar-3}
\hrefCMSnoop {} {{ CMS} Collaboration, ``Measurement of the polarization of {W}
  bosons with large transverse momentum in {W}+jets events at the {LHC}'',}
  \textit{ Phys. Rev. Lett.} \textbf{ 107} (2011) 021802,
\href{http://dx.doi.org/10.1103/PhysRevLett.107.021802}{\doi{10.1103/PhysRevLett.107.021802}}.
%%CITATION = ARXIV:1104.3829;%%.

\bibitem{Aad:2012ky}
\hrefCMSnoop {} {{ ATLAS} Collaboration, ``{Measurement of the {W} boson
  polarization in top quark decays with the {ATLAS} detector}'',} \textit{ J.
  High Energy Phys.} \textbf{ 06} (2012) 088,
\href{http://dx.doi.org/10.1007/JHEP06(2012)088}{\doi{10.1007/JHEP06(2012)088}}.
%%CITATION = ARXIV:1205.2484;%%.

\bibitem{bib-cms-tau}
\hrefCMSnoop {} {{ CMS} Collaboration, ``Performance of $\tau$-lepton
  reconstruction and identification in {CMS}'',} \textit{ Journal of Instrum.}
  \textbf{ 07} (2012) P01001,
  \href{http://dx.doi.org/10.1088/1748-0221/7/01/P01001}{\doi{10.1088/1748-0221/7/01/P01001}}.

\bibitem{bib-johnson-1995}
N.~L. Johnson, S.~Kotz, and N.~N.~Balakrishnan, ``Continuous Univariate
  Distributions'', volume~2.
\newblock Wiley Interscience, 1995.

\bibitem{junk1999}
\hrefCMSnoop {} {T.~Junk, ``{Confidence level computation for combining
  searches with small statistics}'',} \textit{ Nucl. Instr. and Meth. A}
  \textbf{ 434} (1999) 435,
\href{http://dx.doi.org/10.1016/S0168-9002(99)00498-2}{\doi{10.1016/S0168-9002(99)00498-2}}.
%%CITATION = HEP-EX/9902006;%%.

\bibitem{bib-cls}
\hrefCMSnoop {} {A.~L. Read, ``Presentation of search results: the {CL}$_{\rm
  s}$ technique'',} \textit{ J. Phys. G} \textbf{ 28} (2002) 2693,
  \href{http://dx.doi.org/10.1088/0954-3899/28/10/313}{\doi{10.1088/0954-3899/28/10/313}}.

\bibitem{bib-cms-fastsim}
\href {http://cdsweb.cern.ch/record/1309890} {{ CMS} Collaboration,
  ``Comparison of the fast simulation of {CMS} with the first {LHC} data'',}
  CMS Detector Performance Summary CMS-DP-2010-039, (2010).

\bibitem{PDF4LHC}
M.~Botje\hrefCMSnoop {} { {et~al.}, ``The {PDF4LHC} working group interim
  recommendations'',} (2011).
\href{http://www.arXiv.org/abs/1101.0538}{\texttt{ arXiv:1101.0538}}.
%%CITATION = arXiv:1101.0538;%%.

\bibitem{CMS-PAS-SMP-12-008}
\href {http://cdsweb.cern.ch/record/1434360} {{ {CMS}} Collaboration,
  ``Absolute calibration of the luminosity measurement at {CMS}: winter 2012
  update'',} CMS Physics Analysis Summary CMS-PAS-SMP-12-008, (2012).

\bibitem{bib-NLO-NLL}
M.~Kr\"{a}mer\hrefCMSnoop {} { {et~al.}, ``Supersymmetry production cross
  sections in pp collisions at $\sqrt{s}=7$~{TeV}'',} (2012).
\href{http://www.arXiv.org/abs/1206.2892}{\texttt{ arXiv:1206.2892}}.
%%CITATION = arXiv:1206.2892;%%.

\end{thebibliography}\endgroup

\cleardoublepage \appendix\section{The CMS Collaboration \label{app:collab}}\begin{sloppypar}\hyphenpenalty=5000\widowpenalty=500\clubpenalty=5000\textbf{Yerevan Physics Institute,  Yerevan,  Armenia}\\*[0pt]
S.~Chatrchyan, V.~Khachatryan, A.M.~Sirunyan, A.~Tumasyan
\vskip\cmsinstskip
\textbf{Institut f\"{u}r Hochenergiephysik der OeAW,  Wien,  Austria}\\*[0pt]
W.~Adam, E.~Aguilo, T.~Bergauer, M.~Dragicevic, J.~Er\"{o}, C.~Fabjan\cmsAuthorMark{1}, M.~Friedl, R.~Fr\"{u}hwirth\cmsAuthorMark{1}, V.M.~Ghete, J.~Hammer, N.~H\"{o}rmann, J.~Hrubec, M.~Jeitler\cmsAuthorMark{1}, W.~Kiesenhofer, V.~Kn\"{u}nz, M.~Krammer\cmsAuthorMark{1}, I.~Kr\"{a}tschmer, D.~Liko, I.~Mikulec, M.~Pernicka$^{\textrm{\dag}}$, B.~Rahbaran, C.~Rohringer, H.~Rohringer, R.~Sch\"{o}fbeck, J.~Strauss, A.~Taurok, W.~Waltenberger, G.~Walzel, E.~Widl, C.-E.~Wulz\cmsAuthorMark{1}
% ...
\vskip\cmsinstskip
\dag:~Deceased\\
1:~~Also at Vienna University of Technology, Vienna, Austria\\
% ...

\end{sloppypar}
\end{document}